\documentclass{aa}  

\usepackage{graphicx}
\usepackage{txfonts}
\usepackage[colorlinks=true,citecolor=cyan]{hyperref}
\usepackage[singlelinecheck=off]{subcaption}
\usepackage{gensymb}
\usepackage[switch]{lineno}

\newcommand{\abvb}{SN\,2024abvb}	
\newcommand{\kms}{km~s$^{-1}$}

\usepackage[dvipsnames]{xcolor}
\definecolor{mpcol}{rgb}{0, 0.6, 0.7}

\begin{document}

   \title{Nested, asymmetric H–He circumstellar shells in the Type Icn/Ibn SN 2024abvb}


   \author{J. P. Anderson \inst{1}
   \and C. Aster \inst{2}
   \and M. Bulla \inst{3,4,5}
   \and T.-W. Chen \inst{6}
   \and M. Fraser \inst{7}
   \and L. Galbany \inst{8,9} 
   \and C. P. Guti\'errez \inst{8,9}
   \and C. Inserra \inst{2}\fnmsep\thanks{Leading and corresponding author; email: InserraC@cardiff.ac.uk}
   \and T. Killestein \inst{10}
   \and G. Leloudas \inst{11}
   \and J. D. Lyman \inst{10}
   \and K. Maeda \inst{12}
   \and K. Maguire \inst{13}
   \and E. Mason \inst{14} 
   \and T. Moriya \inst{15}
   \and A. Pastorello \inst{16}
   \and S. Taubenberger \inst{17,18}
   \and M. Pursiainen \inst{10} 
   \and H. Wichern \inst{11}      
          \\(The INTEL Collaboration)
          }

   \institute{
   European Southern Observatory, Alonso de Córdova 3107, Vitacura, Casilla 19001, Santiago, Chile
   \and
   Cardiff Hub for Astrophysics Research and Technology, School of Physics \& Astronomy, Cardiff University, Queens Buildings, The Parade, Cardiff, CF24 3AA, UK
\and 
Department of Physics and Earth Science, University of Ferrara, via Saragat 1, I-44122 Ferrara, Italy
\and 
INFN, Sezione di Ferrara, via Saragat 1, I-44122 Ferrara, Italy 
\and 
INAF, Osservatorio Astronomico d’Abruzzo, via Mentore Maggini snc, 64100 Teramo, Italy 
        \and 
        Graduate Institute of Astronomy, National Central University, 300 Jhongda Road, 32001 Jhongli, Taiwan
        \and
        School of Physics, University College Dublin, LMI Main Building, Beech Hill Road, Dublin 4, D04 P7W1
        \and
        Institut d'Estudis Espacials de Catalunya (IEEC), Edifici RDIT, Campus UPC, 08860 Castelldefels (Barcelona), Spain
        \and
        Institute of Space Sciences (ICE, CSIC), Campus UAB, Carrer de Can Magrans, s/n, E-08193 Barcelona, Spain
        \and
        Department of Physics, University of Warwick, Gibbet Hill Road, Coventry CV4 7AL, UK
        \and
        DTU Space, Department of Space Research and Space Technology, Technical University of Denmark, Elektrovej 327, 2800 Kgs. Lyngby, Denmark
        \and
        Department of Astronomy, Kyoto University, Kitashirakawa-Oiwake-cho, Sakyo-ku, Kyoto, Kyoto 606-8502, Japan
        \and
        School of Physics, Trinity College Dublin, College Green, Dublin 2, Ireland
        \and
        INAF-OATS, Via G.B. Tiepolo 11, 34143, Trieste, Italy
        \and
        National Astronomical Observatory of Japan, 2-21-1 Osawa, Mitaka, Tokyo 181-8588, Japan
        \and
        INAF-Osservatorio Astronomico di Padova, Vicolo dell’Osservatorio 5, 35122 Padova, Italy
        \and
        Technical University of Munich, TUM School of Natural Sciences,
Physics Department, James-Franck-Str. 1, 85741 Garching, Germany
\and
        Max-Planck-Institut für Astrophysik, Karl-Schwarzschild Str. 1, D85741 Garching, Germany
             }

   \date{Received January XX, YYYY; accepted March ZZ, WWWW}

 \authorrunning{The INTEL Collaboration}
 
  \abstract
   {Interacting transients probe mass loss in the final stages of stellar evolution; however, the geometry and timing of multi-episode mass loss remain poorly constrained. SN\,2024abvb is a nearby interacting event with transitional Ibn/Icn spectroscopic properties and multi-epoch polarimetry, offering a rare opportunity to study structured circumstellar material (CSM). }
   {We aim to characterise the kinematics, composition and geometry of the CSM around SN\,2024abvb and to identify plausible progenitor/ejection scenarios that can produce the observed spectro-polarimetric evolution.}
   {We present high-resolution (VLT/UVES and VLT/X-Shooter) optical/NIR spectroscopy across several epochs, complemented by broadband polarimetry and spectropolarimetry (VLT/FORS2 and NOT/ALFOSC). Line identifications, velocity decompositions and polarimetric time-series are used to trace multiple kinematic components and changes in scattering geometry.}
   {The high-resolution spectra reveal multiple narrow CSM components composed of He, C and O with absorption minima at $\sim150 - 400$ \kms\/ and additional faster material up to $\sim2000$ \kms. Low-velocity Balmer absorptions are present, indicating distant H-rich material, a first in SNe Ibn/Icn. Polarimetry shows a marked evolution ($P\sim1\%$ near peak, $\lesssim0.5\%$ after $\sim1$ week, rising to $\sim1.5\%$ at $\sim20$\,d with $\sim50^\circ$ position-angle rotation and to $\sim4\%$ at $\sim30$\,d, stronger in the blue), implying a time-variable, wavelength-dependent scattering/obscuration component. The combination of kinematics and polarimetric behaviour is consistent with multiple, concentric toroidal shells with differing orientations and partial dust content.
}
   {}


   \maketitle
%

\section{Introduction}


Understanding the evolution and the ultimate fate of stars remains one of the key questions in astrophysics. The INteracting Transients at ESO-vLt (INTEL) program aims to investigate the role of circumstellar medium (CSM) interaction in various species of stellar transients.
Targets of the INTEL program are stellar transients with observational evidence of slowly expanding CSM, such as tidal-disruption events \citep[TDEs; e.g.,][]{Gezari_2021}, interacting gap transients \citep{Pasto_2019}, interacting SNe \citep{Smith_2014,Fraser_2020}, SNe interacting with a confined CSM \citep[also known as `flash features', e.g.][]{2023ApJ...952..119B, 2024ApJ...970..189J} and super-luminous SNe \citep[SLSNe;][]{GalYam_2019,Inserra_2019}. 
This program uses the Very Large Telescope (VLT) Ultraviolet and Visual Echelle Spectrograph \citep[UVES,][]{UVES} and the X-Shooter spectrograph \citep{XShooter} for high-resolution spectroscopy, while the FOcal Reducer/low dispersion Spectrograph 2  \citep[FORS2,][]{Appenzeller1998} for polarimetric data. This science programme operates in parallel with low-resolution spectroscopy efforts and is intended to complement and enhance them, with a distinct and differing scientific focus.   

SNe with narrow-lined spectra are ideal targets for the INTEL facilities, as a good spectral resolution is necessary to precisely infer the velocity, composition and ionisation state of the stellar wind, while spectropolarimetry is required to study the asymmetries in the distribution of the expanding material. SNe with spectra showing prominent narrow Balmer lines in emission belong to the Type IIn class \citep{Schlegel_1990}. Sometimes the narrow lines sit atop of broader components, suggesting that the different line components likely arise from different kinematic regions, but all rich in hydrogen (H). 

In the past few years, other subclasses of ejecta-CSM interacting SNe have emerged, showing H-free, narrow-lined spectra. Type Ibn SNe have spectra dominated by relatively narrow helium (He) features produced in a H-deprived, He-rich CSM \citep{Pasto_2008a}. Type Icn SNe show totally H- and He-free spectra, while narrow lines of carbon, nitrogen and oxygen (C/N/O) are prominent \citep{Pasto_2021,GalYam_2021,GalYam_2022,Pellegrino_2022}. The newly classified Ien SN shows spectra with narrow emission lines of Si and S \citep{2025Natur.644..634S}. The composite spectral classification scheme of interacting SNe indicates they are produced by progenitor stars that are progressively more stripped throughout the IIn-Ibn-Icn-Ien sequence.
Despite observational studies revealing a wide observational heterogeneity among the different types of interacting SNe, only limited progress has been made in our understanding of their explosion mechanisms and the properties of their progenitors \citep[e.g.,][]{Metzger_2022, Inoue_2025}.

To complicate the overall picture, SNe with intermediate properties between two different types are occasionally observed, including the transitional SNe IIn/Ibn \citep[whose spectra show narrow H and He~I with comparable strengths;][]{Pasto_2008b,2015MNRAS.449.1921P,Smith_2012, Reguitti_2022,Farias_2024,Gangopadhyay_2025} and SNe Ibn/Icn \citep[with spectra showing at the same time He~I and C/N/O lines; e.g.,][]{Pursiainen_2023a}. The recent discovery of the nearby SN~2024abvb provides a rare opportunity to study in detail a SN with transitioning properties between those of a Type Icn and a Type Ibn SN.

\section{Observations}
\begin{figure}
\includegraphics[width=\columnwidth]{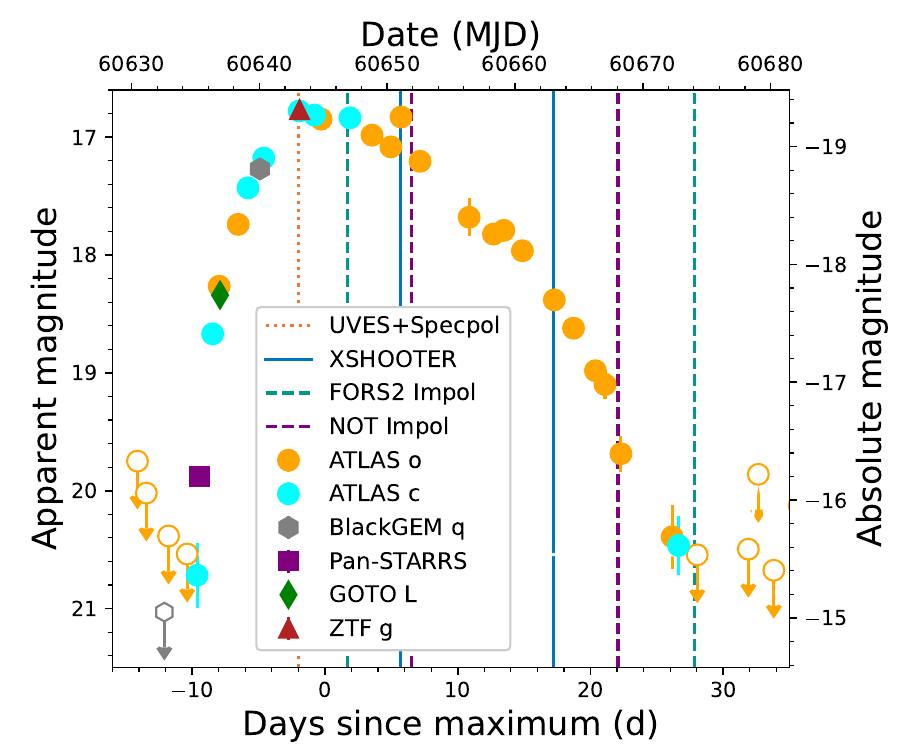}
\caption{ATLAS, Pan-STARRS and public ZTF, BlackGEM and GOTO photometric data, with uncertainties, of \abvb. Open symbols denote limits. Phase is with respect to the maximum light. INTEL data and ancillary polarimetric data are shown with vertical lines. Note that a further XShooter spectrum was taken at $\sim 55$d post maximum when the SN was no longer visible in the optical.}
\label{fig:lc}
\end{figure}

\abvb\/ \citep{2026MNRAS.tmp..363A,2026arXiv260101333H, 2026arXiv260216227S} was discovered by the Asteroid Terrestrial-impact Last Alert System \citep[ATLAS,][]{2020PASP..132h5002S} on 2024 November 22 \citep{24abvb_discovery} and subsequently classified by the Nordic-optical-telescope Un-biased Transient Survey \citep[NUTS,][]{2016ATel.8992....1M} on 2024 November 28 as a Type Icn SN \citep{24abvb_classification} at z~=~0.039 \citep[see][]{2026MNRAS.tmp..363A}, equivalent to a luminosity distance of 164.8 Mpc using the following cosmology measurements $H_0=73.04$ km s$^{-1}$ Mpc$^{-1}$, $\Omega_M=0.326$ and a flat Universe \citep[][]{2022ApJ...934L...7R}. The Galactic reddening toward the SN position is E(B-V)~=~0.164 mag \citep{2011ApJ...737..103S}. None of our high-resolution spectra show Na~{\sc id} lines related to host galaxy reddening, hence we assume the Galactic reddening as the total. This is also consistent with the remote location of \abvb\/ (See Section~\ref{ss:host}). The light curve of \abvb\/ from publicly available ATLAS \citep{2018PASP..130f4505T}, Pan-STARRS \citep{chambers2019panstarrs1surveys}, BlackGEM \citep{2024PASP..136k5003G}, GOTO \citep{Steeghs22} and ZTF \citep{2019PASP..131a8002B} surveys data on the Transient Name Server (TNS\footnote{\url{https://www.wis-tns.org/object/2024abvb})}) is shown in Fig.~\ref{fig:lc} with respect to the maximum light \citep[MJD~60645.13,][and references therein]{2026MNRAS.tmp..363A} to guide the spectroscopic temporal analysis.

\subsection{Data}

Our first spectroscopic observation of \abvb\ was obtained on 2024 Nov. 29 UT at the VLT+UVES. The instrument setup was DIC1$\lambda$346+580, covering the wavelength range $\sim$3050-3859 \AA\ in the blue, and 4789-6788 \AA\footnote{The red setup has a gap between 5748-5849 \AA\ because of the CCD mosaic.} in the red arm. The detectors were unbinned, but the resolution, R$\sim$45000, was dictated by the slit width set to 1\arcsec\ in both arms. The total integration time was 8100s split across 9 exposures. The data were reduced using the ESO Recipe Execution Tool ({\it EsoRex}) UVES pipeline, v~6.1.3, together with {\it Gasgano}, v~2.4.8, and adopting the archive-delivered master calibrations and optimal extraction. The data were flux calibrated using the master response function (obtained from a combination of different spectrophotometric standard star spectra observed in photometric conditions) provided by the observatory. Each exposure was individually reduced and then combined into a sigma-clipped average spectrum.

Three epochs of medium resolution spectroscopy were also obtained using VLT/XShooter, as listed in Tab.~\ref{table:specdata}. These data were reduced in the standard fashion using the XShooter pipeline v.~3.6.1 within the ESO Common Pipeline Library running under the EsoRex tool. Night-specific processed calibration files were downloaded from the ESO archive using the CalSelector service. The NIR arm data were corrected for telluric absorption using the {\sc molecfit} routines within {\it EsoRex}; we do not correct the VIS arm spectra as the pipeline was found to over-correct the absorption, but rather these regions are marked in figures. Unfortunately, the $\phi = 5.7$d (MJD~60651.02) spectra were obtained using a mixture of 1$\times$1 and 1$\times$2 binning. As the 1$\times$2 binning data had significantly higher S/N than the 1$\times$1 binning data, we only used the 1$\times$2 binning data in our analysis. Moreover, the MJD~60701.03, 60703.03 and 60705.03 (median $\phi= 55.7$d) spectra were combined and show the SN signal in the NIR arm only.

We obtained one epoch of spectropolarimetry and two epochs of multi-band imaging polarimetry with VLT/FORS2 and complemented by two $V$-band epochs taken with Alhambra Faint Object Spectrograph and Camera (ALFOSC), mounted on the Nordic Optical Telescope (NOT).
The spectropolarimetry was reduced with a custom pipeline presented in \citet{Wichern2025}. The pipeline optimally extracts the spectra of the ordinary and extraordinary beams within regions of equal size, in this case of widths equal to four times the FWHM of the ordinary beam. Three sets of spectra were obtained for each half-wave plate angle; these were combined by computing a variance-weighted average of the three sets. The spectra were binned by 25\,Å to increase the signal-to-noise.
The broadband polarimetry data of both FORS2 and ALFOSC were reduced using a \texttt{photutils}~\citep{Bradley2024} based pipeline, using an annulus background. We followed the reduction steps for ALFOSC detailed in \citet{Pursiainen2023} using large apertures ($r=2\times\mathrm{FWHM}$). This is because the imaging polarimetry mode of ALFOSC
tends to elongate the point-spread function of the sources \citep[see e.g.][for details]{Pursiainen2023}.  For FORS2 images, we used apertures of $r=1\times\mathrm{FWHM}$. The FORS2 data were also corrected for the small chromatic rotation using the values provided on the instrument website\footnote{\url{www.eso.org/sci/facilities/paranal/instruments/fors}}, and for the instrumental polarisation affecting imaging polarimetry using the polarisation maps derived in \citet{Gonzalez-Gaitan2020}.

\begin{figure}
    \centering
    \includegraphics[width=\linewidth]{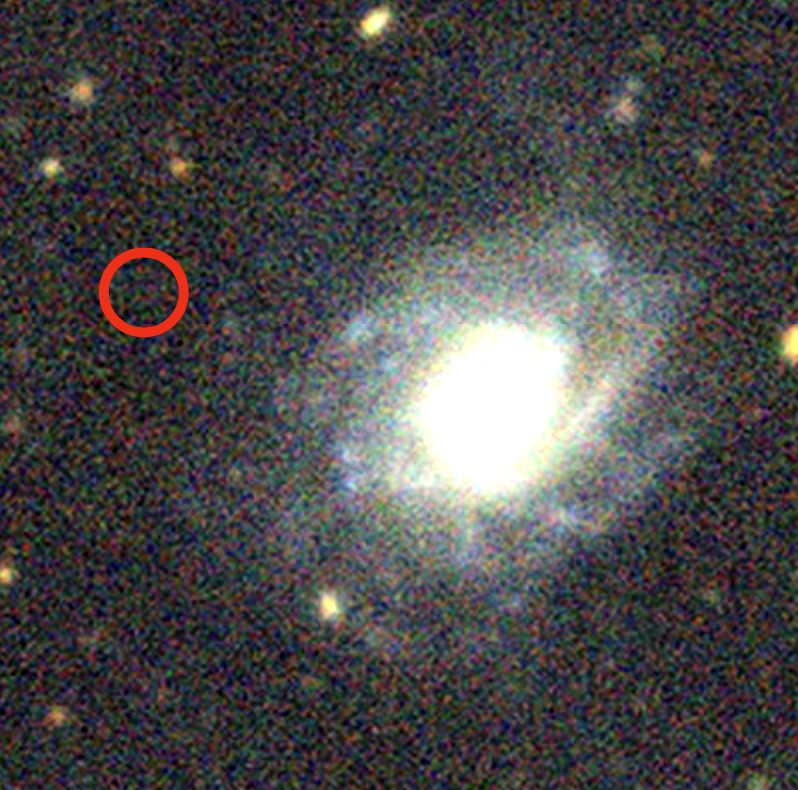}
    \caption{Cutout from the Legacy Surveys covering the position of \abvb\ (in the centre of the red circle) and its host galaxy. The image is oriented North up, East to the left, and covers approximately 1\arcmin\/ on each side.}
    \label{fig:legacy}
\end{figure}

\begin{figure}
    \centering
    \includegraphics[width=\linewidth]{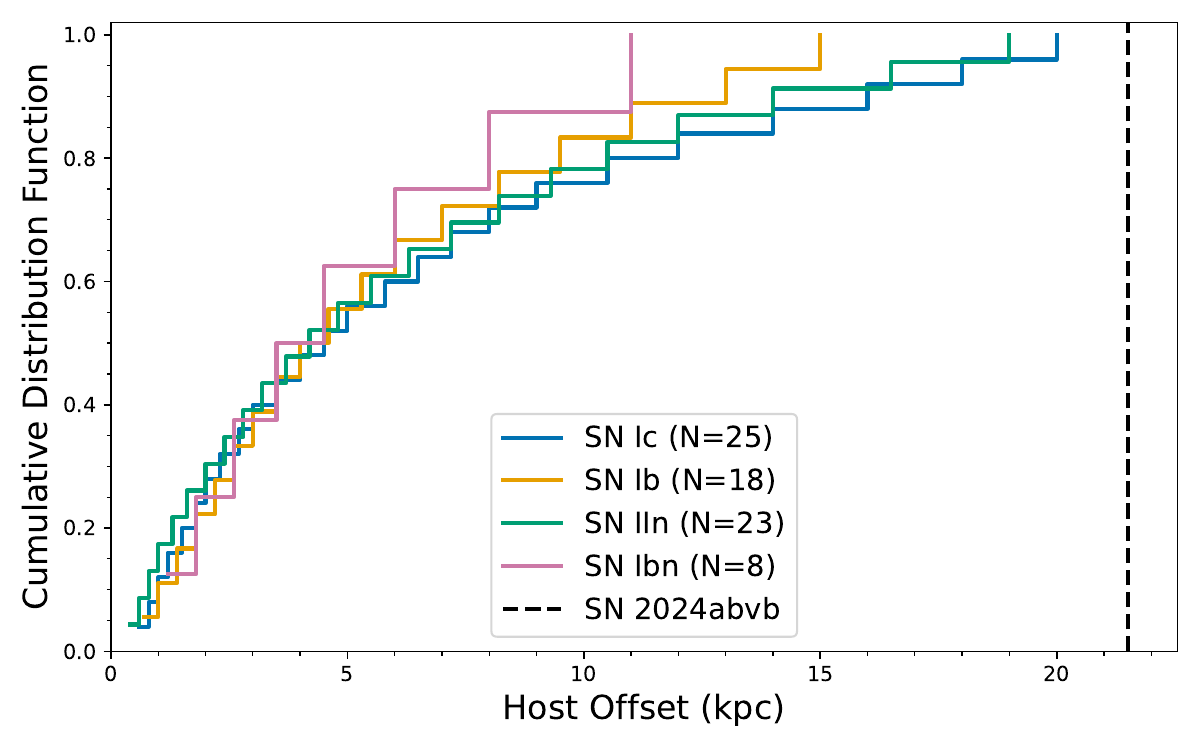}
    \caption{Cumulative Distribution Function of the observed offsets from the host galaxy nucleus for different SN types \citep{2021ApJS..255...29S}. The location of \abvb\ lies far beyond the typical offset seen for SNe Ibn.}
    \label{fig:offset}
\end{figure}

\begin{figure*}
\centering
\includegraphics[width=\textwidth]{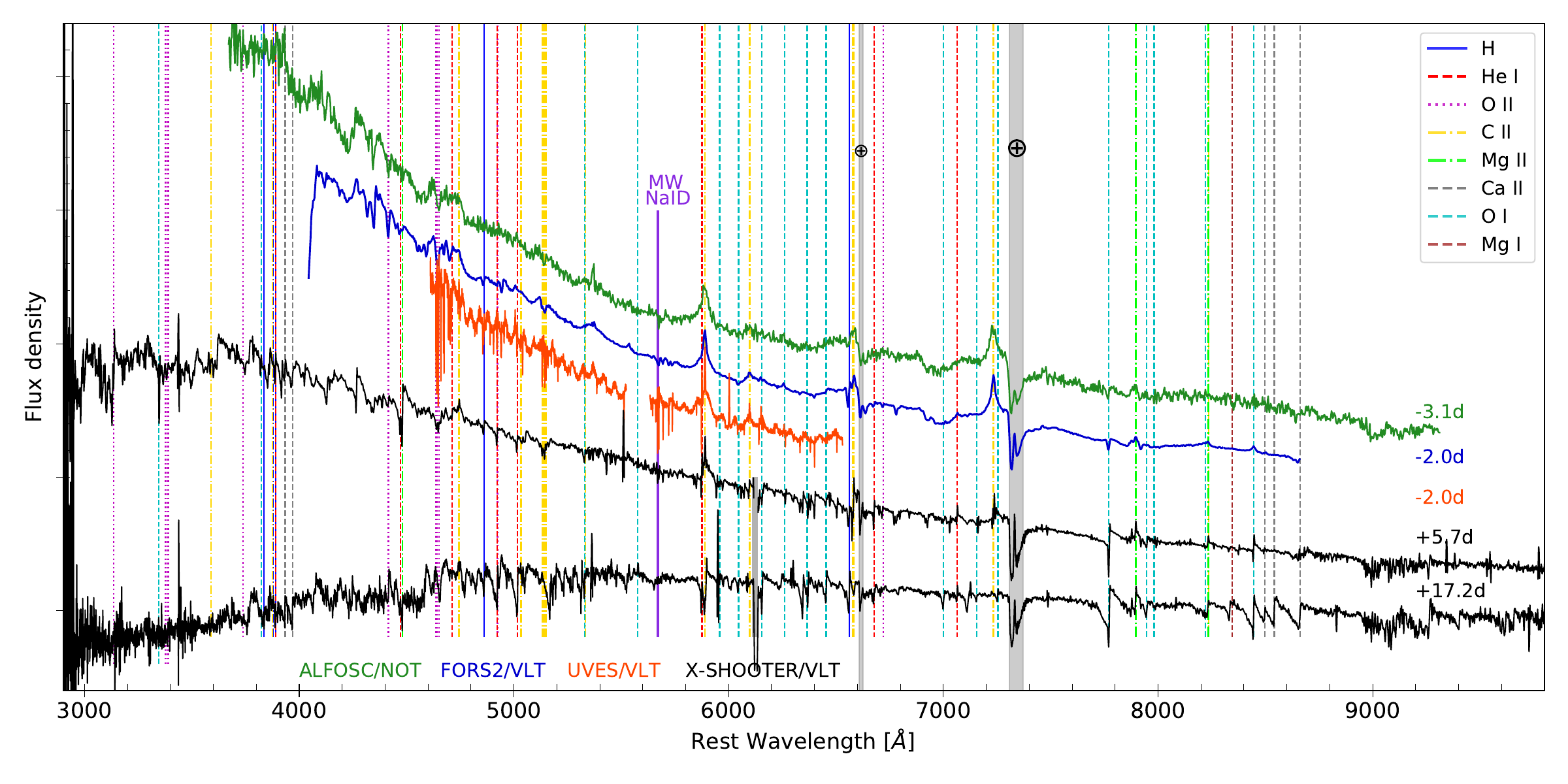}
\caption{Spectral sequence for \abvb. The spectra have been corrected for redshift and Milky Way reddening, and are vertically offset by arbitrary amounts for clarity. The colour of each spectrum indicates the instrument used. Vertical dashed lines mark the rest wavelengths of the strongest spectral features. Caution is advised when interpreting the UVES spectrum, as many apparent features are due to the echelle order pattern.
}
    \label{fig:spec}
\end{figure*}

\subsection{The host and environment of SN~2024abvb}\label{ss:host}

The host galaxy of \abvb, WISEA J011055.76-054416.6, has a magnitude of $g=17.3$, and its nucleus lies 28\arcsec\ from the location of the SN. It appears to be a fairly normal spiral galaxy, close to face-on to our line of sight. The galaxy has the same redshift of the SN ($z=0.039$) as shown by the low-resolution host galaxy spectrum \citep[see][]{2026MNRAS.tmp..363A}, and hence has an absolute magnitude of $g=-18.9$, while the projected separation from the SN is 21.5\,kpc in the plane of sky alone.

Taking the r-band radius of the host from a measurement on Legacy Survey data \citep{Dey19} using the Tractor code \citep{2016ascl.soft04008L}, this corresponds to a separation of 2.4 times the half-light radius $r_e$ of this galaxy.
We see no signs of flux, galaxy emission or cooling filaments at the position of the SN in the same deep Legacy Survey images (Fig. \ref{fig:legacy}). Taking the g-band limit of $\sim25$ from the Legacy Surveys catalogue, the absolute magnitude of any cluster or complex of stars at the location of \abvb\ must be fainter than $g=-11.0$ mag, so roughly 5~mag fainter than a Small Magellanic Cloud-like satellite galaxy. 

However, we cannot exclude the possibility that \abvb\/ is located within a faint tidal stream, as observed for other SNe \citep[e.g.][]{2012MNRAS.420.1135S,2017A&A...606A.111F}. Such streams can reach integrated luminosities as low as $M_V = -6.2$~mag \citep[the Indus system;][]{2018ApJ...862..114S}. Their detectability, however, is primarily governed by surface brightness ($\mu$), as their light is distributed over large areas. The Dark Energy Survey stream catalogue \citep{2018ApJ...862..114S} provides several examples of streams with surface brightnesses as faint as $\mu \approx 29\text{--}30$~mag~arcsec$^{-2}$ \citep{2024A&A...691A.196M}. In contrast, typical galaxies have $\mu_{\rm band} \approx 20\text{--}22$~mag~arcsec$^{-2}$, implying a difference of $\Delta \mu \approx 6\text{--}8$~mag~arcsec$^{-2}$, corresponding to a dimming factor exceeding $\sim 600$ \citep{2022A&A...662A.124S}. Following the approach described by \citet{2024MNRAS.528.4289W}, we measured the surface-brightness limit at the SN position using Legacy Survey images, obtaining $\mu_g = 30.19 \pm 0.37$~mag~arcsec$^{-2}$. For comparison, the host galaxy of \abvb\ has a mean surface brightness of $\mu_g = 22.34 \pm 0.04$~mag~arcsec$^{-2}$. Therefore, if \abvb\/ is indeed located in a tidal stream, such a structure would lie at the very faint end of the known population, or possibly below the surface-brightness regime of streams detected to date. 

We compare the projected host-galaxy offsets of \abvb\/ with the distributions of SNe Ib, Ic, IIn, and Ibn from the PTF sample \citep{2021ApJS..255...29S}. Fig.~\ref{fig:offset} shows the cumulative distributions, demonstrating that our event lies in the extreme tail of the offset distribution, particularly for SNe Ibn. We note that there are insufficient data to perform a similar analysis for SNe Icn. \citet{2021ApJS..255...29S} report that the majority of SNe (92\%) fall within the 80\% light radius ($r_{80}$), which, together with galaxy mass, traces the stellar mass content of galaxies independently of type and star formation activity \citep{2019ApJ...872L..14M,2019ApJ...872L..13M}. A first-order approximation for spiral galaxies gives $r_e / r_{80} \approx 1$ \citep{2001AJ....121..820G}, implying that \abvb\ is indeed located in an extreme region. \citet{2026MNRAS.tmp..363A} report a host-galaxy mass of $\log(M/M_{\odot})=10.33^{+0.11}_{-0.10}$ for \abvb, that combined with the SN position, it implies that the event lies outside the $r_{80}$ radius.

Interestingly, there is another galaxy in the field (WISEA J011040.80-054527.5) that lies 4.4\arcmin~from \abvb\/ (200 kpc), and is also at $z=0.039$. While the physical separation to \abvb\/ is too large for a direct association, the consistent redshift may suggest that it is part of a galaxy cluster. This is intriguing, particularly in light of the environment of a similar interacting SN far from the host, PS1-12sk, where the elliptical host galaxy was also in a galaxy cluster \citep{Sanders13}. In the absence of redshift information for other galaxies in the field, we cannot draw firm conclusions on the existence of any possible cluster. We also note that the host of \abvb\/, a spiral, seems younger than that of PS1-12sk.

\section{Spectral analysis}

\subsection{UV and Optical spectra}

To identify lines in our UVES and XShooter data, we used the line lists in \protect\cite{Moore45} as a reference. Particular care was taken to ensure that all strong components of a given multiplet were seen, supporting the identification. The full spectra are plotted in Appendix~\ref{appendix:uves} with detailed line identification, while the spectral sequence, together with the public ALFOSC classification spectrum, is shown in Fig.~\ref{fig:spec}.

\subsubsection{Pre peak UVES and FORS2 spectra}
Our FORS2 and UVES spectra were taken at $-2$ days from maximum light. While the former covers a wider wavelength range, it does not allow us to resolve the narrow features seen in the UVES spectrum of \abvb ~(Figs. \ref{fig:uves1} and \ref{fig:uves2}). 
They both show a hot blue continuum. We cannot accurately measure the blackbody temperature as in the subsequent XShooter epochs because the peak of the first spectrum is to the blue of 3000 \AA, while in the second epoch, we are unable to fit both the blue and red parts of the spectrum, possibly due to the effect of absorption by metals.  
Most of the features bluer than 4500 \AA\/ observed in the FORS2 spectrum and the ALFOSC classification spectrum can be attributed to O~{\sc ii} and Ne~{\sc ii}, while no lines corresponding to higher ionisation states of carbon are detected, unlike what is observed in SNe~Icn \citep[e.g.][]{Pellegrino_2022,2023A&A...673A..27N}.

The two strongest emission features in the FORS2 spectrum, also visible in the ALFOSC classification spectrum (Fig. \ref{fig:spec}), are at 5890 \AA\/ and 7236 \AA, and we associate these with C~{\sc ii}. While the lines exhibit a broad base (more evident in the UVES data; see, e.g. in Fig. \ref{fig:uves2}), we do not attribute this to ejecta. The wings of the broad feature extend out to 2000 \kms, and these likely arise from electron scattering in the CSM~\citep[e.g.][]{2001MNRAS.326.1448C,2014ApJ...797..118F}. 
This implies an electron temperature of $T_{e}\sim10^4$ K and an electron-scattering optical depth of order unity. This requires a compact, highly ionised CSM with electron densities of $n_e \sim 10^9$–$10^{10}$ cm$^{-3}$ at radii $R\sim10^{14}$–$10^{15}$ cm, roughly matching what inferred from the bolometric luminosity \citep{2026MNRAS.tmp..363A,2026arXiv260101333H}. Such conditions are inconsistent with a steady stellar wind and instead point to eruptive or binary-driven mass-loss episodes prior to explosion. 

We also observe a line profile around 6578 \AA\/ where another C~{\sc ii} line, expected to have a strength comparable to that of $\lambda$7236 due to its role in the recombination cascade, should be present. In addition, the C~{\sc ii} $\lambda$4267 line, which shares the same lower energy level as $\lambda$5890 but has an Einstein coefficient for spontaneous emission ($A_{ki}$) approximately an order of magnitude higher, is also detected in the FORS2 and ALFOSC spectra, albeit with a P~Cygni profile (assuming that the prominent blue-shifted absorption is indeed associated with C~{\sc ii}).

In the UVES spectrum, turning to the other lines, we see multiplets 15 and 16 of C~{\sc ii} at wavelengths $5640-5662~\AA$ and $5133-5151~\AA$, respectively. Although these are at relatively low S/N, we measure a velocity from the P-Cygni absorption component minima of these lines of $\sim$100~\kms. We also see strong lines of O~{\sc ii}, including multiplets 1, 9, 24, 25, 28, and 32. Multiplet 1 is particularly strong, with numerous lines between 4639 and 4696~\AA.  Al~{\sc ii} appears to be present with multiplets 1 and 2.

Besides the weak P-Cygni absorptions, two lines in emission stand out at 5889~\AA\ and 6003~\AA. The bluer line is also shown in the classification spectrum and, as mentioned above, it is consistent with C~{\sc ii}. However, due to the different profile displayed by this line with respect to other C~{\sc ii} lines, we cannot rule out that such a profile belongs to Na~{\sc iD}. The fact that only the emission component is observed would suggest that it arises in a Na-rich CSM shell, similarly to what is observed in the type IIn SN~2011A \citep{2015ApJ...807...63D}, although in that case the Na~{\sc iD} displayed both an emission and blue-shifted absorption. With respect to the 6003\AA\/ line, we also checked for diffuse interstellar bands (DIBs) using the catalogue of \cite{Jenniskens94} with updates as listed online\footnote{\url{https://leonid.arc.nasa.gov/DIBcatalog.html}}, but find no convincing matches. Furthermore, no absorption resonance lines such as Na~{\sc iD} and Ca~H\&K have been observed in any of the spectra, suggesting that none of the observed lines could be associated with the ISM. 

Looking in detail at our high resolution UVES spectrum, multiple He~{\sc i} lines are seen, most clearly the strong $\lambda$5875 line which has a pronounced P-Cygni profile (see Fig. \ref{fig:uves2}), with a velocity minimum at 120~\kms, and a blue edge extending out to $\sim$190~\kms, broadly consistent with the C~{\sc ii}. Such a line is comparable in strength to O~{\sc i} lines, especially those of multiplet 1.  Other He lines are visible, including $\lambda$5015 and $\lambda$4921 (Fig. \ref{fig:uves1}), but these lines are weaker and only the absorption component is observed. This is consistent with what is observed for He~{\sc i~}$\lambda$5875, where the absorption is stronger than the emission component. The presence of narrow He lines, identifiable only thanks to our high resolution, would imply a Type Ibn classification for \abvb, although we note that the weakness of these features could also lead to labelling it as a transitional Ibn/Icn SN (analogous to the transitional Ibn/IIn SNe; \citealp{2015MNRAS.449.1921P}). The possibility of hidden He in Icn SNe was briefly discussed in \citet{GalYam_2022} (Extended Data Figure 10). This suggests that an empirical classification based on low-resolution spectra may be unreliable, especially when interaction with a CSM is present and could have broader implications for the physical interpretation and classification of Icn SNe as a whole.

\subsubsection{XShooter post peak spectra}
Our first XShooter spectrum was taken roughly a week after the UVES and FORS2 ones, corresponding to a post maximum phase ($\phi = 5.7$d; Fig. \ref{fig:XShooter1}), while the second ($\phi = 17.2$d; Fig. \ref{fig:XShooter2}) half-way through the observed decline in the optical light curve (see Fig.~\ref{fig:lc}). Numerous He~{\sc i} lines are detected across the spectra. All He~{\sc i} features exhibit clear P~Cygni profiles with a velocity minimum of 120~\kms, consistent with our UVES measurements. For the high-level transitions, only the absorption components are observed. Among the observed He~{\sc i} lines, only $\lambda$3187 and $\lambda$4387 lack any emission component.

C~{\sc ii} lines are also prominently detected, with all expected strong transitions clearly visible. These lines, like He~{\sc i}, display P~Cygni profiles with a velocity of roughly 120~\kms\/ as measured from the absorption component. In contrast, there is no evidence for the presence of He~{\sc ii}, C~\textsc{iv}, N~{\sc ii} or  N~{\sc i} lines in the spectra.
The presence of C~{\sc iii} and O~{\sc iii} remains uncertain; if present, these features are extremely weak and appear only in absorption. Nevertheless, these tentative identifications, along with the non-detection of higher ionisation lines, place constraints on the ionising radiation field, limiting it to a maximum of approximately 35–40 eV, with the bulk of ionisation radiation likely peaking in the 10–25 eV range \citep{Moore45}. O~{\sc i} lines are also detected and are mostly attributable to recombination processes. 
[O~{\sc i}] $\lambda6300$ is detected in emission at $\phi = 17.2$d with a velocity width of the order of $\sim100~$\kms, suggesting a CSM origin. 

Among other lines, Si~{\sc ii} is identified, with multiplets 1, 2, and 3 being present in the spectra. Multiplet 4 is absent, and the strongest line of multiplet 5, if present, appears extremely weak. Mg~{\sc ii} is observed, particularly in multiplets 4, 6, and possibly 8, which seem to appear solely in emission. Conversely, multiplets 2 and 10 are undetected, which is consistent with their known atomic parameters. Lines at 5696.6 and 5722.7\,\AA\ could be tentatively identified as Al~{\sc iii} (multiplet 2). Al~{\sc ii} (multiplets 1 and 2) is still detected as in the UVES spectrum. This identification excludes the possibility that the feature at 5696.0\,\AA\ is due to C~{\sc iii} (singlet transition). However, the line at 4325.7\,\AA\ remains unidentified, with multiple possible line assignments requiring further investigation. Notably, as in the FORS2 and classification spectra, none of the C~{\sc iii} triplet transitions are detected. Fe~{\sc ii} lines are observed prominently in the XShooter spectra, particularly from multiplet 42, as well as multiplets 49, 48, and 37. These lines show P~Cygni profiles, at times complex due to proximity with other lines, displaying an absorption minimum velocity of 100~\kms. By the time of the second XShooter spectrum ($\phi = 17.2$d; Fig. \ref{fig:XShooter1}), Fe~{\sc ii} features have significantly strengthened, suggesting the possible formation of an ``iron blanket'' in the near-ultraviolet region ($\leq 3500$\AA) similar to the usual pseudo-continuum formed by Fe lines in SNe IIn \citep[e.g.][and references therein]{2017hsn..book..403S}.
In addition to the emission lines at 5889~\AA\ and 6003~\AA\ observed in the UVES spectrum, another emission line at 8235~\AA\ is identified in the XShooter spectra.
For this feature, potential associations include C~{\sc ii}, Mg~{\sc ii} and O~{\sc i}. However, since its profile differs from those of other C~{\sc ii}, Mg~{\sc ii} and O~{\sc i} lines, the precise nature remains uncertain. This feature may represent a blend of contributions from all three species.

The ionisation potentials for Mg~{\sc ii} and Si~{\sc ii}, assuming the observed lines arise from recombination, are approximately 15 and 16 eV, respectively. Their detection thus broadens the inferred ionisation range while slightly lowering the average ionising energy with respect to the analysis of the non-metal elements. This reinforces the conclusion that the source is dominated by relatively soft UV radiation.

\begin{figure*}
    \centering
    \includegraphics[width=\textwidth]{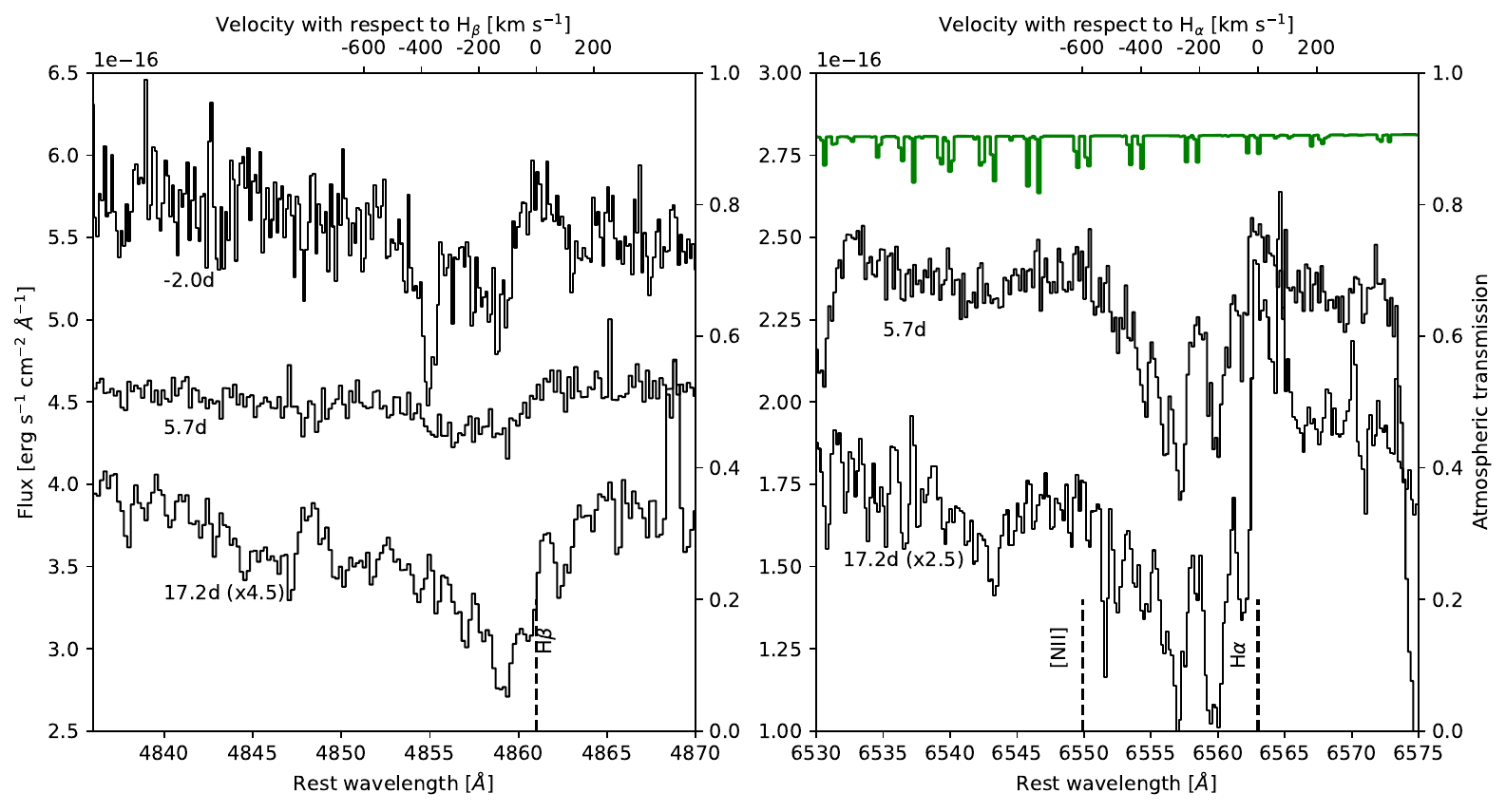}
    \caption{Regions of our spectra covering H$\beta$ (left panel) and H$\alpha$ (right panel). In both cases, we see time evolving narrow absorptions with velocity between 100 and 400 \kms\/ with respect to the rest wavelength of the line. In the right panel, we also plot the atmospheric transmission, demonstrating that these lines are not associated with any telluric features.}
    \label{fig:Halpha}
\end{figure*}

\subsection{NIR spectra}

Our NIR spectra, shown in Fig. \ref{fig:nirspec}, have a lower S/N than the optical spectra, but several lines are visible, especially in the first spectrum ($\phi = 5.7$ d). He~{\sc i} $\lambda 1.083~\mu$m, displaying a P-Cygni profile, is observed at velocities comparable to those of the optical lines. However, the profile is complex, as a line at the location of Pa$\gamma$ is also detected. Pa$\beta$ is also tentatively identified, but Pa$\alpha$ falls in a region of strong telluric absorption and therefore cannot be observed. O~{\sc i} $\lambda1.129~\mu$m is also visible with a P-Cygni profile, as well as Mg~{\sc i} $\lambda1.18~\mu$m. The presence of both the NIR O~{\sc i} line and the O~{\sc i} $\lambda7774$ line suggests that the emission arises from normal recombination. 

Intriguingly, we detect an emission line at 1.933~$\mu$m  at $\phi = 5.7$d and $\phi = 17.2$d with a full width at half maximum corresponding to a velocity of approximately 400 \kms. This feature could be identified as [Ni~{\sc ii}], blueshifted by about 900 \kms, or alternatively as a blend of [N~{\sc ii}] and [Fe~{\sc ii}]. Emission from nickel arises from the production of stable isotopes and has been observed in a handful of SNe \citep[e.g.][]{2015MNRAS.448.2482J,2018A&A...619A.102D,2021ApJ...922..186H}. However, in our case, the observed blueshift and relatively low velocity would instead suggest a CSM origin, which is difficult to explain, as it would require extremely high temperatures ($T \gtrsim 10^9$ K), high density ($\rho \gtrsim 10^7$ g~$cm^{-3}$) and conditions close to nuclear statistical equilibrium. Such conditions are typically achieved only in the inner layers of SN progenitors or during the SN explosion itself \citep{2022A&A...660A..96B}. Unfortunately, the presence of [Ni~{\sc ii}] cannot be independently confirmed, as the other strong stable nickel line at 7378~\AA\/ falls within a region strongly affected by telluric absorption.

The second NIR spectrum shows similar lines and velocities, while the third spectrum displays residual flux in the 1.18–1.20~$\mu$m region; however, due to the low S/N, no conclusive analysis of line identification or velocities can be performed.

Our dataset does not show any clear evidence of dust formation as the flux redward of 1.90~$\mu$m remains constant during the three available epochs. Moreover, it does not cover the wavelengths ($\lambda > 2.20 \mu$m) where the formation of CO molecules due to newly formed dust can be observed \citep[e.g.,][]{Gerardy2000}. Nevertheless, dust formation cannot be excluded as it has been observed in other interacting SNe starting at a similar evolutionary phase as \abvb\/ \citep[e.g.][]{Mattila2008}.

\subsection{Hydrogen in the spectra of \abvb}

Remarkably, we see structured, multi-component low velocity absorption features in both the UVES and XShooter data, which we associate with the Balmer series. This strengthens our Paschen series detection. This feature can be seen most clearly in H$\alpha$, where we see multiple absorptions with velocity minima between 150 and 400 \kms\ in both the +5.7d and +17.2d spectra (Fig. \ref{fig:Halpha}). While there are some small differences between these two epochs in terms of the relative strengths of individual components, the overall line structure persists. We also see a weak, narrow emission feature at the rest wavelength of H$\alpha$, likely to be the emission component of a possible P-Cygni profile with absorption minima at 50 \kms. This appears to grow in strength in the +17.2d spectrum. We used the ESO SkyCalc software \citep{Noll12} to model the sky transmission in the region of H$\alpha$ and verify that there are no strong telluric absorptions that can account for this feature.
A similar complex of low velocity absorptions between 150 and 400 \kms\/ can also be seen in H$\beta$. Our UVES spectrum covers this spectral region, and we clearly see the complex at -2.0 days. The XShooter data is less clear in the region of H$\beta$, although there is still clearly an absorption present.

\subsection{Spectral Evolution}

\begin{figure}
\centering
\includegraphics[width=\columnwidth]{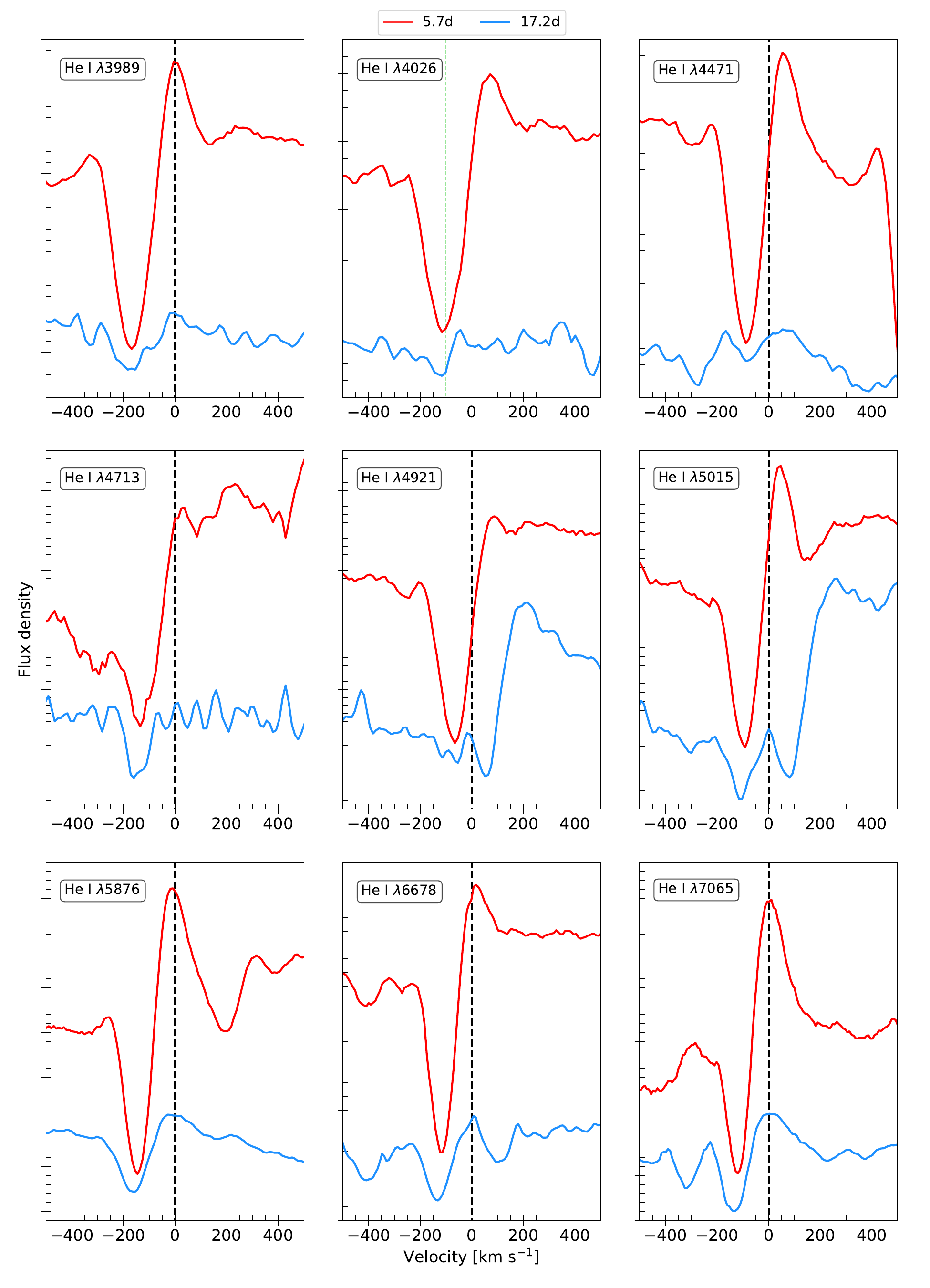}
\caption{He~{\sc i} profiles at $\phi=5.7$ and $\phi=17.2$ days in velocity space (rest-frame). The vertical green dashed lines mark velocities at 100~\kms\/ intervals, ranging from $-400$ to $+200$~\kms, excluding 0~\kms, which is marked with a black dashed line.}
\label{fig:hevel}
\end{figure}

There is clear evidence of spectral evolution in \abvb\ spectra. During the early phases (up to  $\phi=5.7$), the spectra are dominated by prominent He~{\sc i}, C~{\sc ii} and O~{\sc ii} features. Between $\phi=-2.0$ and  $\phi=5.7$, these lines exhibit relatively slow evolution, with their velocities remaining nearly constant over time. The minimum velocities measured for different ions are approximately $\sim130$ \kms\ for O~{\sc ii}, $\sim100$ \kms\ for C~{\sc ii} and around 120 \kms\/ for He~{\sc i}. By  $\phi=17.2$, the C~{\sc ii} and O~{\sc ii} features have disappeared, indicating significant changes in the ionisation conditions and/or chemical composition of the CSM. However, several He~{\sc i} lines, such as $\lambda5015$, $\lambda5876$, $\lambda6678$ and $\lambda7065$ remain detectable, albeit with reduced strength.

Figures~\ref{fig:hevel}, \ref{fig:ovel} and \ref{fig:cvel} show the temporal evolution of these lines at  $\phi=5.7$ and  $\phi=17.2$, while \ref{fig:hvel} also includes the last NIR epoch at $\phi = 55.7$. Despite the overall similarity observed in these P-Cygni profiles at 5.7d, not all emission peaks are centred at the rest-frame wavelength. In several cases, the emission components exhibit redshifted peak offsets of $\sim50-100$ \kms, which could be attributed to asymmetries in the density distribution of ejecta.

Simultaneously, a decline in the strength of Fe~{\sc ii} features, particularly multiplet 42 by 17.2d, indicates a rapid evolution of the circumstellar environment and increasing optical depth in the NUV due to the emergence of iron-rich layers. In contrast, the Ca~{\sc ii} NIR triplet lines are visible at this epoch, displaying expansion velocities of $\sim100$ \kms, roughly consistent with the velocities of He~{\sc i}, C~{\sc ii} and O~{\sc ii} lines.

\begin{figure}
    \centering
    \includegraphics[width=1\linewidth]{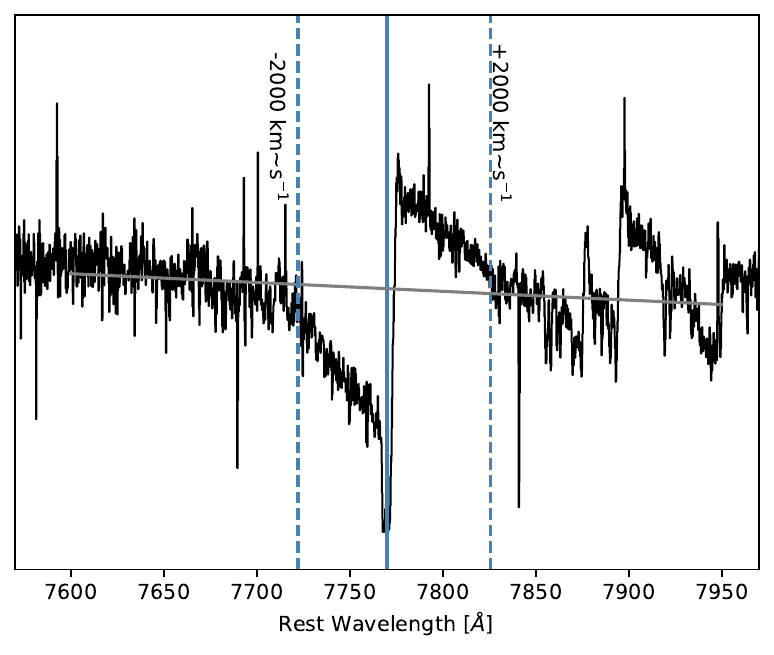}
    \caption{O~I line as seen in the second epoch of XShooter spectroscopy. The blue solid line shows the rest wavelength of O~{\sc i} 7774\AA, while the dashed lines show a velocity of $\pm$2000~\kms. The grey line is an approximate continuum to guide the eye.}
    \label{fig:OI}
\end{figure}

\begin{figure*}
\centering
\includegraphics[width=\textwidth]{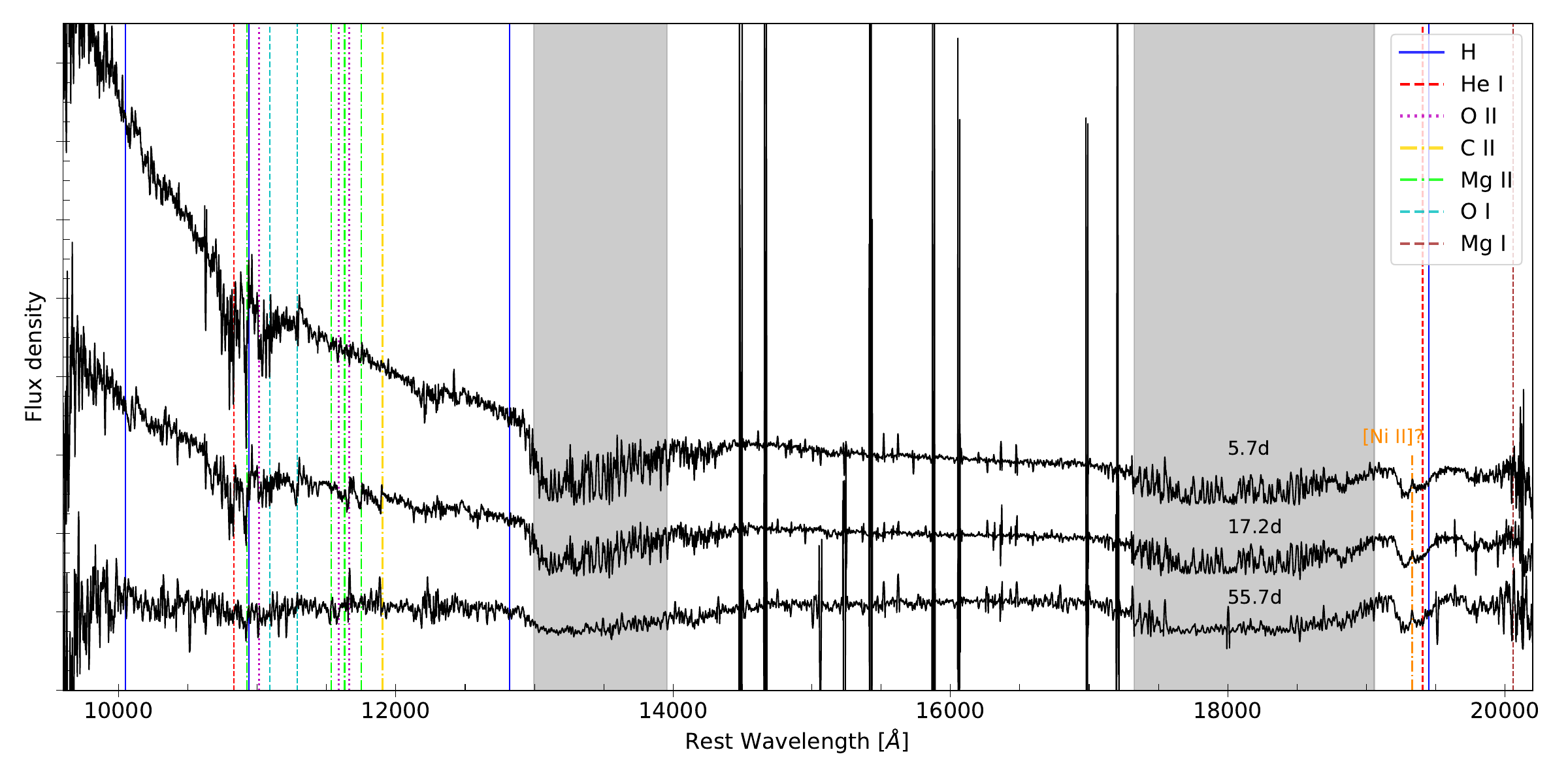}
\caption{Spectral sequence for \abvb\ in the NIR. Spectra have been rebinned to 25\AA\ (35\AA\/ the one at  $\phi = 55.7$d) sampling and regions of strong telluric absorption have been masked.}
\label{fig:nirspec}
\end{figure*}

Although the XShooter spectra at both epochs are primarily dominated by narrow features (aside from broad bases to some lines, which we attribute to electron scattering), there is also evidence of some higher velocity material. In particular, the OI $\lambda$7774 line at 17.2d (Fig. \ref{fig:OI}) shows a distinct absorption component with a blue edge extending up to 2000~\kms. This is also observed for the Ca~{\sc ii} NIR triplet. This broad absorption appears separate from the narrower, low-velocity component and suggests the presence of an outer shell of faster-moving ejecta.
In the Sobolev approximation, the shape of these lines would suggest a density profile that is at least $\rho(r) \propto r^{-2}$ or steeper, as well as a relatively low value for $\tau_0$, the optical depth at the photosphere. The latter is also consistent with the absence of strong emission components to these lines. This would disfavour a dense shell origin for the CSM surrounding \abvb.

In the final epoch of XShooter spectroscopy taken between +53.8 and +57.6 days, we see no continuum in either the UVB or VIS arms. Careful inspection of the two-dimensional spectrum reveals a single weak emission line at $\sim$8726~\AA. Interestingly, we do see a continuum in the NIR arm at this phase.

\subsection{Comparison to Ibn/Icn}

\begin{figure*}
    \centering
    \includegraphics[width=0.98\linewidth]{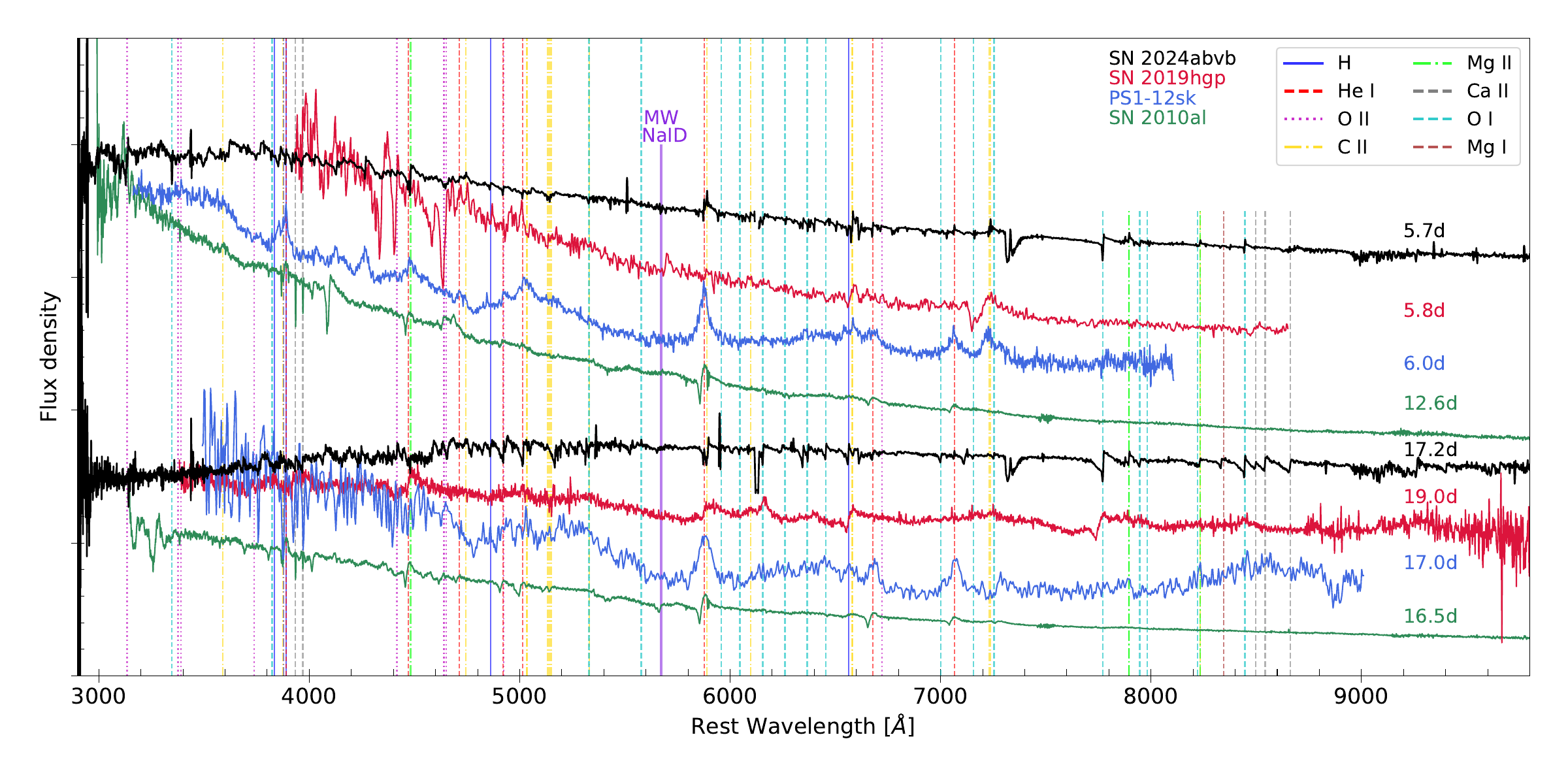}
    \caption{Spectra comparison with Ibn and Icn at a similar evolution phase. }
    \label{fig:spec_cmp}
\end{figure*}

In Fig.~\ref{fig:spec_cmp} we compare two XShooter epochs of \abvb\/ with the prototypical Type Icn SN~2019hgp \citep{GalYam_2022}, the transitional Type Ibn SN~2010al observed with high-resolution spectroscopy \citep{2015MNRAS.449.1921P}, and PS1--12sk, a Type Ibn event that exploded at a large projected distance from its host galaxy \citep{Sanders13}. 
The spectra of \abvb\/ show less prominent O~{\sc ii} features than SN~2019hgp and no highly ionised carbon lines (e.g. C~{\sc iii}), particularly at wavelengths bluer than 5000~\AA\/ and usually observed in Icn events \citep{Pellegrino_2022}. 
The C~{\sc ii} line velocities in \abvb\/ are comparable to those measured in SN~2019hgp, although they are systematically slightly lower at similar evolutionary phases. 
A similar behaviour is observed when comparing He~{\sc i} velocities with those of SN~2010al, which lacks prominent O~{\sc i} and C~{\sc ii} features, or shows equivalent widths less than half of those measured in \abvb\/ at both comparison epochs.

The helium line profiles of \abvb\/ also differ from those typically observed in Type Ibn SNe, including PS1-12sk, as well as from the transitional case SN~2010al. 
In particular, \abvb\/ lacks the broad emission component commonly seen in Ibn events, while the velocity of the narrow P-Cygni absorption is approximately a factor of two-three lower than that observed in typical Type Ibn SNe. 
Moreover, the spectra of \abvb\/ exhibit a redder continuum than both Type Ibn and Icn events \citep[see][]{2026MNRAS.tmp..363A} and do not show the Fe~{\sc ii} pseudo-continuum commonly observed, albeit with varying strength, in Type Icn SNe (e.g. SN~2019hgp) and Type Ibn SNe (e.g. PS1-12sk). 
Taken together, these similarities and differences in ionic species, line strengths, and velocities support, from a phenomenological perspective, the classification of \abvb\/ as a transitional Type Icn/Ibn event.

\begin{figure*}
    \centering
    \begin{subfigure}[b]{0.49\textwidth}
        \centering
        \includegraphics[height=4.425in]{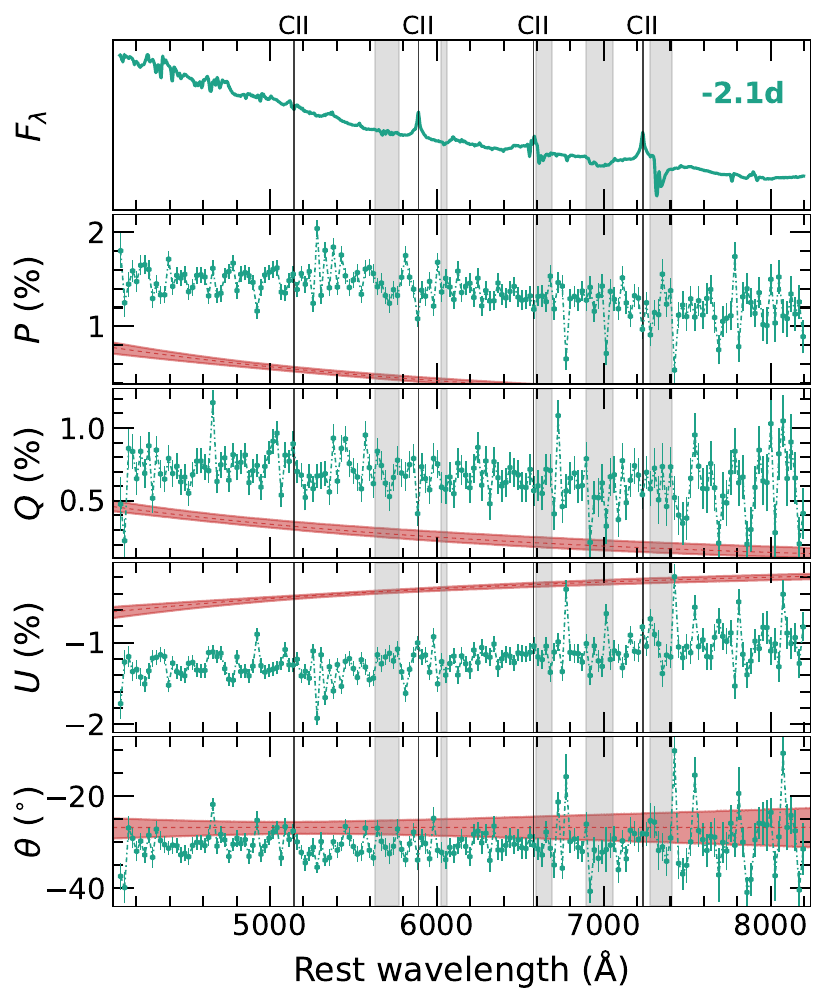}
    \end{subfigure} %
    \begin{subfigure}[b]{0.49\textwidth}
        \centering
        \includegraphics[height=4.425in]{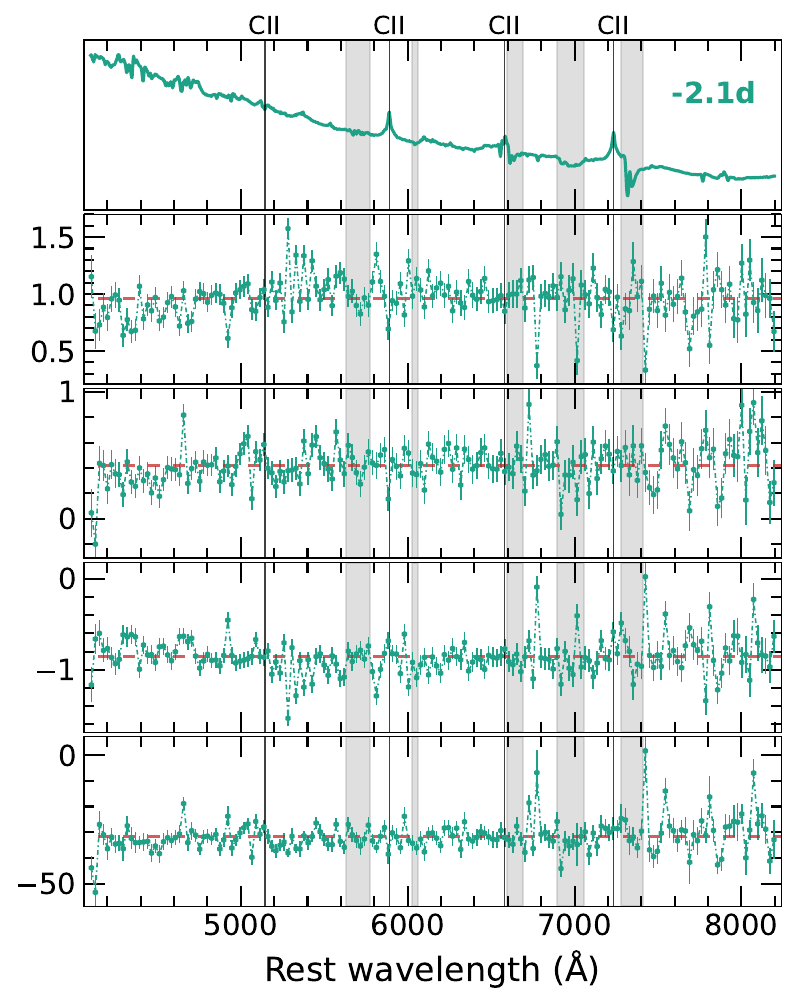}
    \end{subfigure} %
    \caption{The FORS2 spectropolarimetry taken at $-2$\,d, binned to 25\,Å in rest frame. Left: The data before the ISP correction shows polarisation increasing towards the blue, especially in $U$, well-matched with the ISP estimate (shaded region). Right: After the ISP correction, the polarisation is constant at $P\sim1\%$ with no significant variation from the mean of each parameter (dashed line).}
    \label{fig:specpol}
\end{figure*}

\section{Polarimetric Analysis}

\subsection{Interstellar polarisation}
Determining the amount of interstellar polarisation (ISP) caused by dust grains in the line-of-sight is crucial to investigating the intrinsic SN polarisation. As \ion{Na}{ID} absorption at host redshift is not present in the UVES or X-Shooter spectra, the extinction must be very low, as expected from the remote location of SN~2024abvb from its host galaxy (see Section~\ref{ss:host}. As such, the host contribution to the ISP is insignificant; it should be dominated by the Milky Way dust column. 

With our VLT/FORS2 $BVRI$ polarimetry, we can use field stars to estimate the wavelength-dependent ISP. The field is very sparse, with only $\lesssim5$ stars per band that are $\mathrm{S/N}>200$ needed for an accurate ISP estimate. However, most of them are found within 100\,px from the edges of the image, and they appear to have higher polarisation than the ones nearer to the centre of the camera. The known instrumental polarisation increases sharply towards the edges \citep{Gonzalez-Gaitan2020}, and while we correct for the effect, it is possible that the correction is not perfect at such locations. As such, we exclude these stars from measuring the ISP. Were they included the difference would be small as $B$ and $V$ would remain the same, and for $R$ and $I$ the ISP would increase by $P\lesssim0.3\%$.  We verified that the selected stars are suitable for measuring Galactic ISP. Following \citet{Tran1995}, stars must lie  $\gtrsim150$\,pc above the Galactic plane to reliably probe the dust column, and astrometric solutions from Gaia Data Release 3 \citep{GaiaCollaboration2022} show that the selected stars are well above this threshold.

As we also want to correct the spectropolarimetry for the ISP, we fit the \textit{BVRI} broadband photometry ISP estimated based on the Milky Way stars, with the empirical Serkowski Law \citep{Serkowski1973,Serkowski1975}, $P(\lambda)/P_{\mathrm{max}} = \exp{[-K \ln^2{(\lambda_{\mathrm{max}}/\lambda)}]}$, for the full wavelength coverage. $P_{\mathrm{max}}$ is the maximum polarisation at wavelength $\lambda_{\mathrm{max}}$ and $K$ is an arbitrary constant often assumed to be $1.15$. We fit an expanded version of the formula, where we incorporate the Stokes $Q$ and $U$ parameters by adding the polarisation angle $\theta$ as a fit parameter. The resulting fit is shown in Fig.~\ref{fig:serkowski_fit}. For U, the model provides a decent fit, but for $Q$ it does not agree well with the data especially in $B$-band. However, the discrepancy appears to be low-level at $\lesssim0.2\%$ regardless of the band. Such accuracy is sufficient for our purposes. We use the fit to perform the ISP correction on both spectral and imaging polarimetry presented in this paper.

\subsection{Polarimetric evolution}

The one epoch of spectropolarimetry taken at $-2$\,d is presented in Fig.~\ref{fig:specpol}. Before the ISP correction, the data shows a marginally increasing polarisation degree towards the blue. However, this appears to be caused by the ISP, as after the correction, the polarisation is constant at $P\sim1\%$, and no wavelength dependency is present in either $Q$ or $U$. This is a strong indication that the ISP estimated with the Serkowski law (Fig.~\ref{fig:serkowski_fit}), reflects the true ISP in the line-of-sight. 

Post-correction, no significant line features can be identified in either $Q$ or $U$, apart from possible depolarisation corresponding to the emission feature at $\sim5890$\,Å.  For a spectral line to have a significant polarimetric imprint, it will have to be strong in comparison to the continuum. In this situation, the lines are not only fairly weak, but they are also narrow compared to the resolution of 25\,Å needed to achieve a reasonable S/N. For instance, the broad electron scattered wings have been reported to show excess polarisation over the continuum for some SNe \citep[e.g.][]{Hoffman2008, Patat2011}, but in \abvb we cannot determine the presence of such excess due to the limited spectral resolution. Instead, the spectropolarimetry provides a highly accurate estimate of the continuum polarisation arising from an electron-scattering photosphere.

\begin{figure*}
    \centering
    \includegraphics[width=0.98\linewidth]{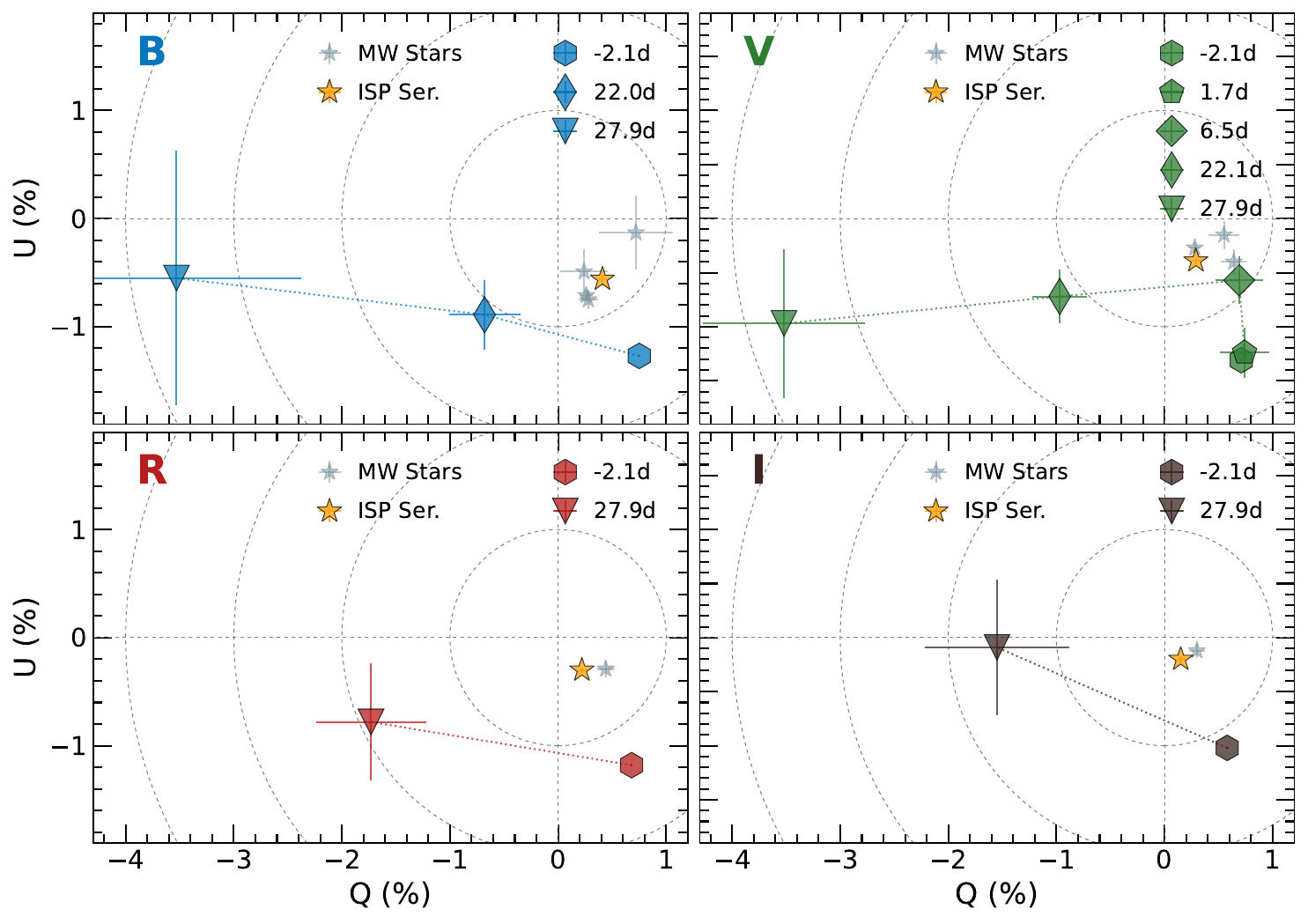}
    \caption{Stokes $Q$\,--\,$U$ planes for $BVRI$ polarimetry covering 30\,d of evolution. MW stars used to derive the ISP are shown. The SN initially exhibited $\sim1\%$ polarisation near-peak, followed by decline to $\lesssim0.5$\% a week after before increasing to high values at late-time ($\sim20$-$30$\,d post-peak).}
    \label{fig:pola_evo}
\end{figure*}

The polarimetric evolution in $BVRI$, based on the VLT/FORS2 spectropolarimetry and broadband polarimetry as well as the NOT/ALFOSC $V$-band polarimetry, is shown in Fig.~\ref{fig:pola_evo}. The SN shows a consistent $P\sim1\%$ near-peak in both FORS2 and ALFOSC data, but the polarisation decreases to $\lesssim0.5$\% at a week post-peak. After this, the polarisation increases again at a rotation of $50\deg$. At $+22$\,d the polarisation degree is $\sim1.5$\% in $BV$, and at $+27$\,d the value appears to have increased drastically to $P\sim4$\% in $BV$ while $RI$ exhibit more moderate values at $P\sim2$\%. The uncertainties of the last epoch are high, but it is clear that they have increased to a high degree at a physical rotation of $50\deg$ compared to near-peak. Further, the fact that both $BV$ are significantly higher than $RI$ indicates wavelength dependency. The observation was taken in prime conditions during 1\% Lunar illumination, and the trend is not caused by significant scattering of Lunar photons in the atmosphere \citep[see e.g.][]{Pursiainen2023}. Given the remote location, the background is also easy to model, and different methods for estimating it do not affect the results.

\subsection{Comparison to Ibn/Icn SN polarimetry}
The polarimetric sequence of \abvb\/ is remarkable for the type, as it is the first multi-epoch dataset. For Type Icn SNe, only SN\,2021csp has spectral polarimetric data in the literature \citep{Perley2022}, and the observation taken at $+3.5$\,d implies continuum polarisation of $P<0.5$\%. While the spectropolarimetry of \abvb\/ taken at $-2$\,d is higher ($P\sim1$\%), possibly implying larger departure from spherical symmetry, the observation is taken at an earlier phase with respect to peak brightness and direct comparison may be misleading. In fact, the low polarisation of \abvb\/ probed by the $V$-band polarimetry taken at $+6.5$\,d is consistent with SN\,2021csp. On the other hand, for Type Ibn SNe only SN\,2015G has conclusive spectropolarimetry showing continuum polarisation up to $P\sim2.7$\%  \citep{Shivvers2017}. While likely high, the exact value is unclear, however, due to the uncertain ISP contribution. Further, comparison to \abvb\/ is difficult as the SN was discovered post-peak, and the phase with respect to peak is not known. While \abvb\/ shows potentially even higher polarisation at late stages, SN\,2015G shows no wavelength dependence, and it is likely that the polarisation in late-phase \abvb\/ broad-band polarimetry is caused by a different mechanism. Three epochs of spectropolarimetry were also obtained for SN\,2015U, but the high Milky Way extinction rendered the observation inconclusive \citep{Shivvers2016}. In addition, there are now two Type Ibn SNe with imaging polarimetry. Both SN\,2023emq \citep{Pursiainen_2023a} and SN\,2023tsz \citep{Warwick2025} showed low polarisation $\sim1$\,--\,2 weeks post-maximum, similar to \abvb.

The sample of Ibn/Icn SNe with optical polarimetry is small, but there appears to be a tendency towards low polaristaion at roughly a week past peak. \abvb\/ is consistent with the sample, but the extensive polarimetric coverage has allowed for the identification of significant polarimetric evolution not possible for the previous events. Given the evolutionary stage, the $1$\% polarisation seen $\pm2$\,d from the peak is possibly related to nearby CSM. Type Ibn SNe have been shown to exhibit significant mass-loss just prior to explosion \citep[e.g.][]{Maeda2022}, and it is possible that the related CSM component has an aspherical structure due to, for example, bi-polar eruptions, explaining the non-zero polarisation. Similar suggestion have been made to for instance Type II SNe before \citep[e.g. SN\,2023ixf;][]{Vasylyev2023,Shrestha2025}. The polarisation decrease to $P\sim0$\% at roughly one week post-peak occurs at a phase when the SN explosion has already blown through a significant amount of more-distant CSM. This does not necessarily imply a transition from asymmetrical to spherical configuration as, for a given asymmetric structure, the decreasing optical depth could lead to a decrease in polarisation. Such a scenario could be achieved when the ejecta has already passed through most of the asymmetric CSM. Alternatively, in case the SN explosion was roughly spherical, the SN photosphere formed at the interaction zone of the CSM and ejecta would follow high spherical symmetry, explaining the low polarisation. The $P\sim0$\% could also be explained with the ejected material colliding with a different shell and geometry than that encountered around peak. 

Finally, the high, late-time wavelength-dependent polarisation likely requires another source of polarisation as electron scattering near an optical photosphere of a SN is an achromatic process. One plausible explanation is scattering by dust as Type Ibn SNe are known to form it in the ejecta or in a cool dense shell \citep[see e.g.][]{Mattila2008}. Dust scattering can create polarisation that naturally increases towards the blue as observed in \abvb. The only stipulation is that the scattering body cannot be perfectly symmetric around the SN (i.e. perfectly spherical shell), as then the scatterings occurring around the SN would nullify each other, resulting in low polarisation. However, even a shallow deviation from spherical symmetry could result in only a fraction of the shell scattering at once, thus producing high polarisation. Similarly, the observed polarisation signal could also be caused by scattering in pre-existing distant dust, formed as a result of a pre-SN eruption. However, in the absence of accurate multi-epoch polarimetry to model the scattering body, it is not possible to constrain the possible distribution of the dust. Nevertheless, we note that nor our NIR continuum or emission lines (e.g. blue-shifting) show sign of dust production at such epochs.

\section{Discussion}
\label{sect:discussion}

In the following, we first briefly summarise the main conclusions from our observations of SN~2024abvb, before considering potential explanations.

\begin{itemize}
    \item \abvb\ lies very far from its host, and is not associated with an obvious region of star formation.
    \item The spectra of \abvb\ show low velocity (a few hundred \kms, presumably circumstellar) material that is composed of He, C and O.
    \item The ions and line profiles suggest a steep density profile with a relatively low value of optical depth at the photosphere/pseudo-photosphere.
    \item We observe low-velocity H absorptions which, based on its line profile, is likely arising from a region distinct from the He, C, and O lines.
    \item The polarisation is around P$\sim$1\% around peak, decreasing then to P$\sim$0\% after one week, similarly to what observed in other Icns, and finally rising up to P$\sim$4\% (stronger in the blue) with a 50$\degree$ rotation with respect to the peak emission.
\end{itemize}

To provide quantitative context for the discussion below, we begin by deriving order-of-magnitude estimates of the characteristic CSM radii and masses implied by the spectroscopic and polarimetric data. Assuming an ionised CSM with electron densities of $n_e \sim 10^9$–$10^{10}$ cm$^{-3}$ and characteristic radii $R \sim 10^{14}$–$10^{15}$ cm, we can estimate the CSM mass at the order-of-magnitude level. Adopting a shell (or toroidal) geometry with a fractional thickness $\Delta R \sim 0.1$–$0.3$~R \citep{Smith_2014,2014A&A...569A..57M,2016MNRAS.458.1618D} and accounting for a limited solid-angle covering factor, we infer CSM masses of $M_{\rm CSM}\sim10^{-2}$~M$_\odot$ per individual structure. The presence of multiple kinematically distinct components implies a total CSM mass of $M_{\rm CSM, total}\sim 10^{-1}$~M$_\odot$, although this value remains highly uncertain and model dependent. Such a value is comparable to the CSM mass from light curve modelling reported by \citet{2026MNRAS.tmp..363A}, \citet{2026arXiv260101333H} and \citet{2026arXiv260216227S}.

\abvb\/, in common with many interacting transients, does not show a broad line that can be clearly identified as being from the underlying SN. Consequently, we must consider whether the Type Ibn/cn \abvb\ is in fact a thermonuclear explosion. Moreover, due to the rapidity of its evolution and remote location, we must also consider rapidly/remote/faint transients such as ultra stripped SNe, as well as transients powered by accretion onto a Black Hole.

\subsection{\abvb\ as a thermonuclear supernova}

The remote location of \abvb\ is suggestive of an old progenitor, and we hence consider whether this could actually be a thermonuclear explosion within a (mostly) He/C/O rich CSM.

\begin{figure}
    \centering
    \includegraphics[width=1\linewidth]{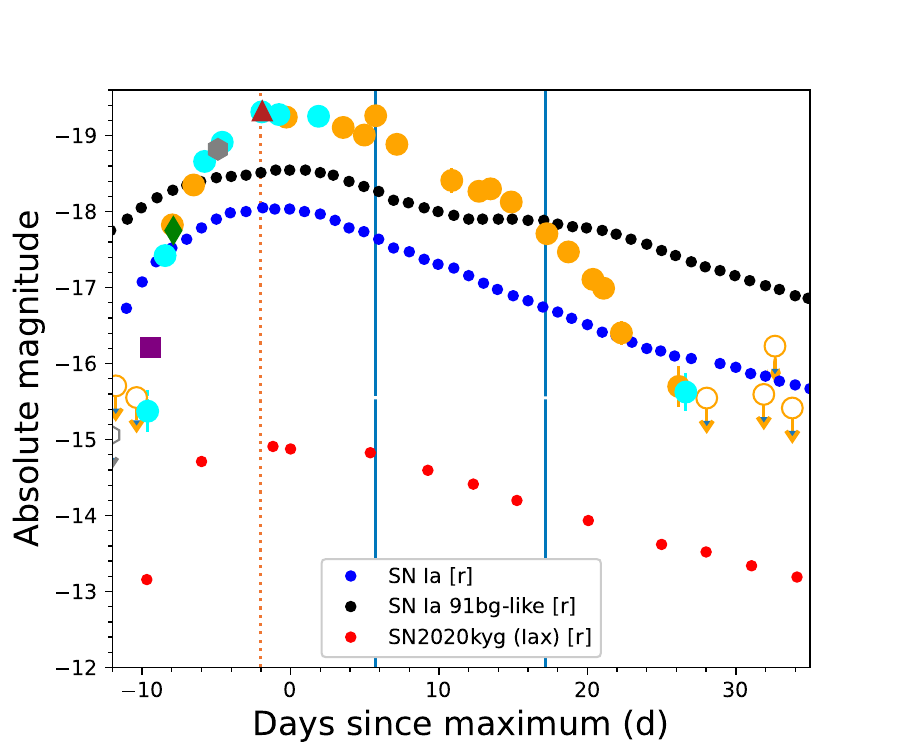}
    \caption{The optical lightcurve of \abvb, compared to template $r$-band lightcurves of both normal and 91bg-like SNe Ia \citep{Nugent02}, as well as the r-band lightcurve of the Iax SN~2020kyg \citep{Srivastav22}. The vertical lines mark the phases of our VLT spectra in which the SN can be detected.}
    \label{fig:Ia_comparison}
\end{figure}

In Fig. \ref{fig:Ia_comparison}, we compare the optical lightcurve of \abvb\ to a number of Type Ia templates \citep{Nugent02}, shifted so that the maxima of their lightcurves coincide. It is clear that a normal Type Ia SN could not be hidden under the light curve of \abvb, as from around 20 days onward, we would see a difference in the light curve. Moreover, the SN would dominate the emission at the epoch of our second XShooter spectrum, and so we would expect to see underlying photospheric features from the SN. From a luminosity point of view, accommodating a sub-luminous SN~1991bg-like SN is more plausible; in such a scenario, the underlying SN would be significantly fainter than our observed light curve for all epochs where we have spectra, and approximately consistent with our upper limits on the SN magnitude at early and late times. Nevertheless, the presence of H in the CSM would prove challenging to explain. 
Similarly, a SN Iax could also be hidden below the CSM interaction as they are several magnitudes below both the detections and upper limits of \abvb\ at all phases \citep[e.g.][]{2013ApJ...767...57F,2020ApJ...892L..24S}  with peak magnitude down to $-13$ mag \citep[SN~2024vjm,][]{2025ApJ...989L..33K,2026arXiv260209096Z} or even fainter \citep[SN~2021fcg,][]{2021ApJ...921L...6K}. However, they are not found in remote locations and show an association to young stellar populations \citep{2018MNRAS.473.1359L}. A similar argument applies to a .Ia SN, whose intrinsic light curve would be hidden beneath the luminosity produced by CSM interaction \citep[e.g.][]{2015ApJ...799L...2I,2022ApJ...934..102D,2024ApJ...964..196P}. Such SNe have been found in remote locations and are thought to be triggered by a detonation of a helium layer on a low-mass CO white dwarf, which would also accommodate the presence of He in the CSM. 

Having established that it is at least possible to hide a sub-luminous Type Ia SN under the light curve of \abvb, the next question is what the progenitor system might be. AM CVn systems \citep{Solheim10, Ramsay18} involve mass transfer of He onto a WD, which is potentially appealing given the He-rich CSM. However, in these cases $\dot{\mathrm{M}}_\odot$ is at most a few times $10^{-8}$~M$_\odot$ per year. It is challenging to see how such a low mass loss rate could lead to the formation of a substantial CSM and strong circumstellar interaction. Another possibility would be that of a WD in a binary system with a Helium donor star, leading to a He-rich CSM mimicking Ibn-like features \citep[e.g. SN~2020eyj,][]{2023Natur.617..477K}. Nevertheless, the presence of H is a challenge, as the progenitor would have had to have lost its H envelope roughly 1 Myr previously \citep[][and references therein]{2025A&A...696A.103E}.

One peculiar channel that has been proposed for Type Ia SNe, and that may be relevant here, is the ``core-degenerate'' scenario \citep{Kashi11}. In this channel, a SN Ia arises from the merger of a WD with the degenerate core of an AGB star. This has been invoked in the past to explain some peculiar Type Ia SNe with H-rich CSM  \citep[e.g.][]{Livio03}, and has the natural advantage that we may expect circumstellar interaction with the ejected common envelope. However, there are significant difficulties in applying this here. Firstly, the core-degenerate scenario requires a relatively massive AGB star, and so the problem of the remote location for \abvb\ remains. More problematic still, in the case of \abvb\ we are seeing interaction with a He-rich CSM (H appears in absorption and presumably comes from material further out). So, the putative ejected common envelope cannot come from an AGB star.

Another potential channel leading to an Icn-like explosion is a WD–WD merger scenario. It has been shown that mergers between O/Ne and C/O WDs can reproduce the overall light-curve evolution of Type Icn SNe \citep{2024ApJ...967L..45W}. In particular, both O/Ne + O/Ne and C/O + O/Ne mergers can form an Fe core that grows to the Chandrasekhar mass, either without or with a carbon shell-burning phase \citep{2025ApJ...982....6S}. 
One of the possible observable outcomes of this scenario is an Fe core-collapse SN interacting with a relatively low-mass, extended H- and He-deficient shell, which would appear observationally as a Type Icn SN. In principle, such a progenitor channel could account for the large projected galactocentric distance of \abvb\/; however, it would require the binary system to be dynamically formed or hardened within a globular cluster \citep{2019ApJ...887..180S}, for which we find no evidence in the host environment (see Sect.~\ref{ss:host}). Moreover, the expected observational properties would more closely resemble those of calcium-rich transients \citep[e.g.][]{2010Natur.465..322P} than those of \abvb\/.

\subsection{\abvb\ as an accretion powered transient}

We note that some fast and luminous transients \citep[e.g. AT2018cow-like events;][]{2019MNRAS.484.1031P} have shown late-time interaction with a He-rich circumstellar medium (CSM), producing spectra reminiscent of Type Ibn SNe. In fact, AT2018cow itself had an absolute magnitude at peak of $r=-20$ mag, comparable to that of \abvb, and a similarly fast decline (both faded to $-16$ mag within 25 days from the maximum). While \abvb\ has a rise time of only 10 days, AT2018cow is somewhat faster. However, the cow-like transients, which are likely powered by accretion onto a black hole, typically exhibit fast, shallow absorption features in their early spectra, followed by broad (several thousand \kms) H or He emission lines about one month after peak brightness. None of these characteristics are observed in \abvb.

\subsection{\abvb\ from a massive star or binary system}

A core-collapse SN should be the natural explanation for the photometric and spectroscopic characteristics of \abvb. The H-poor CSM could arise from a stripped massive star, either a Wolf-Rayet star, or a lower mass He star that has been stripped in a binary. At $\sim$120~\kms, the velocity of the CSM in \abvb\ is much lower than the fast 1000~\kms\ winds that one could expect from either a WR star or a stripped He star. Moreover, the presence of H absorption is hard to explain - presumably this was lost during an earlier stage of evolution, but this cannot have been too long ago, otherwise we would no longer see absorption along the line of sight.

However, the biggest challenge for \abvb\ as a core-collapse SN remains its remote location with no signs of star formation. We compare the location of \abvb\ to the radial distribution of core-collapse supernovae found by \cite{Hakobyan09}. After conversion of these from R25 to effective radii ($r_e$), we find that \abvb\/ is proportionally more distant from its host than any of their sample. This is also true when comparing it to interacting transients and stripped-envelope SNe from the PTF survey \citep{2021ApJS..255...29S}, as the \abvb\/ location is further away than any of their sample (see Fig.~\ref{fig:offset}).

One intriguing channel for a massive star to explode is as an ultra-stripped SN (USSN). Binary systems composed of two massive stars can evolve to form double neutron star (DNS) systems \citep[e.g.][]{2004ApJ...612.1044P,Tauris2017}. When the donor star possesses a deep convective envelope at the onset of mass transfer, corresponding to Case B Roche lobe overflow, a common envelope (CE) phase is initiated. This evolutionary path can ultimately lead to the explosion of the donor star as an ultra-stripped supernova. Such explosions may proceed via either a core-collapse supernova driven by the iron-core collapse (Fe CCSN) or an electron-capture (EC) supernova.

In the case of a USSN, the ejecta mass is typically low \citep[$\leq 0.2$~M$_\odot$,][]{Tauris2015}, which is consistent with the overall luminosity evolution of \abvb\/ \citep[see also]{2026arXiv260101333H}. Moreover, if the explosion is an EC supernova, the amount of helium stripped from the progenitor is expected to be greater than in the case of an Fe CCSN \citep{Tauris2015}. Hence, the explosion is more prone to be a Type Ibn/Icn event or an interacting SN with low He content for the closest CSM. EC SNe tend to be fainter than other stripped SNe and display a risetime consistent with what is observed in \abvb\/ \citep{2009ApJ...705L.138P,2016MNRAS.461.2155M}, meaning that an EC SN lightcurve can be hidden under the light curve of \abvb. An EC supernova originating from a binary system also provides a natural explanation for the remote location of \abvb, as DNS systems are not necessarily found in regions of ongoing star formation. The neutron stars may have drifted from their birth sites, or star formation in the region may have ceased long ago \citep[e.g.][]{2019MNRAS.484..698R,2020ApJ...892L...9A}. Indeed, other rapidly evolving interacting supernovae, such as the Type Icn SN~2019jc \citep{Pellegrino_2022} and the Type Ibn PS1-12sk \citep{Sanders13}, have also occurred in remote environments or in regions with low star formation activity. These cases suggest that the location of \abvb\ is not unique and might represent an intrinsically rare evolutionary path. It is therefore plausible that all three events originated from binary systems that underwent common envelope evolution, albeit with differences in progenitor radius or CSM configuration \citep{Tauris2015,Tauris2017}. 

A CE phase naturally enables repeated mass-loss episodes with a changing symmetry axis but a fixed barycentre. 
The observed expansion velocities and relatively low optical depths are consistent with predominantly wind-like mass loss. 
During a CE phase, evolution of the orbital plane can lead to asymmetric mass-loss episodes occurring in different instantaneous orbital planes, producing concentric but mutually tilted tori \citep{2013A&ARv..21...59I,2023A&A...674A.121G}. 
A similar geometry may also arise from precession of an accretion disk around the companion or the stellar core \citep{2013A&ARv..21...59I,2017ApJ...848L..34M}. 

\begin{figure*}
    \centering
    \includegraphics[width=0.98\linewidth]{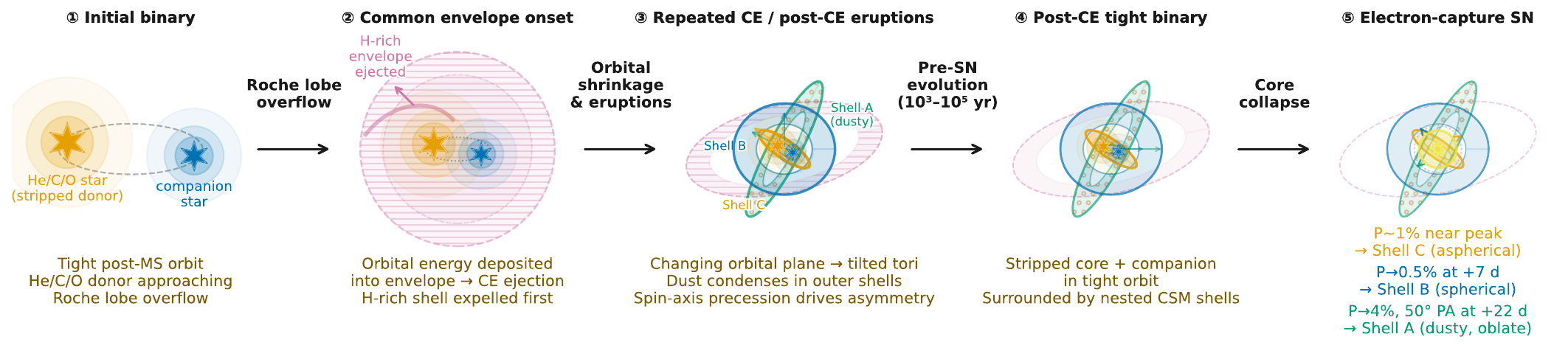}
    \caption{Proposed common-envelope evolutionary pathway leading to the nested CSM shells observed in SN 2024abvb. Successive mass-loss episodes driven by orbital evolution and outflow precession during the CE/post-CE phase produce concentric toroidal shells with changing symmetry axes, culminating in an electron-capture supernova whose ejecta interact with each shell in sequence.}
    \label{fig:CE_schematic}
\end{figure*}

\begin{figure*}
    \centering
    \includegraphics[width=0.98\linewidth]{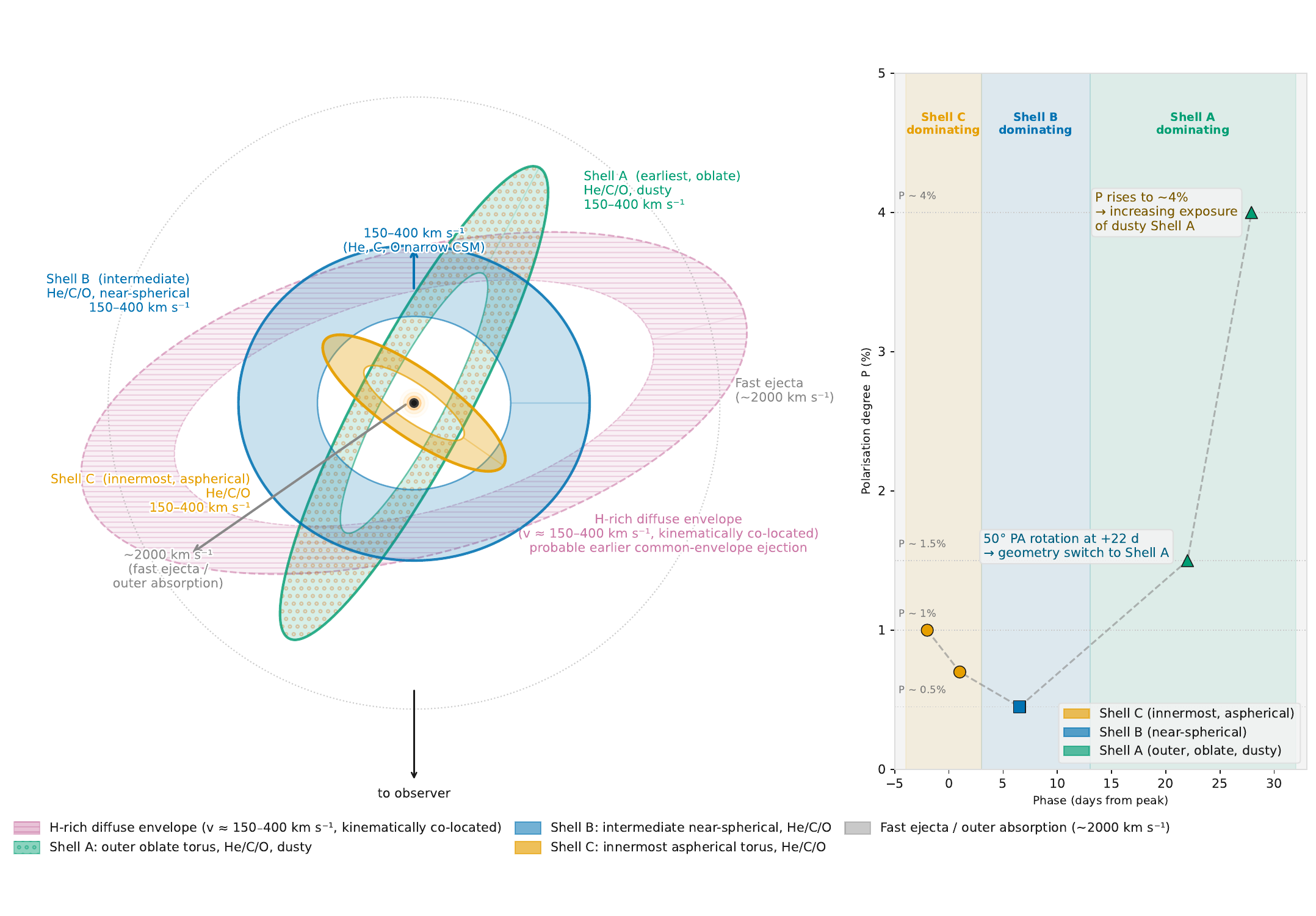}
    \caption{Schematic of the nested, asymmetric circumstellar medium around SN 2024abvb.  Left: the projected geometry of concentric He/C/O toroidal shells at differing orientations, together with the co-located diffuse H-rich envelope. Right: the observed polarimetric evolution, with epoch-coloured markers indicating which shell geometry dominates the scattering at each phase.}
    \label{fig:CSM_schematic}
\end{figure*}

For completeness, we note that nested, ring-like dust shells have been directly imaged around massive colliding-wind binaries (notably WR\,140), demonstrating that repeated, discrete dust-forming episodes can produce concentric shells or rings with distinct radii and orientations \citep{2022NatAs...6.1308L}.  Although the progenitor system of \abvb\/ may differ (e.g. a stripped star rather than a classical WC star), the existence of long-lived, nested dust shells in WR binaries provides an empirical precedent for multi-episode, quasi-periodic mass ejection that naturally yields concentric structures and enables in situ dust formation.  In addition, multi-epoch studies of SNe with evidence for nested shells or interacting shells have been interpreted in terms of repeated mass-loss episodes or collisions between ejecta and pre-existing shells \citep{2023ApJ...956L..31M}.  Thus, the detection of multiple kinematic components in our data (tori with measured line velocities of $\sim150 - 400$ \kms\/ and additional faster components up to $\sim2000$ \kms) is qualitatively consistent with a scenario of episodic shell/toroid formation followed by later, faster interaction.

The wavelength-dependent polarisation evolution we measure is naturally explained if (i) dust forms and survives in some of the earlier-ejected toroidal components and (ii) the line of sight traverses different combinations of dusty and dust-poor shells as the system evolves and as the effective scattering geometry changes.  Dust scattering (and/or differential extinction by newly formed grains) produces a wavelength dependence with larger polarisation at shorter wavelengths, whereas electron scattering does not produce a strong wavelength slope; hence, the blue-weighted rise at late times argues in favour of a dust contribution to the observed polarisation.  The WR\,140 JWST observations demonstrate that colliding-wind binaries can produce long-lived carbonaceous dust shells that survive multiple orbital cycles \citep{2022NatAs...6.1308L}, lending plausibility to the idea that dust formed in earlier tori could later dominate the polarimetric signal when illuminated or partially occulting the SN.

Finally, the relative asymmetries we retrieve can be reconciled with a time-dependent mass-loss axis.  In a binary/common-envelope context, the orientation and degree of asymmetry of successive ejection episodes can change because of orbital evolution, episodic disk formation and dissipation, or torque-induced precession of an accretion disk or outflow-launching structure \citep[e.g.][]{2013A&ARv..21...59I, Tauris2015}.  Such mechanisms naturally produce concentric but non-coaxial tori whose apparent flattening and inclination vary with epoch; the outermost shells then record an earlier geometry, while inner shells trace later, possibly reoriented mass-loss.  We therefore consider a physically conservative interpretation in which (a) repeated, binary-driven ejection episodes produce concentric toroidal shells, (b) transient colliding-wind shocks enable dust condensation in some episodes (providing the wavelength-dependent polarisation), and (c) changes in the instantaneous mass-loss symmetry axis (e.g. driven by precession or secular orbital reorientation) produce the observed sequence of inclinations and asymmetries. Fig.~\ref{fig:CE_schematic} and \ref{fig:CSM_schematic} show a schematic CE pathway leading to the formation of the \abvb\/ CSM configuration and a represetation of the nested, asymmetric CSM surrounding \abvb, respectively.  While WR binaries provide a direct observational analogue for nested dusty shells \citep{2022NatAs...6.1308L}, detailed hydrodynamic modelling targeted at CE/post-CE mass loss in stripped progenitors is required to quantify whether the exact sequence of asymmetries and kinematics seen here can be reproduced. 

However, we would like to note that multiple narrow spectral components do not uniquely imply the presence of concentric, smooth shells; an alternative interpretation is a fragmented or clumpy CSM, as observed in nova ejecta. However, several lines of evidence favour an organised,  multi-episode geometry (concentric shells/tori, possibly clumpy) in the case of \abvb. 
First, the kinematic components are stable across epochs and are seen consistently in multiple ions (He, C, O) at the same velocities, which is naturally produced by coherent shells at different radii rather than by numerous small, stochastic clumps. Second, the polarimetric evolution with a systematic rotation of the position angle, combined with a late, blue-weighted increase in polarisation is most straightforwardly explained by different inclined shells dominating the scattering/occultation at different times. We therefore favour a picture of episodic, binary-driven mass loss producing concentric, mutually inclined toroidal structures. These structures may themselves be inhomogeneous (clumpy), which would naturally reconcile discrete velocity components with the overall toroidal geometry.

\section{Conclusions}

We summarise the observational evidence presented in this work. SN\,2024abvb is located at a large projected offset from its host (well outside $r_{80}$), i.e. in a remote environment. The spectra show multiple narrow, low-velocity CSM components composed of He, C and O with absorption minima in the range $\sim150 - 400$ \kms, an additional faster material up to $\sim2000$ \kms\/ and a low optical depth suggesting a wind-like CSM. Narrow, multi-component Balmer absorptions (150--400 \kms) are detected and persist between epochs. The optical polarisation evolves strongly: $P\approx1\%$ near peak, falling to $\lesssim0.5\%$ after $\sim1$ week, then rising to $\sim1.5\%$ at $\sim20$\,d with a $\sim50^\circ$ rotation and to $\sim4\%$ (stronger in the blue) by $\sim30$\,d. A NIR line at 1.933\,$\mu$m is detected (FWHM $\sim400\,$km\,s$^{-1}$), tentatively associated with [Ni\,{\sc ii}] or a blend, while the NIR continuum does not show clear evidence for in-situ dust formation within our wavelength coverage.

We propose the following conservative interpretation. The data are consistent with episodic, binary-driven mass loss from a common envelope scenario (see Fig.~\ref{fig:CE_schematic}) leading to the explosion of an electron-capture supernova surrounded by multiple, concentric (ring-like) CSM tori with differing orientations. In this picture (see Fig.~\ref{fig:CSM_schematic}), the outer (earlier) tori are more oblate, while intermediate shells are more spherical and the innermost shells are again more aspherical, a sequence that can be produced by changes in the instantaneous mass-loss axis driven by orbital evolution, transient accretion disks, or torque-induced precession during common-envelope / post-common-envelope phases. Dust condensed in earlier, more distant tori can naturally explain the late-time, wavelength-dependent increase in polarisation (blue-weighted) when different dusty shells come into the line of sight or are illuminated by the fading transient. This interpretation is supported by empirical examples of nested dust/ring structures in massive colliding-wind systems and the qualitative consistency of our measured kinematics and polarimetric evolution with episodic tilted ejection. While, to our knowledge, this represents the first SN with spectroscopic evidence consistent with nested, concentric tori, detailed hydrodynamic and radiative-transfer models targeted at CE/post-CE mass loss in stripped progenitors are required to confirm whether the precise sequence of inclinations, asymmetries and kinematics can be reproduced.

\begin{acknowledgements}
Based on observations collected at the European Organisation for Astronomical Research in the Southern Hemisphere under ESO programme 114.2744. 
Based in part on observations made with the Nordic Optical Telescope, owned in collaboration by the University of Turku and Aarhus University, and operated jointly by Aarhus University, the University of Turku and the University of Oslo, representing Denmark, Finland, and Norway, the University of Iceland and Stockholm University at the Observatorio del Roque de los Muchachos, La Palma, Spain, of the Instituto de Astroﬁsica de Canarias. The data presented here were obtained with ALFOSC, which is provided by the Instituto de Astroﬁsica de Andalucia (IAA) under a joint agreement with the University of Copenhagen and NOT.
CI gratefully acknowledges the support received from the MERAC Foundation. 
CPG acknowledges financial support from the Secretary of Universities and Research (Government of Catalonia) and by the Horizon 2020 Research and Innovation Programme of the European Union under the Marie Sk\l{}odowska-Curie and the Beatriu de Pin\'os 2021 BP 00168 programme. 
CPG and LG recognise the support from the Spanish Ministerio de Ciencia e Innovaci\'on (MCIN) and the Agencia Estatal de Investigaci\'on (AEI) 10.13039/501100011033 under the PID2023-151307NB-I00 SNNEXT project, from Centro Superior de Investigaciones Cient\'ificas (CSIC) under the PIE project 20215AT016 and the program Unidad de Excelencia Mar\'ia de Maeztu CEX2020-001058-M, and from the Departament de Recerca i Universitats de la Generalitat de Catalunya through the 2021-SGR-01270 grant.
MB acknowledges the Department of Physics and Earth Science of the University of Ferrara for the financial support through the FIRD 2025 grant.
TLK acknowledges support via an Research Council of Finland grant (340613), support from the Turku University Foundation (grant no. 081810), and a Warwick Astrophysics prize post-doctoral fellowship made possible thanks to a generous philanthropic donation.
JDL and MP acknowledge support from a UK Research and Innovation Future Leaders Fellowship (grant references MR/T020784/1 and UKRI1062).
T.-W.C. acknowledges financial support from the Yushan Fellow Program of the Ministry of Education, Taiwan (MOE-111-YSFMS-0008-001-P1), and from the National Science and Technology Council, Taiwan (NSTC 114-2112-M-008-021-MY3).\\
{\bf Author contributions:} CI is the PI of INTEL, wrote and led the manuscript analysis and interpretation. He is the corresponding author. MF, CPG, MP and AP co-wrote the manuscript. EM reduced the UVES data and contributed to the analysis and discussion. MF reduced the XShooter data. HW reduced the FORS2 spectropolarimetry data. MP led polarimetric triggering, reduced the broadband polarimetry and performed the polarimetric analysis. CA provided insight about the object and host. TK was the PI of the NOT impol programme. JA, EM, MB, GL, KM, KeM, TM, JLD, ST contributed to the analysis and discussion. JA, AP, MB, TWC, LG, GL, JLD, KM, KeM, TM and ST are co-I of the INTEL pilot proposal.
\end{acknowledgements}

\bibliography{references}{}

\begin{thebibliography}{122}
\expandafter\ifx\csname natexlab\endcsname\relax\def\natexlab#1{#1}\fi

\bibitem[{{Andrews} {et~al.}(2020){Andrews}, {Breivik}, {Pankow}, {D'Orazio},
  \& {Safarzadeh}}]{2020ApJ...892L...9A}
{Andrews}, J.~J., {Breivik}, K., {Pankow}, C., {D'Orazio}, D.~J., \&
  {Safarzadeh}, M. 2020, \apjl, 892, L9

\bibitem[{Appenzeller {et~al.}(1998)Appenzeller, Fricke, Fürtig, Gässler,
  Häfner, Harke, Hess, Hummel, Jürgens, Kudritzki, Mantel, Meisl, Muschielok,
  Nicklas, Rupprecht, Seifert, Stahl, Szeifert, \& Tarantik}]{Appenzeller1998}
Appenzeller, I., Fricke, K., Fürtig, W., {et~al.} 1998, The Messenger, 94, 1

\bibitem[{{Aster} {et~al.}(2026){Aster}, {Inserra}, {Pastorello}, {Anderson},
  {Bauer}, {Bostroem}, {Chambers}, {Chen}, {Farah}, {Fraser}, {Fugazza},
  {Gromadzki}, {Guti{\'e}rrez}, {Howell}, {Kankare}, {Killestein}, {Koivisto},
  {Leloudas}, {Lyman}, {Medler}, {Moran}, {M{\"u}ller-Bravo}, {Pignata},
  {Pursiainen}, {Ragosta}, {Reguitti}, {Sollerman}, {Valerin}, {Warwick}, \&
  {Young}}]{2026MNRAS.tmp..363A}
{Aster}, C., {Inserra}, C., {Pastorello}, A., {et~al.} 2026, \mnras, 547,
  stag404

\bibitem[{{Bellm} {et~al.}(2019){Bellm}, {Kulkarni}, {Graham}, {Dekany},
  {Smith}, {Riddle}, {Masci}, {Helou}, {Prince}, {Adams}, {Barbarino},
  {Barlow}, {Bauer}, {Beck}, {Belicki}, {Biswas}, {Blagorodnova}, {Bodewits},
  {Bolin}, {Brinnel}, {Brooke}, {Bue}, {Bulla}, {Burruss}, {Cenko}, {Chang},
  {Connolly}, {Coughlin}, {Cromer}, {Cunningham}, {De}, {Delacroix}, {Desai},
  {Duev}, {Eadie}, {Farnham}, {Feeney}, {Feindt}, {Flynn}, {Franckowiak},
  {Frederick}, {Fremling}, {Gal-Yam}, {Gezari}, {Giomi}, {Goldstein},
  {Golkhou}, {Goobar}, {Groom}, {Hacopians}, {Hale}, {Henning}, {Ho}, {Hover},
  {Howell}, {Hung}, {Huppenkothen}, {Imel}, {Ip}, {Ivezi{\'c}}, {Jackson},
  {Jones}, {Juric}, {Kasliwal}, {Kaspi}, {Kaye}, {Kelley}, {Kowalski},
  {Kramer}, {Kupfer}, {Landry}, {Laher}, {Lee}, {Lin}, {Lin}, {Lunnan},
  {Giomi}, {Mahabal}, {Mao}, {Miller}, {Monkewitz}, {Murphy}, {Ngeow},
  {Nordin}, {Nugent}, {Ofek}, {Patterson}, {Penprase}, {Porter}, {Rauch},
  {Rebbapragada}, {Reiley}, {Rigault}, {Rodriguez}, {van Roestel}, {Rusholme},
  {van Santen}, {Schulze}, {Shupe}, {Singer}, {Soumagnac}, {Stein}, {Surace},
  {Sollerman}, {Szkody}, {Taddia}, {Terek}, {Van Sistine}, {van Velzen},
  {Vestrand}, {Walters}, {Ward}, {Ye}, {Yu}, {Yan}, \&
  {Zolkower}}]{2019PASP..131a8002B}
{Bellm}, E.~C., {Kulkarni}, S.~R., {Graham}, M.~J., {et~al.} 2019, \pasp, 131,
  018002

\bibitem[{{Blondin} {et~al.}(2022){Blondin}, {Bravo}, {Timmes}, {Dessart}, \&
  {Hillier}}]{2022A&A...660A..96B}
{Blondin}, S., {Bravo}, E., {Timmes}, F.~X., {Dessart}, L., \& {Hillier}, D.~J.
  2022, \aap, 660, A96

\bibitem[{Bradley {et~al.}(2024)Bradley, Sipőcz, Robitaille, Tollerud,
  Vinícius, Deil, Barbary, Wilson, Busko, Donath, Günther, Cara, Lim,
  Meßlinger, Burnett, Conseil, Droettboom, Bostroem, Bray, Bratholm, Jamieson,
  Ginsburg, Barentsen, Craig, Pascual, Rathi, Perrin, Morris, \&
  Perren}]{Bradley2024}
Bradley, L., Sipőcz, B., Robitaille, T., {et~al.} 2024, astropy/photutils:
  1.12.0

\bibitem[{{Bruch} {et~al.}(2023){Bruch}, {Gal-Yam}, {Yaron}, {Chen},
  {Strotjohann}, {Irani}, {Zimmerman}, {Schulze}, {Yang}, {Kim}, {Bulla},
  {Sollerman}, {Rigault}, {Ofek}, {Soumagnac}, {Masci}, {Fremling}, {Perley},
  {Nordin}, {Cenko}, {Ho}, {Adams}, {Adreoni}, {Bellm}, {Blagorodnova},
  {Burdge}, {De}, {Dekany}, {Dhawan}, {Drake}, {Duev}, {Graham}, {Graham},
  {Jencson}, {Karamehmetoglu}, {Kasliwal}, {Kulkarni}, {Miller}, {Neill},
  {Prince}, {Riddle}, {Rusholme}, {Sharma}, {Smith}, {Sravan}, {Taggart},
  {Walters}, \& {Yan}}]{2023ApJ...952..119B}
{Bruch}, R.~J., {Gal-Yam}, A., {Yaron}, O., {et~al.} 2023, \apj, 952, 119

\bibitem[{Chambers {et~al.}(2019)Chambers, Magnier, Metcalfe, Flewelling,
  Huber, Waters, Denneau, Draper, Farrow, Finkbeiner, Holmberg, Koppenhoefer,
  Price, Rest, Saglia, Schlafly, Smartt, Sweeney, Wainscoat, Burgett, Chastel,
  Grav, Heasley, Hodapp, Jedicke, Kaiser, Kudritzki, Luppino, Lupton, Monet,
  Morgan, Onaka, Shiao, Stubbs, Tonry, White, Bañados, Bell, Bender, Bernard,
  Boegner, Boffi, Botticella, Calamida, Casertano, Chen, Chen, Cole, Deacon,
  Frenk, Fitzsimmons, Gezari, Gibbs, Goessl, Goggia, Gourgue, Goldman, Grant,
  Grebel, Hambly, Hasinger, Heavens, Heckman, Henderson, Henning, Holman, Hopp,
  Ip, Isani, Jackson, Keyes, Koekemoer, Kotak, Le, Liska, Long, Lucey, Liu,
  Martin, Masci, McLean, Mindel, Misra, Morganson, Murphy, Obaika, Narayan,
  Nieto-Santisteban, Norberg, Peacock, Pier, Postman, Primak, Rae, Rai, Riess,
  Riffeser, Rix, Röser, Russel, Rutz, Schilbach, Schultz, Scolnic, Strolger,
  Szalay, Seitz, Small, Smith, Soderblom, Taylor, Thomson, Taylor, Thakar,
  Thiel, Thilker, Unger, Urata, Valenti, Wagner, Walder, Walter, Watters,
  Werner, Wood-Vasey, \& Wyse}]{chambers2019panstarrs1surveys}
Chambers, K.~C., Magnier, E.~A., Metcalfe, N., {et~al.} 2019, The Pan-STARRS1
  Surveys

\bibitem[{{Chugai}(2001)}]{2001MNRAS.326.1448C}
{Chugai}, N.~N. 2001, \mnras, 326, 1448

\bibitem[{Collaboration {et~al.}(2022)Collaboration, Vallenari, Brown, \&
  Prusti}]{GaiaCollaboration2022}
Collaboration, G., Vallenari, A., Brown, A., \& Prusti, T. 2022, Astronomy \&
  Astrophysics, 9, 10

\bibitem[{{de Jaeger} {et~al.}(2015){de Jaeger}, {Anderson}, {Pignata},
  {Hamuy}, {Kankare}, {Stritzinger}, {Benetti}, {Bufano}, {Elias-Rosa},
  {Folatelli}, {F{\"o}rster}, {Gonz{\'a}lez-Gait{\'a}n}, {Guti{\'e}rrez},
  {Inserra}, {Kotak}, {Lira}, {Morrell}, {Taddia}, \&
  {Tomasella}}]{2015ApJ...807...63D}
{de Jaeger}, T., {Anderson}, J.~P., {Pignata}, G., {et~al.} 2015, \apj, 807, 63

\bibitem[{{Dekker} {et~al.}(2000){Dekker}, {D'Odorico}, {Kaufer}, {Delabre}, \&
  {Kotzlowski}}]{UVES}
{Dekker}, H., {D'Odorico}, S., {Kaufer}, A., {Delabre}, B., \& {Kotzlowski}, H.
  2000, in Society of Photo-Optical Instrumentation Engineers (SPIE) Conference
  Series, Vol. 4008, Optical and IR Telescope Instrumentation and Detectors,
  ed. M.~{Iye} \& A.~F. {Moorwood}, 534--545

\bibitem[{{Dessart} {et~al.}(2016){Dessart}, {Hillier}, {Woosley}, {Livne},
  {Waldman}, {Yoon}, \& {Langer}}]{2016MNRAS.458.1618D}
{Dessart}, L., {Hillier}, D.~J., {Woosley}, S., {et~al.} 2016, \mnras, 458,
  1618

\bibitem[{{Dey} {et~al.}(2019){Dey}, {Schlegel}, {Lang}, {Blum}, {Burleigh},
  {Fan}, {Findlay}, {Finkbeiner}, {Herrera}, {Juneau}, {Landriau}, {Levi},
  {McGreer}, {Meisner}, {Myers}, {Moustakas}, {Nugent}, {Patej}, {Schlafly},
  {Walker}, {Valdes}, {Weaver}, {Y{\`e}che}, {Zou}, {Zhou}, {Abareshi},
  {Abbott}, {Abolfathi}, {Aguilera}, {Alam}, {Allen}, {Alvarez}, {Annis},
  {Ansarinejad}, {Aubert}, {Beechert}, {Bell}, {BenZvi}, {Beutler}, {Bielby},
  {Bolton}, {Brice{\~n}o}, {Buckley-Geer}, {Butler}, {Calamida}, {Carlberg},
  {Carter}, {Casas}, {Castander}, {Choi}, {Comparat}, {Cukanovaite}, {Delubac},
  {DeVries}, {Dey}, {Dhungana}, {Dickinson}, {Ding}, {Donaldson}, {Duan},
  {Duckworth}, {Eftekharzadeh}, {Eisenstein}, {Etourneau}, {Fagrelius},
  {Farihi}, {Fitzpatrick}, {Font-Ribera}, {Fulmer}, {G{\"a}nsicke},
  {Gaztanaga}, {George}, {Gerdes}, {Gontcho}, {Gorgoni}, {Green}, {Guy},
  {Harmer}, {Hernandez}, {Honscheid}, {Huang}, {James}, {Jannuzi}, {Jiang},
  {Joyce}, {Karcher}, {Karkar}, {Kehoe}, {Kneib}, {Kueter-Young}, {Lan},
  {Lauer}, {Le Guillou}, {Le Van Suu}, {Lee}, {Lesser}, {Perreault Levasseur},
  {Li}, {Mann}, {Marshall}, {Mart{\'\i}nez-V{\'a}zquez}, {Martini}, {du Mas des
  Bourboux}, {McManus}, {Meier}, {M{\'e}nard}, {Metcalfe},
  {Mu{\~n}oz-Guti{\'e}rrez}, {Najita}, {Napier}, {Narayan}, {Newman}, {Nie},
  {Nord}, {Norman}, {Olsen}, {Paat}, {Palanque-Delabrouille}, {Peng},
  {Poppett}, {Poremba}, {Prakash}, {Rabinowitz}, {Raichoor}, {Rezaie},
  {Robertson}, {Roe}, {Ross}, {Ross}, {Rudnick}, {Safonova}, {Saha},
  {S{\'a}nchez}, {Savary}, {Schweiker}, {Scott}, {Seo}, {Shan}, {Silva},
  {Slepian}, {Soto}, {Sprayberry}, {Staten}, {Stillman}, {Stupak}, {Summers},
  {Sien Tie}, {Tirado}, {Vargas-Maga{\~n}a}, {Vivas}, {Wechsler}, {Williams},
  {Yang}, {Yang}, {Yapici}, {Zaritsky}, {Zenteno}, {Zhang}, {Zhang}, {Zhou}, \&
  {Zhou}}]{Dey19}
{Dey}, A., {Schlegel}, D.~J., {Lang}, D., {et~al.} 2019, \aj, 157, 168

\bibitem[{{Dhawan} {et~al.}(2018){Dhawan}, {Fl{\"o}rs}, {Leibundgut},
  {Maguire}, {Kerzendorf}, {Taubenberger}, {Van Kerkwijk}, \&
  {Spyromilio}}]{2018A&A...619A.102D}
{Dhawan}, S., {Fl{\"o}rs}, A., {Leibundgut}, B., {et~al.} 2018, \aap, 619, A102

\bibitem[{{Dong} {et~al.}(2022){Dong}, {Valenti}, {Polin}, {Boyle},
  {Fl{\"o}rs}, {Vogl}, {Kerzendorf}, {Sand}, {Jha}, {Wyrzykowski}, {Bostroem},
  {Pearson}, {McCully}, {Andrews}, {Benetti}, {Blondin}, {Galbany},
  {Gromadzki}, {Hosseinzadeh}, {Howell}, {Inserra}, {Jencson}, {Lundquist},
  {Lyman}, {Magee}, {Maguire}, {Meza}, {Srivastav}, {Taubenberger}, {Terwel},
  {Wyatt}, \& {Young}}]{2022ApJ...934..102D}
{Dong}, Y., {Valenti}, S., {Polin}, A., {et~al.} 2022, \apj, 934, 102

\bibitem[{{Ercolino} {et~al.}(2025){Ercolino}, {Jin}, {Langer}, \&
  {Dessart}}]{2025A&A...696A.103E}
{Ercolino}, A., {Jin}, H., {Langer}, N., \& {Dessart}, L. 2025, \aap, 696, A103

\bibitem[{{Farias} {et~al.}(2024){Farias}, {Gall}, {Narayan}, {Rest}, {Villar},
  {Angus}, {Auchettl}, {Davis}, {Foley}, {Gagliano}, {Hjorth}, {Izzo},
  {Kilpatrick}, {Perkins}, {Ramirez-Ruiz}, {Ransome}, {Sarangi}, {Yarza},
  {Coulter}, {Jones}, {Khetan}, {Rest}, {Siebert}, {Swift}, {Taggart},
  {Tinyanont}, {Wrubel}, {de Boer}, {Clever}, {Dhara}, {Gao}, \&
  {Lin}}]{Farias_2024}
{Farias}, D., {Gall}, C., {Narayan}, G., {et~al.} 2024, \apj, 977, 152

\bibitem[{{Ferretti} {et~al.}(2017){Ferretti}, {Amanullah}, {Goobar},
  {Petrushevska}, {Borthakur}, {Bulla}, {Fox}, {Freeland}, {Fremling},
  {Hangard}, \& {Hayes}}]{2017A&A...606A.111F}
{Ferretti}, R., {Amanullah}, R., {Goobar}, A., {et~al.} 2017, \aap, 606, A111

\bibitem[{{Foley} {et~al.}(2013){Foley}, {Challis}, {Chornock},
  {Ganeshalingam}, {Li}, {Marion}, {Morrell}, {Pignata}, {Stritzinger},
  {Silverman}, {Wang}, {Anderson}, {Filippenko}, {Freedman}, {Hamuy}, {Jha},
  {Kirshner}, {McCully}, {Persson}, {Phillips}, {Reichart}, \&
  {Soderberg}}]{2013ApJ...767...57F}
{Foley}, R.~J., {Challis}, P.~J., {Chornock}, R., {et~al.} 2013, \apj, 767, 57

\bibitem[{{Fransson} {et~al.}(2014){Fransson}, {Ergon}, {Challis}, {Chevalier},
  {France}, {Kirshner}, {Marion}, {Milisavljevic}, {Smith}, {Bufano},
  {Friedman}, {Kangas}, {Larsson}, {Mattila}, {Benetti}, {Chornock}, {Czekala},
  {Soderberg}, \& {Sollerman}}]{2014ApJ...797..118F}
{Fransson}, C., {Ergon}, M., {Challis}, P.~J., {et~al.} 2014, \apj, 797, 118

\bibitem[{{Fraser}(2020)}]{Fraser_2020}
{Fraser}, M. 2020, Royal Society Open Science, 7, 200467

\bibitem[{{Gagnier} \& {Pejcha}(2023)}]{2023A&A...674A.121G}
{Gagnier}, D. \& {Pejcha}, O. 2023, \aap, 674, A121

\bibitem[{{Gal-Yam}(2019)}]{GalYam_2019}
{Gal-Yam}, A. 2019, \araa, 57, 305

\bibitem[{{Gal-Yam} {et~al.}(2022){Gal-Yam}, {Bruch}, {Schulze}, {Yang},
  {Perley}, {Irani}, {Sollerman}, {Kool}, {Soumagnac}, {Yaron}, {Strotjohann},
  {Zimmerman}, {Barbarino}, {Kulkarni}, {Kasliwal}, {De}, {Yao}, {Fremling},
  {Yan}, {Ofek}, {Fransson}, {Filippenko}, {Zheng}, {Brink}, {Copperwheat},
  {Foley}, {Brown}, {Siebert}, {Leloudas}, {Cabrera-Lavers}, {Garcia-Alvarez},
  {Marante-Barreto}, {Frederick}, {Hung}, {Wheeler}, {Vink{\'o}}, {Thomas},
  {Graham}, {Duev}, {Drake}, {Dekany}, {Bellm}, {Rusholme}, {Shupe},
  {Andreoni}, {Sharma}, {Riddle}, {van Roestel}, \& {Knezevic}}]{GalYam_2022}
{Gal-Yam}, A., {Bruch}, R., {Schulze}, S., {et~al.} 2022, \nat, 601, 201

\bibitem[{{Gal-Yam} {et~al.}(2021){Gal-Yam}, {Yaron}, {Pastorello},
  {Taubenberger}, {Fraser}, \& {Perley}}]{GalYam_2021}
{Gal-Yam}, A., {Yaron}, O., {Pastorello}, A., {et~al.} 2021, Transient Name
  Server AstroNote, 76, 1

\bibitem[{{Gangopadhyay} {et~al.}(2025){Gangopadhyay}, {Dukiya}, {Moriya},
  {Tanaka}, {Maeda}, {Howell}, {Singh}, {Singh}, {Sollerman}, {Kawabata},
  {Brennan}, {Pellegrino}, {Dastidar}, {Nakaoka}, {Kawabata}, {Misra},
  {Schulze}, {Chandra}, {Taguchi}, {Sahu}, {McCully}, {Bostroem}, {Gonzalez},
  {Newsome}, {Hiramatsu}, {Takei}, {Yamanaka}, {Tajitsu}, \&
  {Isogai}}]{Gangopadhyay_2025}
{Gangopadhyay}, A., {Dukiya}, N., {Moriya}, T.~J., {et~al.} 2025, \mnras, 537,
  2898

\bibitem[{{Gerardy} {et~al.}(2000){Gerardy}, {Fesen}, {H{\"o}flich}, \&
  {Wheeler}}]{Gerardy2000}
{Gerardy}, C.~L., {Fesen}, R.~A., {H{\"o}flich}, P., \& {Wheeler}, J.~C. 2000,
  \aj, 119, 2968

\bibitem[{{Gezari}(2021)}]{Gezari_2021}
{Gezari}, S. 2021, \araa, 59, 21

\bibitem[{González-Gaitán {et~al.}(2020)González-Gaitán, Mourão, Patat,
  Anderson, Cikota, Wiersema, Higgins, \& Silva}]{Gonzalez-Gaitan2020}
González-Gaitán, S., Mourão, A.~M., Patat, F., {et~al.} 2020, A\&A, 634

\bibitem[{{Graham}(2001)}]{2001AJ....121..820G}
{Graham}, A.~W. 2001, \aj, 121, 820

\bibitem[{{Groot} {et~al.}(2024){Groot}, {Bloemen}, {Vreeswijk}, {van Roestel},
  {Jonker}, {Nelemans}, {Klein-Wolt}, {Lepoole}, {Pieterse}, {Rodenhuis},
  {Boland}, {Haverkorn}, {Aerts}, {Bakker}, {Balster}, {Bekema}, {Dijkstra},
  {Dolron}, {Elswijk}, {van Elteren}, {Engels}, {Fokker}, {de Haan}, {Hahn},
  {ter Horst}, {Lesman}, {Kragt}, {Morren}, {Nillissen}, {Pessemier}, {Raskin},
  {de Rijke}, {Scheers}, {Schuil}, {Timmer}, {Antunes Amaral},
  {Arancibia-Rojas}, {Arcavi}, {Blagorodnova}, {Biswas}, {Breton}, {Dawson},
  {Dayal}, {De Wet}, {Duffy}, {Faris}, {Fausnaugh}, {Gal-Yam}, {Geier},
  {Horesh}, {Johnston}, {Katusiime}, {Kelley}, {Kosakowski}, {Kupfer},
  {Leloudas}, {Levan}, {Modiano}, {Mogawana}, {Munday}, {Paice}, {Patat},
  {Pelisoli}, {Ramsay}, {Ranaivomanana}, {Ruiz-Carmona}, {Schaffenroth},
  {Scaringi}, {Stoppa}, {Street}, {Tranin}, {Uzundag}, {Valenti},
  {Veresvarska}, {Vuc̆kovi{\'c}}, {Wichern}, {Wijers}, {Wijnands}, \&
  {Zimmerman}}]{2024PASP..136k5003G}
{Groot}, P.~J., {Bloemen}, S., {Vreeswijk}, P.~M., {et~al.} 2024, \pasp, 136,
  115003

\bibitem[{{Hakobyan} {et~al.}(2009){Hakobyan}, {Mamon}, {Petrosian}, {Kunth},
  \& {Turatto}}]{Hakobyan09}
{Hakobyan}, A.~A., {Mamon}, G.~A., {Petrosian}, A.~R., {Kunth}, D., \&
  {Turatto}, M. 2009, \aap, 508, 1259

\bibitem[{{Hoeflich} {et~al.}(2021){Hoeflich}, {Ashall}, {Bose}, {Baron},
  {Stritzinger}, {Davis}, {Shahbandeh}, {Anand}, {Baade}, {Burns}, {Collins},
  {Diamond}, {Fisher}, {Galbany}, {Hristov}, {Hsiao}, {Phillips}, {Shappee},
  {Suntzeff}, \& {Tucker}}]{2021ApJ...922..186H}
{Hoeflich}, P., {Ashall}, C., {Bose}, S., {et~al.} 2021, \apj, 922, 186

\bibitem[{Hoffman {et~al.}(2008)Hoffman, Leonard, Chornock, Filippenko, Barth,
  \& Matheson}]{Hoffman2008}
Hoffman, J.~L., Leonard, D.~C., Chornock, R., {et~al.} 2008, The Astrophysical
  Journal, 688, 1186

\bibitem[{{Hu} {et~al.}(2026){Hu}, {Yan}, {Wang}, {Iskandar}, {Zhang}, {Li},
  {Esamdin}, {Wang}, {Wang}, {Filippenko}, {Brink}, {Chen}, {Huang}, \&
  {Wang}}]{2026arXiv260101333H}
{Hu}, M., {Yan}, S., {Wang}, X., {et~al.} 2026, arXiv e-prints,
  arXiv:2601.01333

\bibitem[{{Inoue} \& {Maeda}(2025)}]{Inoue_2025}
{Inoue}, Y. \& {Maeda}, K. 2025, \apj, 980, 86

\bibitem[{{Inserra}(2019)}]{Inserra_2019}
{Inserra}, C. 2019, Nature Astronomy, 3, 697

\bibitem[{{Inserra} {et~al.}(2015){Inserra}, {Sim}, {Wyrzykowski}, {Smartt},
  {Fraser}, {Nicholl}, {Shen}, {Jerkstrand}, {Gal-Yam}, {Howell}, {Maguire},
  {Mazzali}, {Valenti}, {Taubenberger}, {Benitez-Herrera}, {Bersier},
  {Blagorodnova}, {Campbell}, {Chen}, {Elias-Rosa}, {Hillebrandt},
  {Kostrzewa-Rutkowska}, {Koz{\l}owski}, {Kromer}, {Lyman}, {Polshaw},
  {R{\"o}pke}, {Ruiter}, {Smith}, {Spiro}, {Sullivan}, {Yaron}, {Young}, \&
  {Yuan}}]{2015ApJ...799L...2I}
{Inserra}, C., {Sim}, S.~A., {Wyrzykowski}, L., {et~al.} 2015, \apjl, 799, L2

\bibitem[{{Ivanova} {et~al.}(2013){Ivanova}, {Justham}, {Chen}, {De Marco},
  {Fryer}, {Gaburov}, {Ge}, {Glebbeek}, {Han}, {Li}, {Lu}, {Marsh},
  {Podsiadlowski}, {Potter}, {Soker}, {Taam}, {Tauris}, {van den Heuvel}, \&
  {Webbink}}]{2013A&ARv..21...59I}
{Ivanova}, N., {Justham}, S., {Chen}, X., {et~al.} 2013, \aapr, 21, 59

\bibitem[{{Jacobson-Gal{\'a}n} {et~al.}(2024){Jacobson-Gal{\'a}n}, {Dessart},
  {Davis}, {Kilpatrick}, {Margutti}, {Foley}, {Chornock}, {Terreran},
  {Hiramatsu}, {Newsome}, {Padilla Gonzalez}, {Pellegrino}, {Howell},
  {Filippenko}, {Anderson}, {Angus}, {Auchettl}, {Bostroem}, {Brink},
  {Cartier}, {Coulter}, {de Boer}, {Drout}, {Earl}, {Ertini}, {Farah},
  {Farias}, {Gall}, {Gao}, {Gerlach}, {Guo}, {Haynie}, {Hosseinzadeh}, {Ibik},
  {Jha}, {Jones}, {Langeroodi}, {LeBaron}, {Magnier}, {Piro}, {Raimundo},
  {Rest}, {Rest}, {Rich}, {Rojas-Bravo}, {Sears}, {Taggart}, {Villar},
  {Wainscoat}, {Wang}, {Wasserman}, {Yan}, {Yang}, {Zhang}, \&
  {Zheng}}]{2024ApJ...970..189J}
{Jacobson-Gal{\'a}n}, W.~V., {Dessart}, L., {Davis}, K.~W., {et~al.} 2024,
  \apj, 970, 189

\bibitem[{{Jenniskens} \& {Desert}(1994)}]{Jenniskens94}
{Jenniskens}, P. \& {Desert}, F.~X. 1994, \aaps, 106, 39

\bibitem[{{Jerkstrand} {et~al.}(2015){Jerkstrand}, {Smartt}, {Sollerman},
  {Inserra}, {Fraser}, {Spyromilio}, {Fransson}, {Chen}, {Barbarino},
  {Dall'Ora}, {Botticella}, {Della Valle}, {Gal-Yam}, {Valenti}, {Maguire},
  {Mazzali}, \& {Tomasella}}]{2015MNRAS.448.2482J}
{Jerkstrand}, A., {Smartt}, S.~J., {Sollerman}, J., {et~al.} 2015, \mnras, 448,
  2482

\bibitem[{{Karambelkar} {et~al.}(2021){Karambelkar}, {Kasliwal}, {Maguire},
  {Anand}, {Andreoni}, {De}, {Drake}, {Duev}, {Graham}, {Kool}, {Laher},
  {Magee}, {Mahabal}, {Medford}, {Perley}, {Rigault}, {Rusholme}, {Schulze},
  {Sharma}, {Sollerman}, {Tzanidakis}, {Walters}, \&
  {Yao}}]{2021ApJ...921L...6K}
{Karambelkar}, V.~R., {Kasliwal}, M.~M., {Maguire}, K., {et~al.} 2021, \apjl,
  921, L6

\bibitem[{{Kashi} \& {Soker}(2011)}]{Kashi11}
{Kashi}, A. \& {Soker}, N. 2011, \mnras, 417, 1466

\bibitem[{{Kool} {et~al.}(2023){Kool}, {Johansson}, {Sollerman}, {Mold{\'o}n},
  {Moriya}, {Mattila}, {Schulze}, {Chomiuk}, {P{\'e}rez-Torres}, {Harris},
  {Lundqvist}, {Graham}, {Yang}, {Perley}, {Strotjohann}, {Fremling},
  {Gal-Yam}, {Lezmy}, {Maguire}, {Omand}, {Smith}, {Andreoni}, {Bellm},
  {Bloom}, {De}, {Groom}, {Kasliwal}, {Masci}, {Medford}, {Park}, {Purdum},
  {Reynolds}, {Riddle}, {Robert}, {Ryder}, {Sharma}, \&
  {Stern}}]{2023Natur.617..477K}
{Kool}, E.~C., {Johansson}, J., {Sollerman}, J., {et~al.} 2023, \nat, 617, 477

\bibitem[{{Kwok} {et~al.}(2025){Kwok}, {Singh}, {Jha}, {Blondin}, {Dastidar},
  {Larison}, {Miller}, {Andrews}, {Andrews}, {Anupama}, {Auchettl},
  {B{\'a}nhidi}, {Barna}, {Bostroem}, {Brink}, {Cartier}, {Chen}, {Christy},
  {Coulter}, {Covarrubias}, {Davis}, {Dickinson}, {Dong}, {Farah},
  {Filippenko}, {Fl{\"o}rs}, {Foley}, {Franz}, {Fremling}, {Galbany},
  {Gangopadhyay}, {Garg}, {Garnavich}, {Gates}, {Graur}, {Gordon}, {Hiramatsu},
  {Hoang}, {Howell}, {Hsu}, {Johansson}, {Joshi}, {Kahinga}, {Kaur}, {Kumar},
  {Kumnurdmanee}, {Kuncarayakti}, {LeBaron}, {Liu}, {Maeda}, {Maguire},
  {McCully}, {Mehta}, {Menotti}, {Metevier}, {Misra}, {Murphey}, {Newsome},
  {Padilla Gonzalez}, {Patra}, {Pearson}, {Piro}, {Polin}, {Ravi}, {Rest},
  {Rehemtulla}, {Meza Retamal}, {Robinson}, {Rojas-Bravo}, {Sahu}, {Sand},
  {Schmidt}, {Schulze}, {Schwab}, {Shrestha}, {Siebert}, {Simha}, {Smith},
  {Sollerman}, {Subrayan}, {Szalai}, {Taggart}, {Teja}, {Temim}, {Terwel},
  {Tinyanont}, {Valenti}, {Anais Vilchez}, {Vink{\'o}}, {Westerling}, {Yang},
  \& {Zheng}}]{2025ApJ...989L..33K}
{Kwok}, L.~A., {Singh}, M., {Jha}, S.~W., {et~al.} 2025, \apjl, 989, L33

\bibitem[{{Lang} {et~al.}(2016){Lang}, {Hogg}, \&
  {Mykytyn}}]{2016ascl.soft04008L}
{Lang}, D., {Hogg}, D.~W., \& {Mykytyn}, D. 2016, {The Tractor: Probabilistic
  astronomical source detection and measurement}, Astrophysics Source Code
  Library, record ascl:1604.008

\bibitem[{{Lau} {et~al.}(2022){Lau}, {Hankins}, {Han}, {Argyriou}, {Corcoran},
  {Eldridge}, {Endo}, {Fox}, {Garcia Marin}, {Gull}, {Jones}, {Hamaguchi},
  {Lamberts}, {Law}, {Madura}, {Marchenko}, {Matsuhara}, {Moffat}, {Morris},
  {Morris}, {Onaka}, {Ressler}, {Richardson}, {Russell}, {Sanchez-Bermudez},
  {Smith}, {Soulain}, {Stevens}, {Tuthill}, {Weigelt}, {Williams}, \&
  {Yamaguchi}}]{2022NatAs...6.1308L}
{Lau}, R.~M., {Hankins}, M.~J., {Han}, Y., {et~al.} 2022, Nature Astronomy, 6,
  1308

\bibitem[{{Livio} \& {Riess}(2003)}]{Livio03}
{Livio}, M. \& {Riess}, A.~G. 2003, \apjl, 594, L93

\bibitem[{{Lyman} {et~al.}(2018){Lyman}, {Taddia}, {Stritzinger}, {Galbany},
  {Leloudas}, {Anderson}, {Eldridge}, {James}, {Kr{\"u}hler}, {Levan},
  {Pignata}, \& {Stanway}}]{2018MNRAS.473.1359L}
{Lyman}, J.~D., {Taddia}, F., {Stritzinger}, M.~D., {et~al.} 2018, \mnras, 473,
  1359

\bibitem[{Maeda \& Moriya(2022)}]{Maeda2022}
Maeda, K. \& Moriya, T.~J. 2022, The Astrophysical Journal, 927, 25

\bibitem[{{Mattila} {et~al.}(2016){Mattila}, {Elias-Rosa}, {Lundqvist},
  {Stritzinger}, {Kuncarayakti}, {Harmanen}, {Pastorello}, {Benetti},
  {Cappellaro}, {Blagorodnova}, {Davis}, {Dong}, {Fraser}, {Gall}, {Harrison},
  {Hodgkin}, {Hsiao}, {Jonker}, {Kangas}, {Kankare}, {Kostrzewa-Rutkowska},
  {Nielsen}, {Ochner}, {Prieto}, {Reynolds}, {Romero-Canizales}, {Taddia},
  {Tartaglia}, {Terreran}, {Tomasella}, \& {Wyrzykowski}}]{2016ATel.8992....1M}
{Mattila}, S., {Elias-Rosa}, N., {Lundqvist}, P., {et~al.} 2016, The
  Astronomer's Telegram, 8992, 1

\bibitem[{Mattila {et~al.}(2008)Mattila, Meikle, Lundqvist, Pastorello, Kotak,
  Eldridge, Smartt, Adamson, Gerardy, Rizzi, Stephens, \& Dyk}]{Mattila2008}
Mattila, S., Meikle, W.~P., Lundqvist, P., {et~al.} 2008, Monthly Notices of
  the Royal Astronomical Society, 389, 141

\bibitem[{{Metzger}(2022)}]{Metzger_2022}
{Metzger}, B.~D. 2022, \apj, 932, 84

\bibitem[{{Miller} {et~al.}(2019){Miller}, {van Dokkum}, {Mowla}, \& {van der
  Wel}}]{2019ApJ...872L..14M}
{Miller}, T.~B., {van Dokkum}, P., {Mowla}, L., \& {van der Wel}, A. 2019,
  \apjl, 872, L14

\bibitem[{{Mir{\'o}-Carretero} {et~al.}(2024){Mir{\'o}-Carretero},
  {Mart{\'\i}nez-Delgado}, {G{\'o}mez-Flechoso}, {Cooper}, {Akhlaghi},
  {Donatiello}, {Kuijken}, {Lang}, {Makarov}, {Laine}, \&
  {Roca-F{\`a}brega}}]{2024A&A...691A.196M}
{Mir{\'o}-Carretero}, J., {Mart{\'\i}nez-Delgado}, D., {G{\'o}mez-Flechoso},
  M.~A., {et~al.} 2024, \aap, 691, A196

\bibitem[{{Moore}(1945)}]{Moore45}
{Moore}, C.~E. 1945, Contributions from the Princeton University Observatory,
  20, 1

\bibitem[{{Moore} {et~al.}(2023){Moore}, {Smartt}, {Nicholl}, {Srivastav},
  {Stevance}, {Jess}, {Grant}, {Fulton}, {Rhodes}, {Sim}, {Hirai},
  {Podsiadlowski}, {Anderson}, {Ashall}, {Bate}, {Fender}, {Guti{\'e}rrez},
  {Howell}, {Huber}, {Inserra}, {Leloudas}, {Monard}, {M{\"u}ller-Bravo},
  {Shappee}, {Smith}, {Terreran}, {Tonry}, {Tucker}, {Young}, {Aamer}, {Chen},
  {Ragosta}, {Galbany}, {Gromadzki}, {Harvey}, {Hoeflich}, {McCully},
  {Newsome}, {Gonzalez}, {Pellegrino}, {Ramsden}, {P{\'e}rez-Torres}, {Ridley},
  {Sheng}, \& {Weston}}]{2023ApJ...956L..31M}
{Moore}, T., {Smartt}, S.~J., {Nicholl}, M., {et~al.} 2023, \apjl, 956, L31

\bibitem[{{Moriya} \& {Eldridge}(2016)}]{2016MNRAS.461.2155M}
{Moriya}, T.~J. \& {Eldridge}, J.~J. 2016, \mnras, 461, 2155

\bibitem[{{Moriya} {et~al.}(2014){Moriya}, {Tominaga}, {Langer}, {Nomoto},
  {Blinnikov}, \& {Sorokina}}]{2014A&A...569A..57M}
{Moriya}, T.~J., {Tominaga}, N., {Langer}, N., {et~al.} 2014, \aap, 569, A57

\bibitem[{{Mowla} {et~al.}(2019){Mowla}, {van der Wel}, {van Dokkum}, \&
  {Miller}}]{2019ApJ...872L..13M}
{Mowla}, L., {van der Wel}, A., {van Dokkum}, P., \& {Miller}, T.~B. 2019,
  \apjl, 872, L13

\bibitem[{{Murguia-Berthier} {et~al.}(2017){Murguia-Berthier}, {Ramirez-Ruiz},
  {Kilpatrick}, {Foley}, {Kasen}, {Lee}, {Piro}, {Coulter}, {Drout}, {Madore},
  {Shappee}, {Pan}, {Prochaska}, {Rest}, {Rojas-Bravo}, {Siebert}, \&
  {Simon}}]{2017ApJ...848L..34M}
{Murguia-Berthier}, A., {Ramirez-Ruiz}, E., {Kilpatrick}, C.~D., {et~al.} 2017,
  \apjl, 848, L34

\bibitem[{{Nagao} {et~al.}(2023){Nagao}, {Kuncarayakti}, {Maeda}, {Moore},
  {Pastorello}, {Mattila}, {Uno}, {Smartt}, {Sim}, {Ferrari}, {Tomasella},
  {Anderson}, {Chen}, {Galbany}, {Gao}, {Gromadzki}, {Guti{\'e}rrez},
  {Inserra}, {Kankare}, {Magnier}, {M{\"u}ller-Bravo}, {Reguitti}, \&
  {Young}}]{2023A&A...673A..27N}
{Nagao}, T., {Kuncarayakti}, H., {Maeda}, K., {et~al.} 2023, \aap, 673, A27

\bibitem[{{Noll} {et~al.}(2012){Noll}, {Kausch}, {Barden}, {Jones}, {Szyszka},
  {Kimeswenger}, \& {Vinther}}]{Noll12}
{Noll}, S., {Kausch}, W., {Barden}, M., {et~al.} 2012, \aap, 543, A92

\bibitem[{{Nugent} {et~al.}(2002){Nugent}, {Kim}, \& {Perlmutter}}]{Nugent02}
{Nugent}, P., {Kim}, A., \& {Perlmutter}, S. 2002, \pasp, 114, 803

\bibitem[{{Padilla Gonzalez} {et~al.}(2024){Padilla Gonzalez}, {Howell},
  {Terreran}, {McCully}, {Newsome}, {Burke}, {Farah}, {Pellegrino}, {Bostroem},
  {Hosseinzadeh}, {Pearson}, {Sand}, {Shrestha}, {Smith}, {Dong}, {Meza
  Retamal}, {Valenti}, {Boos}, {Shen}, {Townsley}, {Galbany}, {Piscarreta},
  {Foley}, {Bustamante-Rosell}, {Coulter}, {Chornock}, {Davis}, {Dickinson},
  {Jones}, {Kutcka}, {Le Saux}, {Rojas-Bravo}, {Taggart}, {Tinyanont}, {Yang},
  {Jha}, \& {Margutti}}]{2024ApJ...964..196P}
{Padilla Gonzalez}, E., {Howell}, D.~A., {Terreran}, G., {et~al.} 2024, \apj,
  964, 196

\bibitem[{{Pastorello} {et~al.}(2015){Pastorello}, {Benetti}, {Brown},
  {Tsvetkov}, {Inserra}, {Taubenberger}, {Tomasella}, {Fraser}, {Rich},
  {Botticella}, {Bufano}, {Cappellaro}, {Ergon}, {Gorbovskoy}, {Harutyunyan},
  {Huang}, {Kotak}, {Lipunov}, {Magill}, {Miluzio}, {Morrell}, {Ochner},
  {Smartt}, {Sollerman}, {Spiro}, {Stritzinger}, {Turatto}, {Valenti}, {Wang},
  {Wright}, {Yurkov}, {Zampieri}, \& {Zhang}}]{2015MNRAS.449.1921P}
{Pastorello}, A., {Benetti}, S., {Brown}, P.~J., {et~al.} 2015, \mnras, 449,
  1921

\bibitem[{{Pastorello} \& {Fraser}(2019)}]{Pasto_2019}
{Pastorello}, A. \& {Fraser}, M. 2019, Nature Astronomy, 3, 676

\bibitem[{{Pastorello} {et~al.}(2008{\natexlab{a}}){Pastorello}, {Mattila},
  {Zampieri}, {Della Valle}, {Smartt}, {Valenti}, {Agnoletto}, {Benetti},
  {Benn}, {Branch}, {Cappellaro}, {Dennefeld}, {Eldridge}, {Gal-Yam},
  {Harutyunyan}, {Hunter}, {Kjeldsen}, {Lipkin}, {Mazzali}, {Milne},
  {Navasardyan}, {Ofek}, {Pian}, {Shemmer}, {Spiro}, {Stathakis},
  {Taubenberger}, {Turatto}, \& {Yamaoka}}]{Pasto_2008a}
{Pastorello}, A., {Mattila}, S., {Zampieri}, L., {et~al.} 2008{\natexlab{a}},
  \mnras, 389, 113

\bibitem[{{Pastorello} {et~al.}(2008{\natexlab{b}}){Pastorello}, {Quimby},
  {Smartt}, {Mattila}, {Navasardyan}, {Crockett}, {Elias-Rosa}, {Mondol},
  {Wheeler}, \& {Young}}]{Pasto_2008b}
{Pastorello}, A., {Quimby}, R.~M., {Smartt}, S.~J., {et~al.}
  2008{\natexlab{b}}, \mnras, 389, 131

\bibitem[{{Pastorello} {et~al.}(2021){Pastorello}, {Vogl}, {Taubenberger},
  {Floers}, {Csoernyei}, {Cudmani}, {Holas}, {Hillebrandt}, {Suyu}, {Blondin},
  {Leibundgut}, {Spyromilio}, {Smartt}, {Dobson}, {Kotak}, {Bruch}, {Gal-Yam},
  \& {Lemon}}]{Pasto_2021}
{Pastorello}, A., {Vogl}, C., {Taubenberger}, S., {et~al.} 2021, Transient Name
  Server AstroNote, 71, 1

\bibitem[{Patat {et~al.}(2011)Patat, Taubenberger, Benetti, Pastorello, \&
  Harutyunyan}]{Patat2011}
Patat, F., Taubenberger, S., Benetti, S., Pastorello, A., \& Harutyunyan, A.
  2011, Astronomy and Astrophysics, 527

\bibitem[{{Pellegrino} {et~al.}(2022){Pellegrino}, {Howell}, {Terreran},
  {Arcavi}, {Bostroem}, {Brown}, {Burke}, {Dong}, {Gilkis}, {Hiramatsu},
  {Hosseinzadeh}, {McCully}, {Modjaz}, {Newsome}, {Gonzalez}, {Pritchard},
  {Sand}, {Valenti}, \& {Williamson}}]{Pellegrino_2022}
{Pellegrino}, C., {Howell}, D.~A., {Terreran}, G., {et~al.} 2022, \apj, 938, 73

\bibitem[{{Perets} {et~al.}(2010){Perets}, {Gal-Yam}, {Mazzali}, {Arnett},
  {Kagan}, {Filippenko}, {Li}, {Arcavi}, {Cenko}, {Fox}, {Leonard}, {Moon},
  {Sand}, {Soderberg}, {Anderson}, {James}, {Foley}, {Ganeshalingam}, {Ofek},
  {Bildsten}, {Nelemans}, {Shen}, {Weinberg}, {Metzger}, {Piro}, {Quataert},
  {Kiewe}, \& {Poznanski}}]{2010Natur.465..322P}
{Perets}, H.~B., {Gal-Yam}, A., {Mazzali}, P.~A., {et~al.} 2010, \nat, 465, 322

\bibitem[{{Perley} {et~al.}(2019){Perley}, {Mazzali}, {Yan}, {Cenko}, {Gezari},
  {Taggart}, {Blagorodnova}, {Fremling}, {Mockler}, {Singh}, {Tominaga},
  {Tanaka}, {Watson}, {Ahumada}, {Anupama}, {Ashall}, {Becerra}, {Bersier},
  {Bhalerao}, {Bloom}, {Butler}, {Copperwheat}, {Coughlin}, {De}, {Drake},
  {Duev}, {Frederick}, {Gonz{\'a}lez}, {Goobar}, {Heida}, {Ho}, {Horst},
  {Hung}, {Itoh}, {Jencson}, {Kasliwal}, {Kawai}, {Khanam}, {Kulkarni},
  {Kumar}, {Kumar}, {Kutyrev}, {Lee}, {Maeda}, {Mahabal}, {Murata}, {Neill},
  {Ngeow}, {Penprase}, {Pian}, {Quimby}, {Ramirez-Ruiz}, {Richer},
  {Rom{\'a}n-Z{\'u}{\~n}iga}, {Sahu}, {Srivastav}, {Socia}, {Sollerman},
  {Tachibana}, {Taddia}, {Tinyanont}, {Troja}, {Ward}, {Wee}, \&
  {Yu}}]{2019MNRAS.484.1031P}
{Perley}, D.~A., {Mazzali}, P.~A., {Yan}, L., {et~al.} 2019, \mnras, 484, 1031

\bibitem[{Perley {et~al.}(2022)Perley, Sollerman, Schulze, Yao, Fremling,
  Gal-Yam, Ho, Yang, Kool, Irani, Yan, Andreoni, Baade, Bellm, Brink, Chen,
  Cikota, Coughlin, Dahiwale, Dekany, Duev, Filippenko, Hoeflich, Kasliwal,
  Kulkarni, Lunnan, Masci, Maund, Medford, Riddle, Rosnet, Shupe, Strotjohann,
  Tzanidakis, \& Zheng}]{Perley2022}
Perley, D.~A., Sollerman, J., Schulze, S., {et~al.} 2022, The Astrophysical
  Journal, 927, 180

\bibitem[{{Podsiadlowski} {et~al.}(2004){Podsiadlowski}, {Langer},
  {Poelarends}, {Rappaport}, {Heger}, \& {Pfahl}}]{2004ApJ...612.1044P}
{Podsiadlowski}, P., {Langer}, N., {Poelarends}, A.~J.~T., {et~al.} 2004, \apj,
  612, 1044

\bibitem[{{Pumo} {et~al.}(2009){Pumo}, {Turatto}, {Botticella}, {Pastorello},
  {Valenti}, {Zampieri}, {Benetti}, {Cappellaro}, \&
  {Patat}}]{2009ApJ...705L.138P}
{Pumo}, M.~L., {Turatto}, M., {Botticella}, M.~T., {et~al.} 2009, \apjl, 705,
  L138

\bibitem[{Pursiainen {et~al.}(2023)Pursiainen, Leloudas, Cikota, Bulla,
  Inserra, Patat, Wheeler, Aamer, Gal-Yam, Maund, Nicholl, Schulze, Sollerman,
  \& Yang}]{Pursiainen2023}
Pursiainen, M., Leloudas, G., Cikota, A., {et~al.} 2023, A\&A, 674, A81

\bibitem[{{Pursiainen} {et~al.}(2023){Pursiainen}, {Leloudas}, {Schulze},
  {Charalampopoulos}, {Angus}, {Anderson}, {Bauer}, {Chen}, {Galbany},
  {Gromadzki}, {Guti{\'e}rrez}, {Inserra}, {Lyman}, {M{\"u}ller-Bravo},
  {Nicholl}, {Smartt}, {Tartaglia}, {Wiseman}, \& {Young}}]{Pursiainen_2023a}
{Pursiainen}, M., {Leloudas}, G., {Schulze}, S., {et~al.} 2023, \apjl, 959, L10

\bibitem[{{Ramsay} {et~al.}(2018){Ramsay}, {Green}, {Marsh}, {Kupfer},
  {Breedt}, {Korol}, {Groot}, {Knigge}, {Nelemans}, {Steeghs}, {Woudt}, \&
  {Aungwerojwit}}]{Ramsay18}
{Ramsay}, G., {Green}, M.~J., {Marsh}, T.~R., {et~al.} 2018, \aap, 620, A141

\bibitem[{{Reguitti} {et~al.}(2022){Reguitti}, {Pastorello}, {Pignata},
  {Fraser}, {Stritzinger}, {Brennan}, {Cai}, {Elias-Rosa}, {Fugazza},
  {Gutierrez}, {Kankare}, {Kotak}, {Lundqvist}, {Mazzali}, {Moran}, {Salmaso},
  {Tomasella}, {Valerin}, \& {Kuncarayakti}}]{Reguitti_2022}
{Reguitti}, A., {Pastorello}, A., {Pignata}, G., {et~al.} 2022, \aap, 662, L10

\bibitem[{{Riess} {et~al.}(2022){Riess}, {Yuan}, {Macri}, {Scolnic}, {Brout},
  {Casertano}, {Jones}, {Murakami}, {Anand}, {Breuval}, {Brink}, {Filippenko},
  {Hoffmann}, {Jha}, {D'arcy Kenworthy}, {Mackenty}, {Stahl}, \&
  {Zheng}}]{2022ApJ...934L...7R}
{Riess}, A.~G., {Yuan}, W., {Macri}, L.~M., {et~al.} 2022, \apjl, 934, L7

\bibitem[{{Ruiter} {et~al.}(2019){Ruiter}, {Ferrario}, {Belczynski},
  {Seitenzahl}, {Crocker}, \& {Karakas}}]{2019MNRAS.484..698R}
{Ruiter}, A.~J., {Ferrario}, L., {Belczynski}, K., {et~al.} 2019, \mnras, 484,
  698

\bibitem[{{Sanders} {et~al.}(2013){Sanders}, {Soderberg}, {Foley}, {Chornock},
  {Milisavljevic}, {Margutti}, {Drout}, {Moe}, {Berger}, {Brown}, {Lunnan},
  {Smartt}, {Fraser}, {Kotak}, {Magill}, {Smith}, {Wright}, {Huang}, {Urata},
  {Mulchaey}, {Rest}, {Sand}, {Chomiuk}, {Friedman}, {Kirshner}, {Marion},
  {Tonry}, {Burgett}, {Chambers}, {Hodapp}, {Kudritzki}, \&
  {Price}}]{Sanders13}
{Sanders}, N.~E., {Soderberg}, A.~M., {Foley}, R.~J., {et~al.} 2013, \apj, 769,
  39

\bibitem[{{Schlafly} \& {Finkbeiner}(2011)}]{2011ApJ...737..103S}
{Schlafly}, E.~F. \& {Finkbeiner}, D.~P. 2011, \apj, 737, 103

\bibitem[{{Schlegel}(1990)}]{Schlegel_1990}
{Schlegel}, E.~M. 1990, \mnras, 244, 269

\bibitem[{{Schulze} {et~al.}(2025){Schulze}, {Gal-Yam}, {Dessart}, {Miller},
  {Woosley}, {Yang}, {Bulla}, {Yaron}, {Sollerman}, {Filippenko}, {Hinds},
  {Perley}, {Tsuna}, {Lunnan}, {Sarin}, {Brennan}, {Brink}, {Bruch}, {Chen},
  {Das}, {Dhawan}, {Fransson}, {Fremling}, {Gangopadhyay}, {Irani},
  {Jerkstrand}, {Kne{\v{z}}evi{\'c}}, {Kushnir}, {Maeda}, {Maguire}, {Ofek},
  {Omand}, {Qin}, {Sharma}, {Sit}, {Srinivasaragavan}, {Strothjohann}, {Takei},
  {Waxman}, {Yan}, {Yao}, {Zheng}, {Zimmerman}, {Bellm}, {Coughlin}, {Masci},
  {Purdum}, {Rigault}, {Wold}, \& {Kulkarni}}]{2025Natur.644..634S}
{Schulze}, S., {Gal-Yam}, A., {Dessart}, L., {et~al.} 2025, \nat, 644, 634

\bibitem[{{Schulze} {et~al.}(2021){Schulze}, {Yaron}, {Sollerman}, {Leloudas},
  {Gal}, {Wright}, {Lunnan}, {Gal-Yam}, {Ofek}, {Perley}, {Filippenko},
  {Kasliwal}, {Kulkarni}, {Neill}, {Nugent}, {Quimby}, {Sullivan},
  {Strotjohann}, {Arcavi}, {Ben-Ami}, {Bianco}, {Bloom}, {De}, {Fraser},
  {Fremling}, {Horesh}, {Johansson}, {Kelly}, {Kne{\v{z}}evi{\'c}},
  {Kne{\v{z}}evi{\'c}}, {Maguire}, {Nyholm}, {Papadogiannakis}, {Petrushevska},
  {Rubin}, {Yan}, {Yang}, {Adams}, {Bufano}, {Clubb}, {Foley}, {Green},
  {Harmanen}, {Ho}, {Hook}, {Hosseinzadeh}, {Howell}, {Kong}, {Kotak},
  {Matheson}, {McCully}, {Milisavljevic}, {Pan}, {Poznanski}, {Shivvers}, {van
  Velzen}, \& {Verbeek}}]{2021ApJS..255...29S}
{Schulze}, S., {Yaron}, O., {Sollerman}, J., {et~al.} 2021, \apjs, 255, 29

\bibitem[{Serkowski(1973)}]{Serkowski1973}
Serkowski, K. 1973, in IAU Symposium Vol. 52 Interstellar Dust and Related
  Topics, ed. J.~M. Greenberg \& H.~C. van~de Hulst, 145

\bibitem[{Serkowski {et~al.}(1975)Serkowski, Mathewson, \&
  Ford}]{Serkowski1975}
Serkowski, K., Mathewson, D.~L., \& Ford, V.~L. 1975, ApJ, 196, 261

\bibitem[{{Shen}(2025)}]{2025ApJ...982....6S}
{Shen}, K.~J. 2025, \apj, 982, 6

\bibitem[{{Shen} {et~al.}(2019){Shen}, {Quataert}, \&
  {Pakmor}}]{2019ApJ...887..180S}
{Shen}, K.~J., {Quataert}, E., \& {Pakmor}, R. 2019, \apj, 887, 180

\bibitem[{{Shi} {et~al.}(2026){Shi}, {Auchettl}, {Hoogendam}, {Farias},
  {Sarin}, {Davis}, {Morrell}, {Hinkle}, {Jones}, {Lidman}, {Angus}, {Ashall},
  {Burns}, {Desai}, {Do}, {Galbany}, {Hsiao}, {Huber}, {Kong}, {Martin},
  {Medler}, {M{\"o}ller}, {Pfeffer}, {Polin}, {Rauf}, {Romagnoli}, {Schmidt},
  {Shappee}, {Stritzinger}, {Syncatto}, {Tucker}, \&
  {Tucker}}]{2026arXiv260216227S}
{Shi}, J., {Auchettl}, K., {Hoogendam}, W.~B., {et~al.} 2026, arXiv e-prints,
  arXiv:2602.16227

\bibitem[{{Shipp} {et~al.}(2018){Shipp}, {Drlica-Wagner}, {Balbinot},
  {Ferguson}, {Erkal}, {Li}, {Bechtol}, {Belokurov}, {Buncher}, {Carollo},
  {Carrasco Kind}, {Kuehn}, {Marshall}, {Pace}, {Rykoff}, {Sevilla-Noarbe},
  {Sheldon}, {Strigari}, {Vivas}, {Yanny}, {Zenteno}, {Abbott}, {Abdalla},
  {Allam}, {Avila}, {Bertin}, {Brooks}, {Burke}, {Carretero}, {Castander},
  {Cawthon}, {Crocce}, {Cunha}, {D'Andrea}, {da Costa}, {Davis}, {De Vicente},
  {Desai}, {Diehl}, {Doel}, {Evrard}, {Flaugher}, {Fosalba}, {Frieman},
  {Garc{\'\i}a-Bellido}, {Gaztanaga}, {Gerdes}, {Gruen}, {Gruendl}, {Gschwend},
  {Gutierrez}, {Hartley}, {Honscheid}, {Hoyle}, {James}, {Johnson}, {Krause},
  {Kuropatkin}, {Lahav}, {Lin}, {Maia}, {March}, {Martini}, {Menanteau},
  {Miller}, {Miquel}, {Nichol}, {Plazas}, {Romer}, {Sako}, {Sanchez},
  {Santiago}, {Scarpine}, {Schindler}, {Schubnell}, {Smith}, {Smith},
  {Sobreira}, {Suchyta}, {Swanson}, {Tarle}, {Thomas}, {Tucker}, {Walker},
  {Wechsler}, \& {DES Collaboration}}]{2018ApJ...862..114S}
{Shipp}, N., {Drlica-Wagner}, A., {Balbinot}, E., {et~al.} 2018, \apj, 862, 114

\bibitem[{Shivvers {et~al.}(2016)Shivvers, Zheng, Mauerhan, Kleiser, Dyk,
  Silverman, Graham, Kelly, Filippenko, \& Kumar}]{Shivvers2016}
Shivvers, I., Zheng, W., Mauerhan, J., {et~al.} 2016, Monthly Notices of the
  Royal Astronomical Society, 461, 3057

\bibitem[{Shivvers {et~al.}(2017)Shivvers, Zheng, Dyk, Mauerhan, Filippenko,
  Smith, Foley, Mazzali, Kamble, Kilpatrick, Margutti, Yuk, Graham, Kelly,
  Andrews, Matheson, Wood-Vasey, Ponder, Brown, Chevalier, Milisavljevic,
  Drout, Parrent, Soderberg, Ashall, Piascik, \& Prentice}]{Shivvers2017}
Shivvers, I., Zheng, W.~K., Dyk, S. D.~V., {et~al.} 2017, Monthly Notices of
  the Royal Astronomical Society, 471, 4381

\bibitem[{Shrestha {et~al.}(2025)Shrestha, DeSoto, Sand, Williams, Hoffman,
  Smith, McCall, Maund, Steele, Wiersema, Andrews, Smith, Bilinski, Milne,
  Anche, Bostroem, Hosseinzadeh, Pearson, Leonard, Hsu, 董, Hoang, Janzen,
  Jencson, Jha, Lundquist, Mehta, Retamal, Valenti, Farah, Howell, McCully,
  Newsome, Gonzalez, Pellegrino, \& Terreran}]{Shrestha2025}
Shrestha, M., DeSoto, S., Sand, D.~J., {et~al.} 2025, ApJ, 982, L32

\bibitem[{{Smith} {et~al.}(2020){Smith}, {Smartt}, {Young}, {Tonry}, {Denneau},
  {Flewelling}, {Heinze}, {Weiland}, {Stalder}, {Rest}, {Stubbs}, {Anderson},
  {Chen}, {Clark}, {Do}, {F{\"o}rster}, {Fulton}, {Gillanders}, {McBrien},
  {O'Neill}, {Srivastav}, \& {Wright}}]{2020PASP..132h5002S}
{Smith}, K.~W., {Smartt}, S.~J., {Young}, D.~R., {et~al.} 2020, \pasp, 132,
  085002

\bibitem[{{Smith}(2014)}]{Smith_2014}
{Smith}, N. 2014, \araa, 52, 487

\bibitem[{{Smith}(2017)}]{2017hsn..book..403S}
{Smith}, N. 2017, in Handbook of Supernovae, ed. A.~W. {Alsabti} \&
  P.~{Murdin}, 403

\bibitem[{{Smith} {et~al.}(2012{\natexlab{a}}){Smith}, {Cenko}, {Butler},
  {Bloom}, {Kasliwal}, {Horesh}, {Kulkarni}, {Law}, {Nugent}, {Ofek},
  {Poznanski}, {Quimby}, {Sesar}, {Ben-Ami}, {Arcavi}, {Gal-Yam}, {Polishook},
  {Xu}, {Yaron}, {Frail}, \& {Sullivan}}]{2012MNRAS.420.1135S}
{Smith}, N., {Cenko}, S.~B., {Butler}, N., {et~al.} 2012{\natexlab{a}}, \mnras,
  420, 1135

\bibitem[{{Smith} {et~al.}(2012{\natexlab{b}}){Smith}, {Mauerhan}, {Silverman},
  {Ganeshalingam}, {Filippenko}, {Cenko}, {Clubb}, \&
  {Kandrashoff}}]{Smith_2012}
{Smith}, N., {Mauerhan}, J.~C., {Silverman}, J.~M., {et~al.}
  2012{\natexlab{b}}, \mnras, 426, 1905

\bibitem[{{Sola} {et~al.}(2022){Sola}, {Duc}, {Richards}, {Paiement}, {Urbano},
  {Klehammer}, {B{\'\i}lek}, {Cuillandre}, {Gwyn}, \&
  {McConnachie}}]{2022A&A...662A.124S}
{Sola}, E., {Duc}, P.-A., {Richards}, F., {et~al.} 2022, \aap, 662, A124

\bibitem[{{Solheim}(2010)}]{Solheim10}
{Solheim}, J.~E. 2010, \pasp, 122, 1133

\bibitem[{{Srivastav} {et~al.}(2022){Srivastav}, {Smartt}, {Huber}, {Chambers},
  {Angus}, {Chen}, {Callan}, {Gillanders}, {McBrien}, {Sim}, {Fulton},
  {Hjorth}, {Smith}, {Young}, {Auchettl}, {Anderson}, {Pignata}, {de Boer},
  {Lin}, \& {Magnier}}]{Srivastav22}
{Srivastav}, S., {Smartt}, S.~J., {Huber}, M.~E., {et~al.} 2022, \mnras, 511,
  2708

\bibitem[{{Srivastav} {et~al.}(2020){Srivastav}, {Smartt}, {Leloudas}, {Huber},
  {Chambers}, {Malesani}, {Hjorth}, {Gillanders}, {Schultz}, {Sim}, {Auchettl},
  {Fynbo}, {Gall}, {McBrien}, {Rest}, {Smith}, {Wojtak}, \&
  {Young}}]{2020ApJ...892L..24S}
{Srivastav}, S., {Smartt}, S.~J., {Leloudas}, G., {et~al.} 2020, \apjl, 892,
  L24

\bibitem[{{Steeghs} {et~al.}(2022){Steeghs}, {Galloway}, {Ackley}, {Dyer},
  {Lyman}, {Ulaczyk}, {Cutter}, {Mong}, {Dhillon}, {O'Brien}, {Ramsay},
  {Poshyachinda}, {Kotak}, {Nuttall}, {Pall{\'e}}, {Breton}, {Pollacco},
  {Thrane}, {Aukkaravittayapun}, {Awiphan}, {Burhanudin}, {Chote}, {Chrimes},
  {Daw}, {Duffy}, {Eyles-Ferris}, {Gompertz}, {Heikkil{\"a}}, {Irawati},
  {Kennedy}, {Killestein}, {Kuncarayakti}, {Levan}, {Littlefair},
  {Makrygianni}, {Marsh}, {Mata-Sanchez}, {Mattila}, {Maund}, {McCormac},
  {Mkrtichian}, {Mullaney}, {Noysena}, {Patel}, {Rol}, {Sawangwit}, {Stanway},
  {Starling}, {Str{\o}m}, {Tooke}, {West}, {White}, \& {Wiersema}}]{Steeghs22}
{Steeghs}, D., {Galloway}, D.~K., {Ackley}, K., {et~al.} 2022, \mnras, 511,
  2405

\bibitem[{{Stritzinger} {et~al.}(2024){Stritzinger}, {Fraser}, {Pastorello},
  {Hoogendam}, {Morrell}, \& {Guti{\'e}rrez}}]{24abvb_classification}
{Stritzinger}, M., {Fraser}, M., {Pastorello}, A., {et~al.} 2024, Transient
  Name Server Classification Report, 2024-4674, 1

\bibitem[{{Tauris} {et~al.}(2017){Tauris}, {Kramer}, {Freire}, {Wex}, {Janka},
  {Langer}, {Podsiadlowski}, {Bozzo}, {Chaty}, {Kruckow}, {van den Heuvel},
  {Antoniadis}, {Breton}, \& {Champion}}]{Tauris2017}
{Tauris}, T.~M., {Kramer}, M., {Freire}, P.~C.~C., {et~al.} 2017, \apj, 846,
  170

\bibitem[{{Tauris} {et~al.}(2015){Tauris}, {Langer}, \&
  {Podsiadlowski}}]{Tauris2015}
{Tauris}, T.~M., {Langer}, N., \& {Podsiadlowski}, P. 2015, \mnras, 451, 2123

\bibitem[{{Tonry} {et~al.}(2024){Tonry}, {Denneau}, {Weiland}, {Siverd},
  {Erasmus}, {Koorts}, {Jordan}, {Suc}, {Smartt}, {Smith}, {Young}, {Nicholl},
  {Fulton}, {McCollum}, {Moore}, {Weston}, {Sheng}, {Angus}, {Wilson}, {Aamer},
  {Magill}, {Ramsden}, {Shingles}, {Srivastav}, {Gillanders}, {Stevance},
  {Cooper}, {Stoppa}, {Rhodes}, {Rest}, {Chen}, {Stubbs}, {Sommer}, \&
  {Schmidt}}]{24abvb_discovery}
{Tonry}, J., {Denneau}, L., {Weiland}, H., {et~al.} 2024, Transient Name Server
  Discovery Report, 2024-4579, 1

\bibitem[{{Tonry} {et~al.}(2018){Tonry}, {Denneau}, {Heinze}, {Stalder},
  {Smith}, {Smartt}, {Stubbs}, {Weiland}, \& {Rest}}]{2018PASP..130f4505T}
{Tonry}, J.~L., {Denneau}, L., {Heinze}, A.~N., {et~al.} 2018, \pasp, 130,
  064505

\bibitem[{Tran(1995)}]{Tran1995}
Tran, H.~D. 1995, The Astrophysical Journal, 440, 565

\bibitem[{Vasylyev {et~al.}(2023)Vasylyev, Yang, Filippenko, Patra, Brink,
  Wang, Chornock, Margutti, Gates, Burgasser, Karpoor, LeBaron, Softich,
  Theissen, Wiston, \& Zheng}]{Vasylyev2023}
Vasylyev, S.~S., Yang, Y., {et~al.} 2023, ApJ, 955, L37

\bibitem[{{Vernet} {et~al.}(2011){Vernet}, {Dekker}, {D'Odorico}, {Kaper},
  {Kjaergaard}, {Hammer}, {Randich}, {Zerbi}, {Groot}, {Hjorth}, {Guinouard},
  {Navarro}, {Adolfse}, {Albers}, {Amans}, {Andersen}, {Andersen}, {Binetruy},
  {Bristow}, {Castillo}, {Chemla}, {Christensen}, {Conconi}, {Conzelmann},
  {Dam}, {de Caprio}, {de Ugarte Postigo}, {Delabre}, {di Marcantonio},
  {Downing}, {Elswijk}, {Finger}, {Fischer}, {Flores}, {Fran{\c{c}}ois},
  {Goldoni}, {Guglielmi}, {Haigron}, {Hanenburg}, {Hendriks}, {Horrobin},
  {Horville}, {Jessen}, {Kerber}, {Kern}, {Kiekebusch}, {Kleszcz}, {Klougart},
  {Kragt}, {Larsen}, {Lizon}, {Lucuix}, {Mainieri}, {Manuputy}, {Martayan},
  {Mason}, {Mazzoleni}, {Michaelsen}, {Modigliani}, {Moehler}, {M{\o}ller},
  {Norup S{\o}rensen}, {N{\o}rregaard}, {P{\'e}roux}, {Patat}, {Pena}, {Pragt},
  {Reinero}, {Rigal}, {Riva}, {Roelfsema}, {Royer}, {Sacco}, {Santin},
  {Schoenmaker}, {Spano}, {Sweers}, {Ter Horst}, {Tintori}, {Tromp}, {van
  Dael}, {van der Vliet}, {Venema}, {Vidali}, {Vinther}, {Vola}, {Winters},
  {Wistisen}, {Wulterkens}, \& {Zacchei}}]{XShooter}
{Vernet}, J., {Dekker}, H., {D'Odorico}, S., {et~al.} 2011, \aap, 536, A105

\bibitem[{Warwick {et~al.}(2025)Warwick, Lyman, Pursiainen, Coppejans, Galbany,
  Jones, Killestein, Kumar, Oates, Ackley, Anderson, Aryan, Breton, Chen,
  Clark, Dhillon, Dyer, Gal-Yam, Galloway, Gutiérrez, Gromadzki, Inserra,
  Jiménez-Ibarra, Kelsey, Kotak, Kravtsov, Kuncarayakti, Magee, Matilainen,
  Mattila, Müller-Bravo, Nicholl, Noysena, Nuttall, O'Brien, O'Neill, Pallé,
  Pessi, Petrushevska, Pignata, Pollacco, Ragosta, Ramsay, Sahu, Sahu, Singh,
  Sollerman, Stanway, Starling, Steeghs, Teja, \& Ulaczyk}]{Warwick2025}
Warwick, B., Lyman, J., Pursiainen, M., {et~al.} 2025, Monthly Notices of the
  Royal Astronomical Society, 536, 3588

\bibitem[{{Watkins} {et~al.}(2024){Watkins}, {Kaviraj}, {Collins}, {Knapen},
  {Kelvin}, {Duc}, {Rom{\'a}n}, \& {Mihos}}]{2024MNRAS.528.4289W}
{Watkins}, A.~E., {Kaviraj}, S., {Collins}, C.~C., {et~al.} 2024, \mnras, 528,
  4289

\bibitem[{Wichern {et~al.}(2025)Wichern, Leloudas, Pursiainen, Cikota,
  Jaisawal, Charalampopoulos, Bulla, Dai, Anderson, Gromadzki, Gutiérrez,
  Müller-Bravo, \& Nicholl}]{Wichern2025}
Wichern, H. C.~I., Leloudas, G., Pursiainen, M., {et~al.} 2025, arXiv,
  arXiv:2510.27007

\bibitem[{{Wu} {et~al.}(2024){Wu}, {Zha}, {Cai}, {Zhang}, {Yang}, {Xiang},
  {Lin}, {Wang}, \& {Wang}}]{2024ApJ...967L..45W}
{Wu}, C., {Zha}, S., {Cai}, Y., {et~al.} 2024, \apjl, 967, L45

\bibitem[{{Zimmerman} {et~al.}(2026){Zimmerman}, {Gal-Yam}, {Groot}, {Ofek},
  {van Roestel}, {Pastorello}, {Valenti}, {Ravi}, {Chen}, {Schulze},
  {Blagorodnova}, {Wavasseur}, {Gomez-Munoz}, {Tranin}, {de Wet}, {Leloudas},
  {Vreeswijk}, {Kwok}, {Schwab}, {Jha}, {Maguire}, {Sand}, {Stringer},
  {Kupfer}, {Faran}, {Anderson}, {Andrews}, {Andrews}, {Badash}, {Bloemen},
  {Bostroem}, {Chen}, {Della Valle}, {Dimitriadis}, {Dong}, {Farah},
  {Gillanders}, {Godson}, {Gromadzki}, {Hiramatsu}, {Hoang}, {Howell},
  {Janzen}, {Kuncarayakti}, {Li}, {Lyman}, {Maeda}, {Magee}, {McCully},
  {Mehta}, {Milligan}, {Moran}, {Ni}, {O'Neill}, {Pearson}, {Pieterse},
  {Pignata}, {Reguitti}, {Reichart}, {Meza Retamal}, {Santos}, {Scaringi},
  {Shrestha}, {Srivastav}, {Stoppa}, {Subrayan}, {Valerin}, {Wang}, {Wynn},
  {Yaron}, \& {Zang}}]{2026arXiv260209096Z}
{Zimmerman}, E.~A., {Gal-Yam}, A., {Groot}, P.~J., {et~al.} 2026, arXiv
  e-prints, arXiv:2602.09096

\end{thebibliography}
\bibliographystyle{aa} 

\begin{appendix} 

\section{Additional Tables and Figures}

\begin{figure}[h!]
    \centering
    \includegraphics[width=\columnwidth]{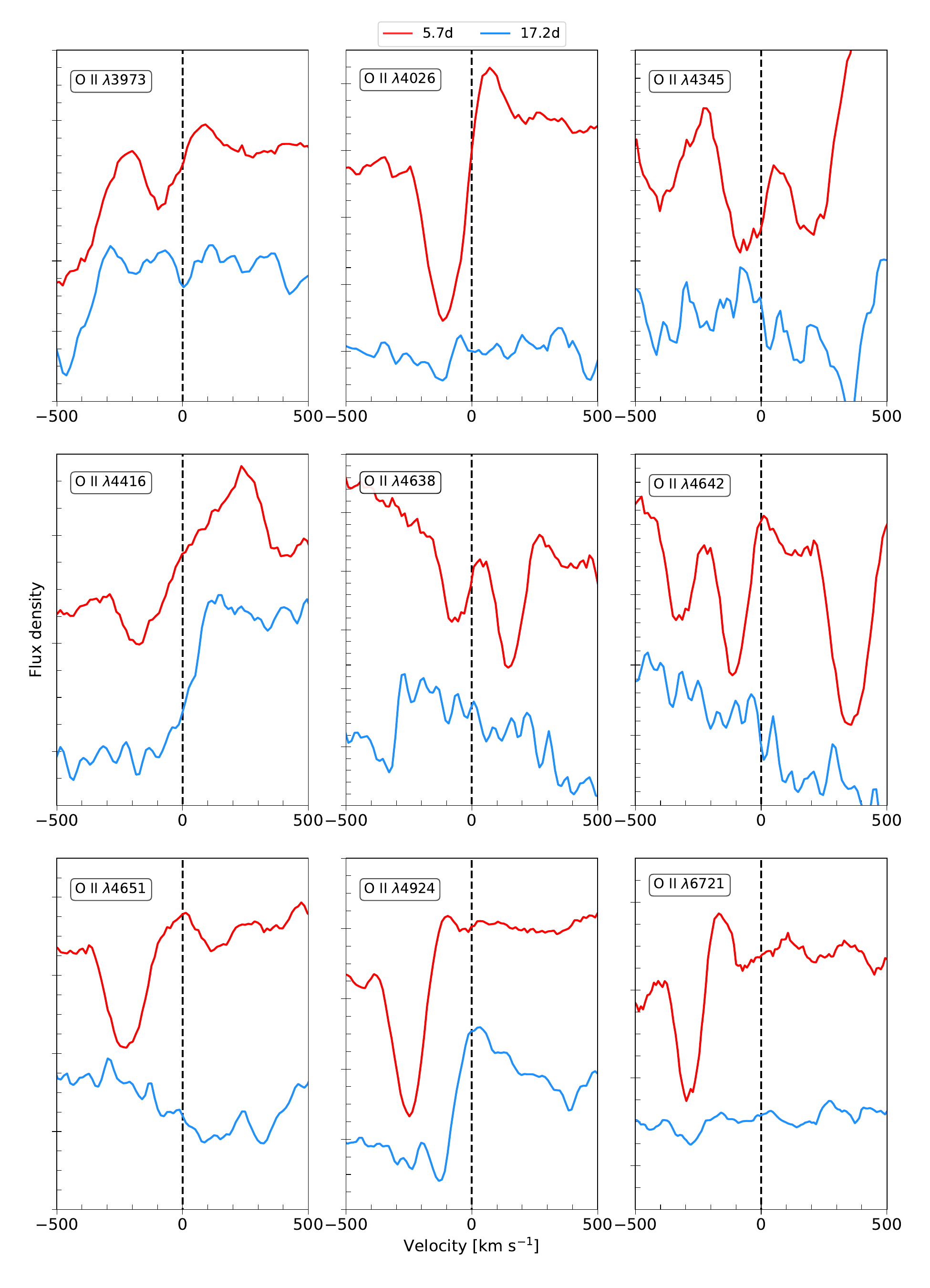}
    \caption{Same as Fig.\ref{fig:hevel} but for  O~{\sc ii}.}
    \label{fig:ovel}
\end{figure}

\begin{figure}[h!]
    \centering
    \includegraphics[width=\columnwidth]{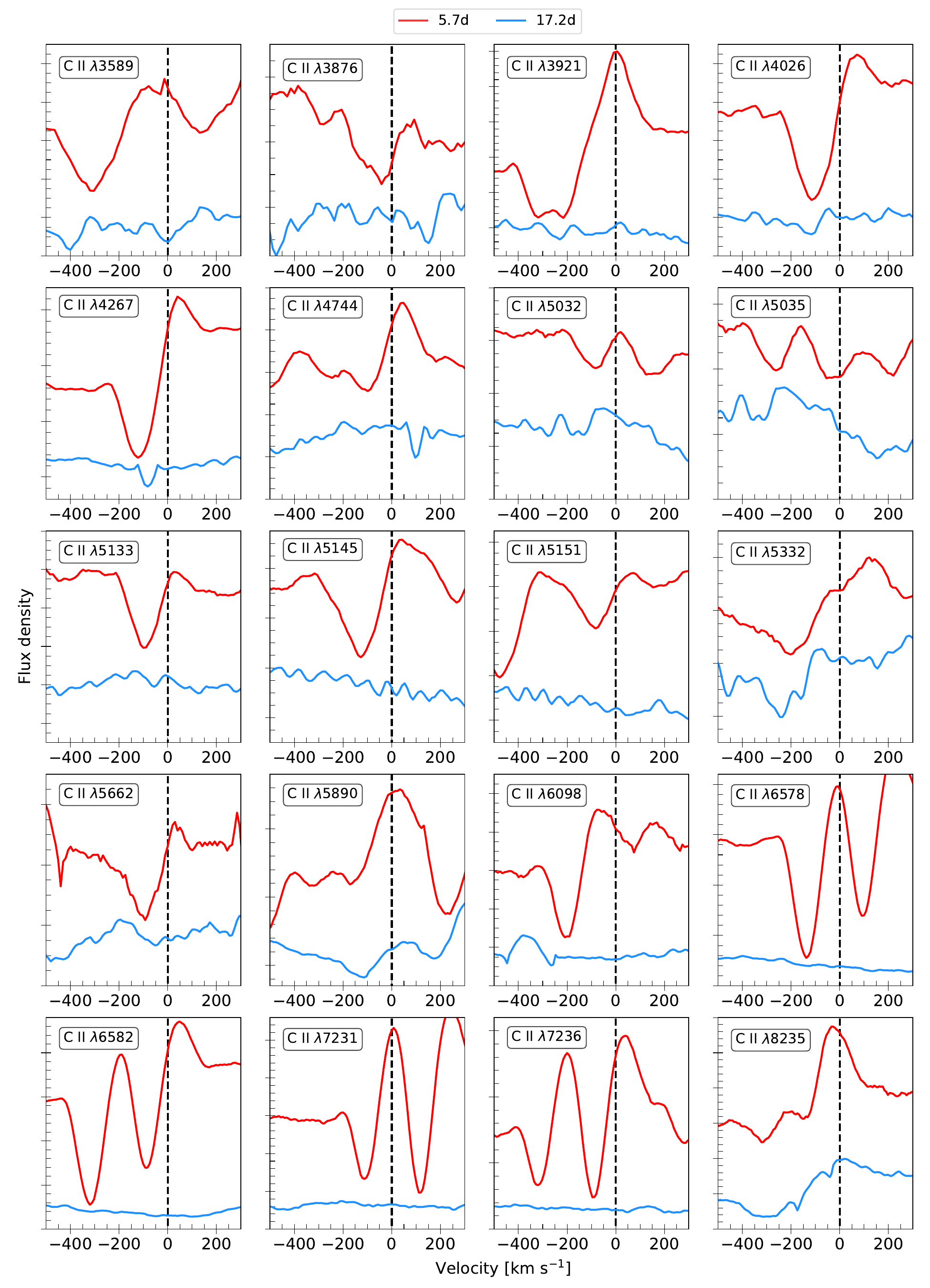}
    \caption{Same as Fig.\ref{fig:hevel} but for  C~{\sc ii}.}
    \label{fig:cvel}
\end{figure}

\begin{figure}[h!]
\centering
\includegraphics[width=\columnwidth]{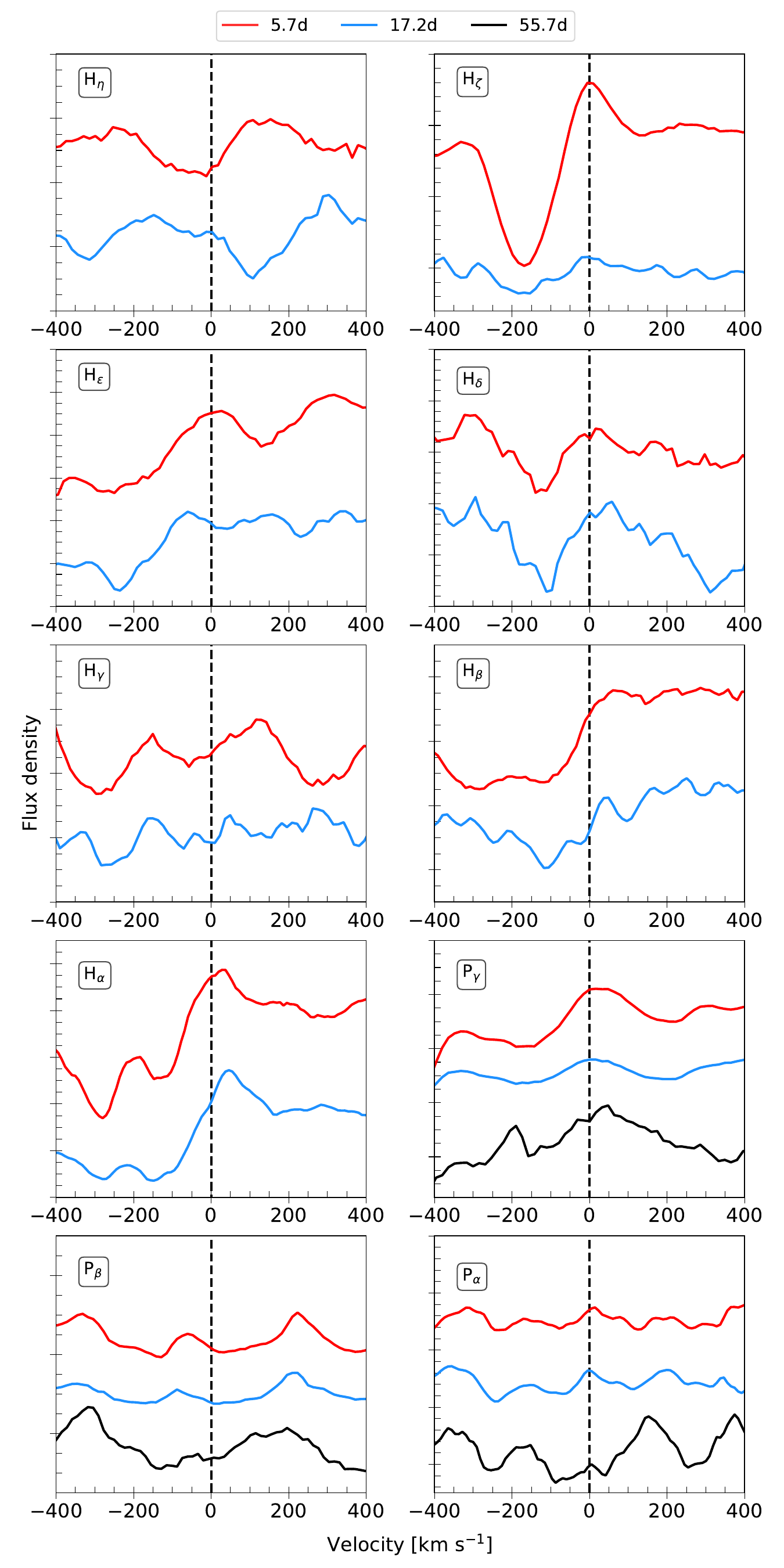}
\caption{Same as Fig.\ref{fig:hevel} but for H and including the $\phi=55.7$ spectrum for NIR lines.}
\label{fig:hvel}
\end{figure}

\begin{table}[h!]
\caption{Spectroscopic Observations}             
\label{table:specdata}      
\centering          
\begin{tabular}{l c c c l }  
\hline\hline       
 MJD & Phase$^{*}$ (days) & Range (\AA) & Resolution (\AA) & Instrumental configuration\\ 
\hline                    
    60643.05 & -2.0 & 2940 -- 3715 / 4610 -- 5535 \&\/ 5630 -- 6535 & 0.11 / 0.18 & UVES (346+580)\\  
    60643.06 & -2.0 & 4280 -- 8325    & 22.40  & FORS2 (300V+GG345)\\
    60651.02 & 5.7 & 2890 -- 19830    & 0.80 / 0.88 / 3.12  & XShooter (UVB+VIS+NIR)\\
    60663.03 & 17.2 & 2890 -- 19830    & 0.80 / 0.88 / 3.12   & XShooter (UVB+VIS+NIR)\\
    \begin{tabular}{@{}c@{}}60701.03 \\ 60703.03 \\ 60705.03 \end{tabular} & 55.7 (median) & 2890 -- 19830    & 0.80 / 0.88 / 3.12   & XShooter (UVB+VIS+NIR)\\
\hline                  
\end{tabular}
\\$^{*}$ Phase with respect to the maximum light
\end{table}

\begin{figure}[h!]
    \centering
    \includegraphics[width=0.98\linewidth]{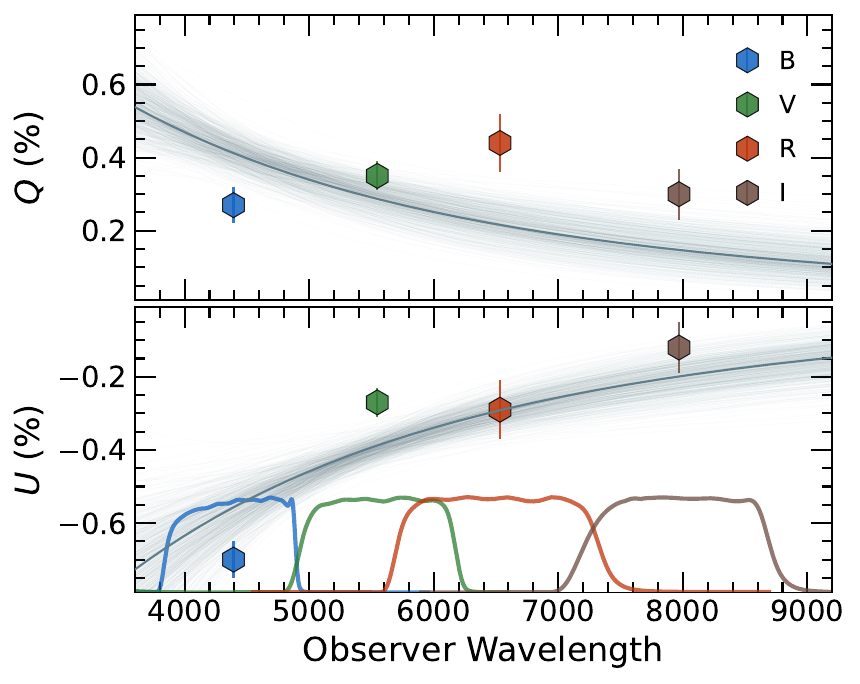}
    \caption{The Serkowski fit to the $BVRI$ ISP estimated based on the FORS2 broadband polarimetry. The best-fitting model (thick solid line), as well as model fits to randomised 1000 Monte Carlo resamplings, are shown. }
    \label{fig:serkowski_fit}
\end{figure}

\section{UVES and XShooter spectra line identifications}
\label{appendix:uves}

Line IDs taken from Moore 1945
\begin{figure*}[h!]
\includegraphics[width=\textwidth]{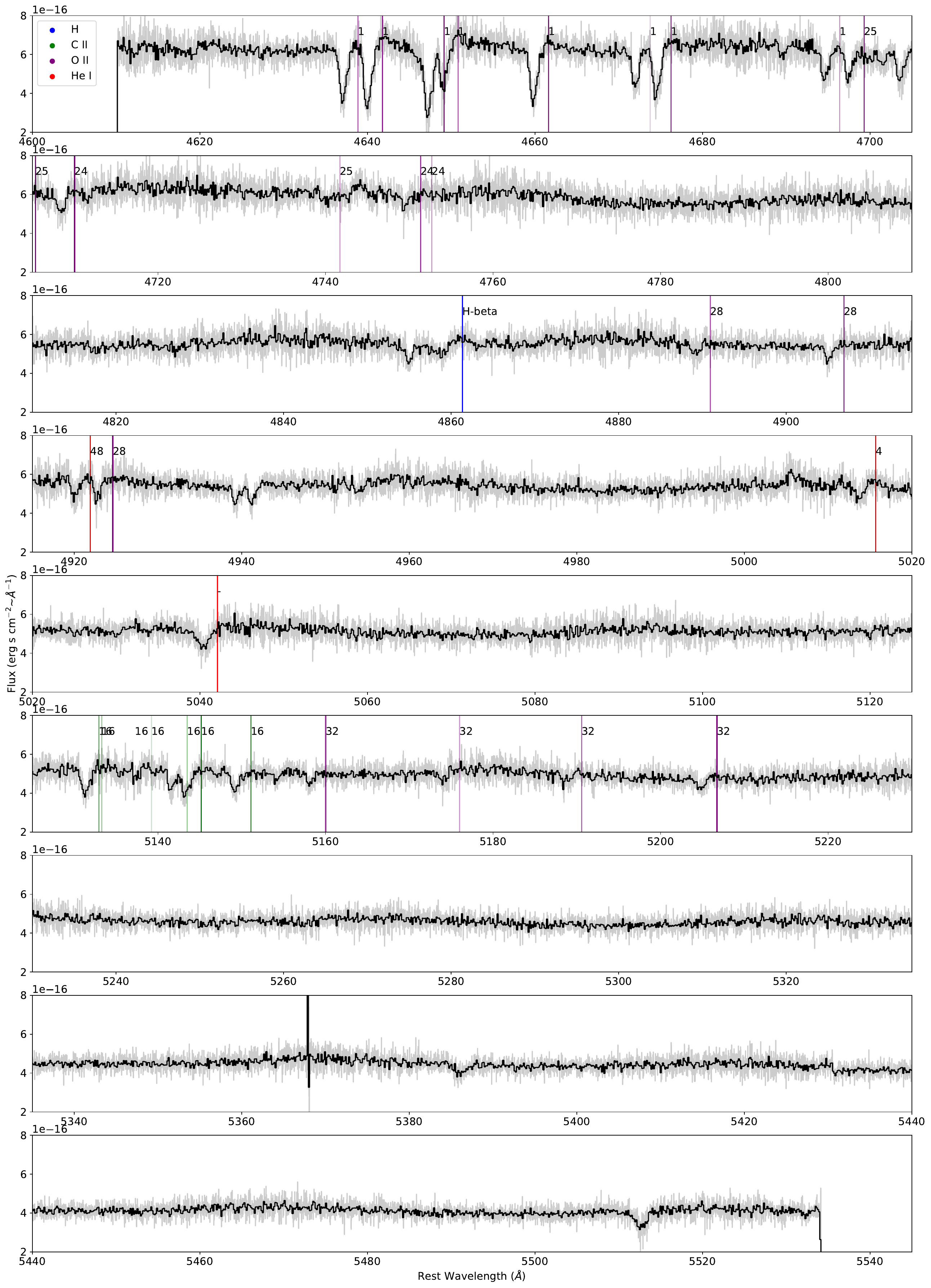}
\caption{Zoom-in view of the UVES spectrum. Identified lines are marked at their rest wavelength, and for metallic lines labeled with the multiplet according to \cite{Moore45}. The transparency of the coloured line marking each component within a multiplet is scaled according to the relative strength of the transition with respect to other components in the multiplet.}
\label{fig:uves1}
\end{figure*}

\begin{figure*}[h!]
\includegraphics[width=\textwidth]{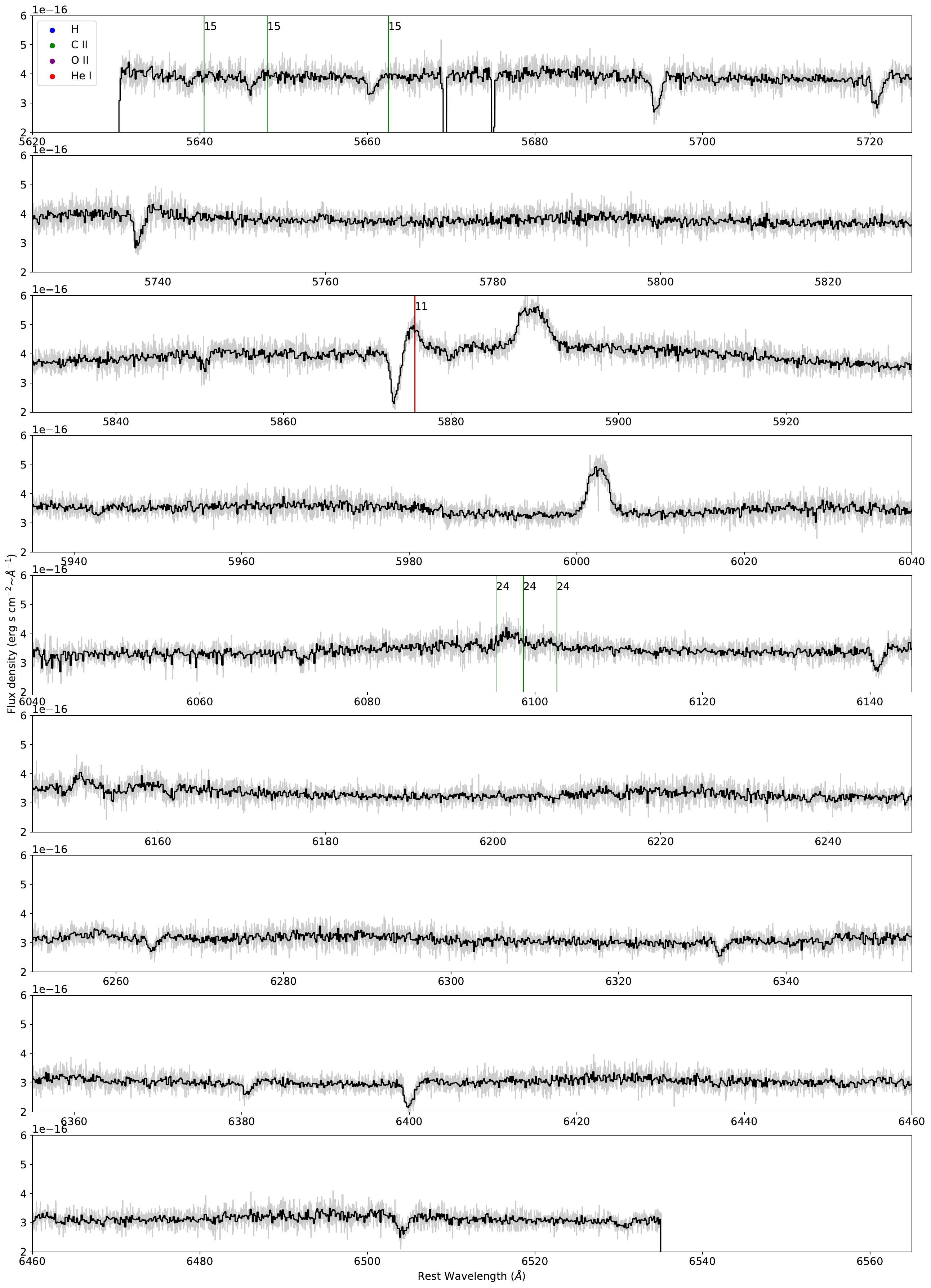}
\caption{Same as Fig. \ref{fig:uves1}.}
\label{fig:uves2}
\end{figure*}

\begin{figure*}[h!]
\centering
\includegraphics[width=0.8\textwidth]{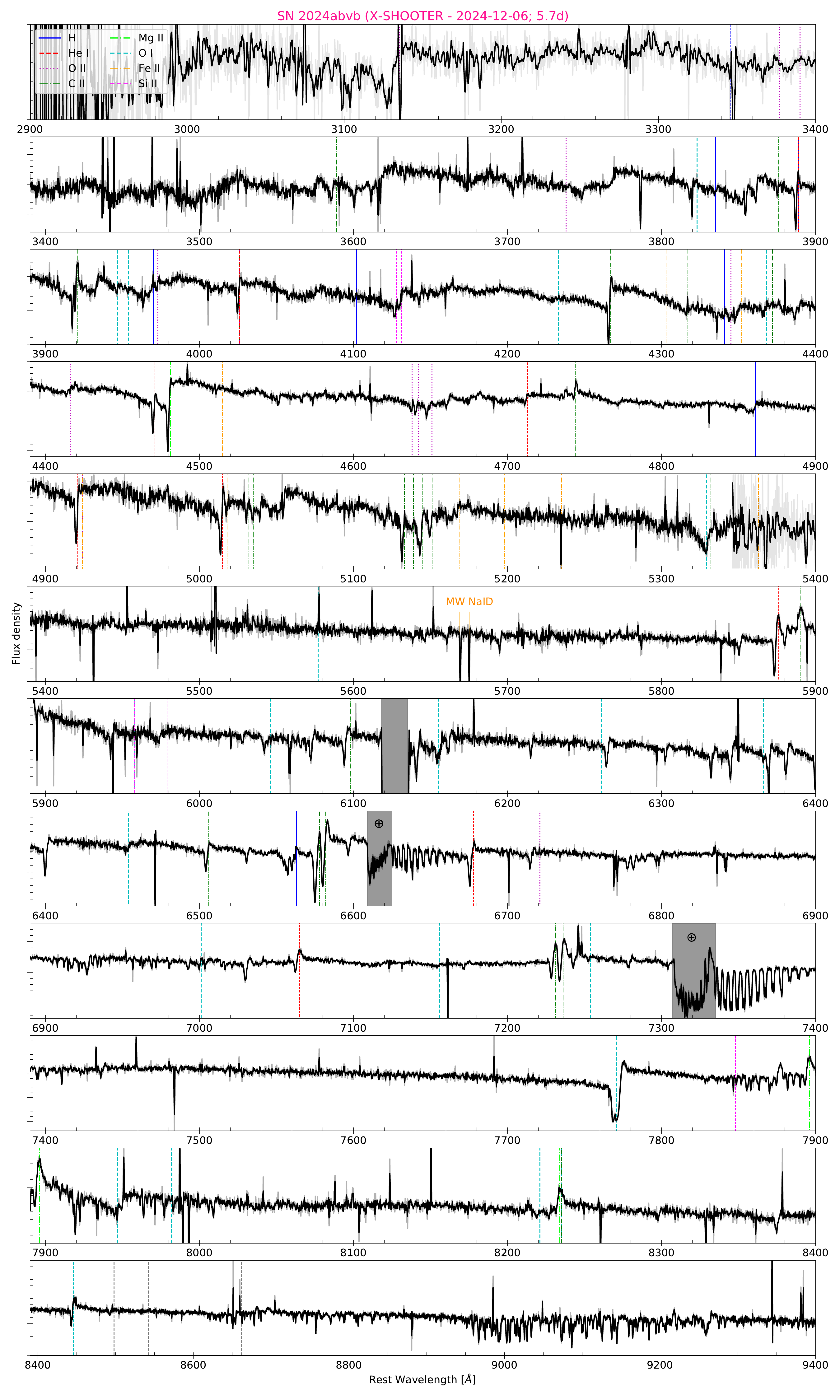}
\caption{Zoom-in view of the XShooter ($\phi = 5.7$d) spectrum. Identified lines are marked at their rest wavelength.}
\label{fig:XShooter1}
\end{figure*}

\begin{figure*}[h!]
\centering
\includegraphics[width=0.8\textwidth]{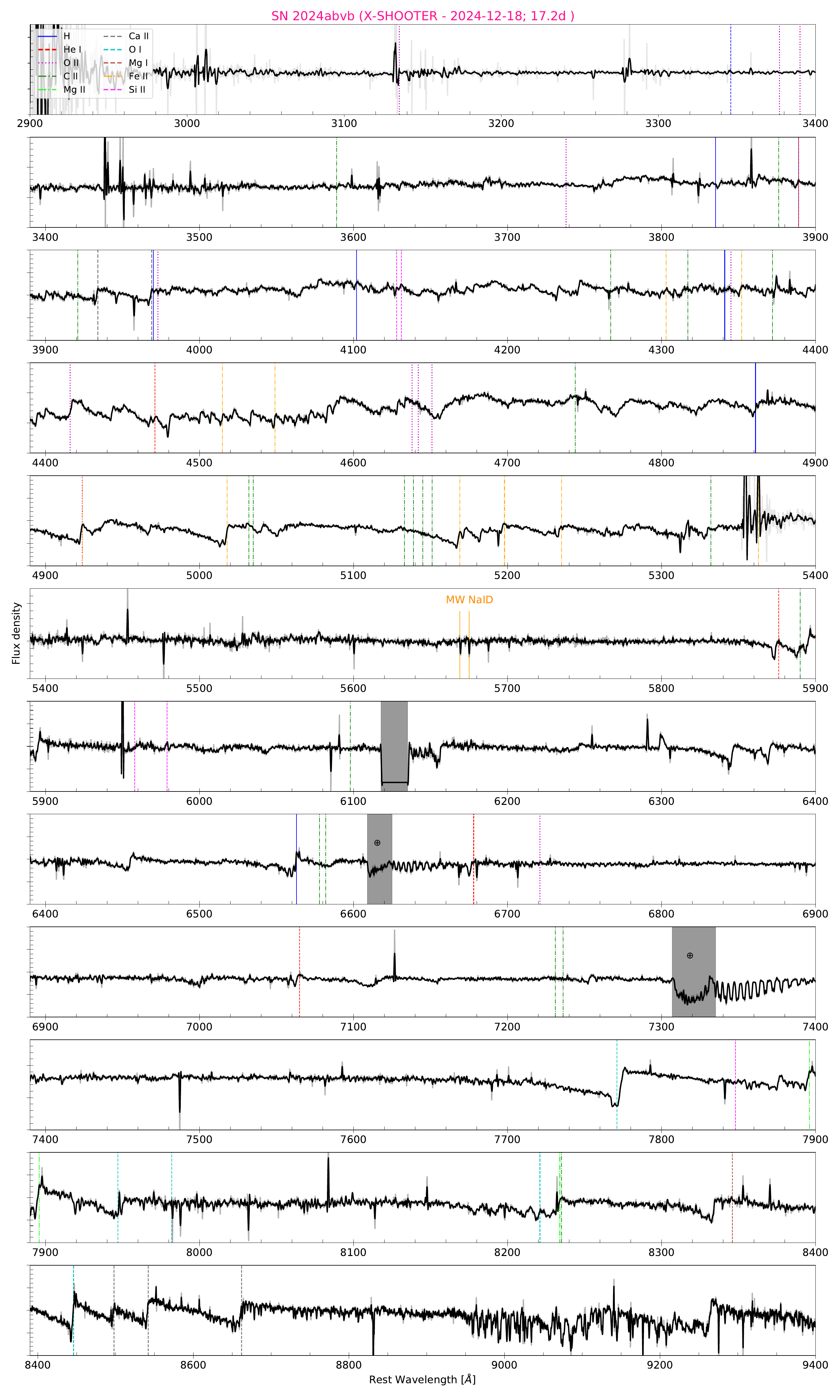}
\caption{Zoom-in view of the XShooter ($\phi = 17.2$d). Identified lines are marked at their rest wavelength.}
\label{fig:XShooter2}
\end{figure*}

\begin{figure*}[h!]
    \centering
    \includegraphics[width=\textwidth]{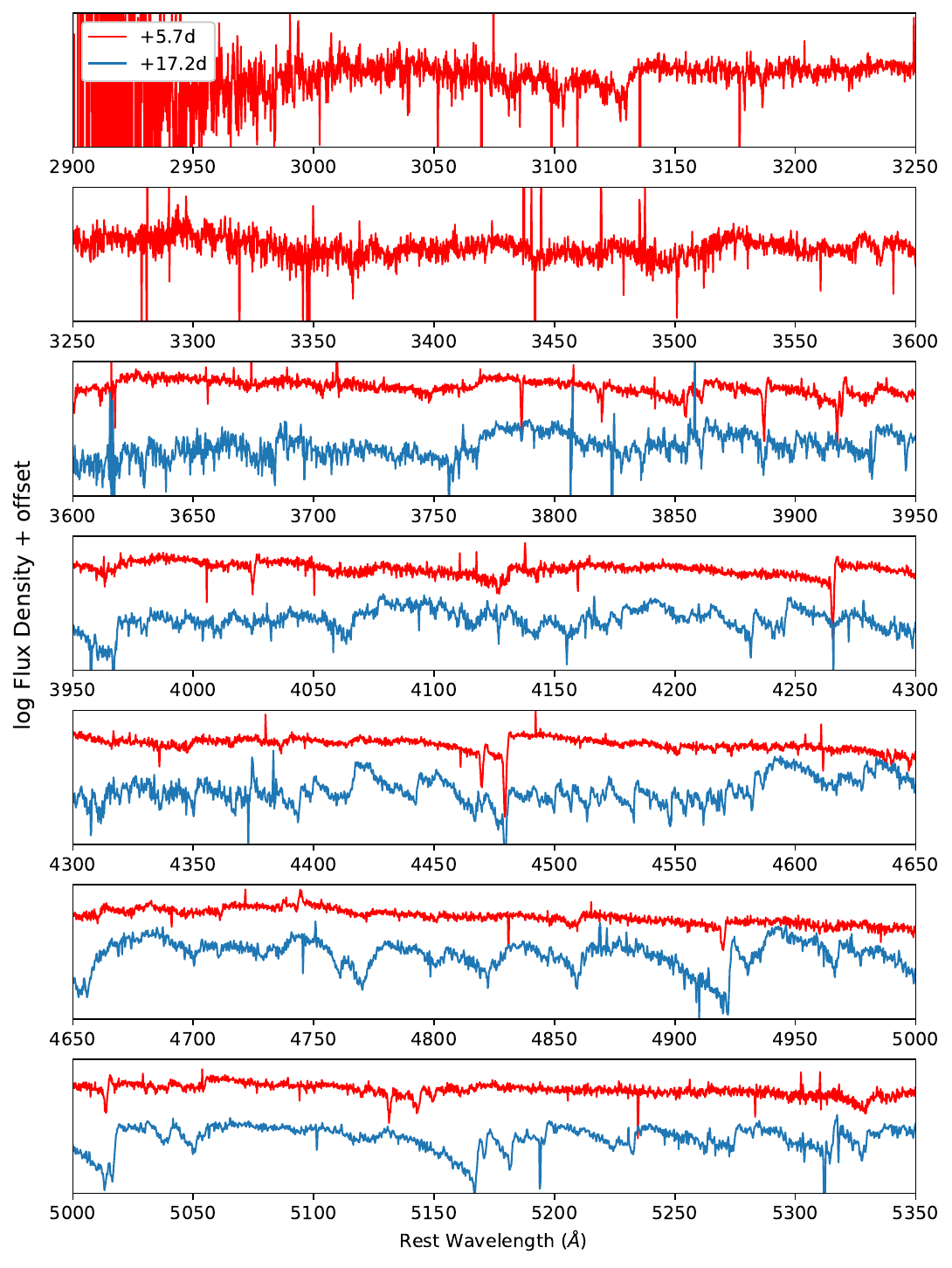}
    \caption{Spectral comparison of UVB arms.}
    \label{fig:spec_uvb}
\end{figure*}

\end{appendix}

\end{document}